\documentclass[a4paper,11pt,USenglish,runningheads,xcolor]{llncs}
\usepackage{graphicx}
 
\usepackage{tikz}
\usetikzlibrary{positioning,shapes,chains,backgrounds,arrows,chains,fit,snakes,patterns,shapes.geometric,shapes.symbols}
\usepackage{pgfplots}
\usepackage{subfigure}
\usepackage[T1]{fontenc}
\usepackage{amsmath}
\usepackage{amsfonts}
\usepackage{amssymb}
\usepackage{flexisym}
\usepackage{amsfonts}
\usepackage{bm}

\newcommand{\comment}[1]{}
\usepackage{afterpage}
\usepackage{lscape}
\usepackage{longtable}

\newcommand{\written}[1]{\mathcal{#1}}
\newcommand{\Nat}{\mathbb{I\!\!N}}
\newcommand{\bound}{\mathbb{K}}

\newcommand{\bond}{\!-\!}

\newcommand{\connected}{\mathsf{con}}
\newcommand{\es}{\emptyset}
\newcommand{\state}[2]{\langle {#1}, {#2}\rangle}
\newcommand{\bondins}[2]{\langle {#1},{#2}\rangle}

\newcommand{\trans}[1]{\ensuremath{\stackrel{#1}{\longrightarrow}}}
\newcommand{\guard}[1]{\mathsf{pre}(#1)}
\newcommand{\effects}[1]{\mathsf{post}(#1)}

\newcommand{\effect}[1]{\mathsf{effect}(#1)}
\newcommand{\beffect}[1]{\mathsf{b.effect}(#1)}
\newcommand{\first}[1]{\mathsf{last}(#1)}

\newcommand{\RPN}{reversing Petri net } 
\newcommand{\btrans}[1]{\ensuremath{\stackrel{#1}{\rightsquigarrow}_{b}}}
\newcommand{\ctrans}[1]{\ensuremath{\stackrel{#1}{\rightsquigarrow}_{c}}}
\newcommand{\otrans}[1]{\ensuremath{\stackrel{#1}{\rightsquigarrow}_{o}}}

\newcommand{\idle}{(\emptyset,\emptyset)}
\newcommand{\trord}{\prec} % Transitive order
\newcommand{\type}{\ell} %TODO Consider new notation
\newcommand{\bondpair}[2]{\langle #1,#2\rangle}

\begin{document}
\title{Formal Translation from Reversing Petri Nets \\to Coloured Petri Nets}

\author{
Kamila Barylska\inst{1},
Anna Gogoli{\'n}ska\inst{1},
{\L}ukasz Mikulski\inst{1},\\
Anna Philippou\inst{2},  Marcin Piatkowski\inst{1}, and
Kyriaki Psara\inst{2}
}
\institute{
Faculty of Mathematics and Computer Science \\
Nicolaus Copernicus University, 87-100 Toru\'n, Poland \\
\email{\scriptsize \{khama,anna.gogolinska,lukasz.mikulski,marcin.piatkowski\}@mat.umk.pl}
\and
Department of Computer Science, University of Cyprus\\
\email{\scriptsize \{anna.philippou,psara.kyriaki\}@ucy.ac.cy}
}

\maketitle

\begin{abstract}
Reversible computation is an emerging computing paradigm that allows any sequence of 
operations to be executed in reverse order at any point during computation. Its appeal 
lies in its potential for low-power computation and its relevance to a wide array of 
applications such as chemical reactions, quantum computation, robotics, and distributed systems. Reversing
Petri nets are a recently-proposed extension of Petri nets that implements the three 
main forms of reversibility, namely, backtracking, causal reversing, and out-of-causal-order
reversing. Their distinguishing feature is the use of named tokens that can be combined together to form bonds. Named tokens
along with a history function, constitute the means of \emph{remembering} past
behaviour, thus, enabling reversal. In recent work, we have proposed a~structural
translation from a subclass of RPNs to the model of Coloured Petri Nets (CPNs), an 
extension of traditional Petri nets where tokens carry data values. In this 
paper, we extend the translation to handle RPNs with token
multiplicity under the individual-token interpretation, a model which allows multiple tokens of the same type to exist in a
system. To support the three types of reversibility, tokens are associated with their causal history and, while tokens
of the same type are equally eligible to fire a transition when going forward, when 
going backwards they are able to reverse only the transitions they have previously 
fired. The new translation, in addition to lifting the restriction on 
token uniqueness, presents a refined approach for transforming RPNs to CPNs through a unifying approach that allows instantiating each of the three types of reversibility.  The paper also reports on 
a tool that implements this translation, paving the way for automated 
translations and analysis of reversible systems using CPN Tools. 
\end{abstract}

%----------------------------------------------------------
% Introduction and state of the art

\section{Introduction}
Reversible computation is a form of computing that allows 
operations to be seamlessly reversed at any point during the 
computational process. This unique capability has been attracting 
growing attention due to its potential applications in low-power 
computing, robotics, and distributed systems as well as applications
which naturally embed reversible behavior such as biological
systems and quantum mechanics.

Aiming to understand the foundations of reversibility, in recent
years work has been carried out towards the development of reversible
models and frameworks. This study has brought forward three main
forms of reversibility in the context of concurrent and distributed systems, namely, {\em backtracking}, allowing actions to be reversed in the exact
order in which they were executed,
\emph{causal-order} reversibility, a form of reversing where an action can 
be undone provided that all of its effects 
(if any) have been undone beforehand, and \emph{out-of-causal-order} 
reversibility, a form of reversing featured most notably in biochemical systems. 
These concepts have been studied 
within a variety of formalisms~\cite{RCCS,TransactionsRCCS,phillips2007reversing,LaneseMS16,ConRev,netsWithBonds,RPlaceTrans,CardelliL11}.

One line of research among these approaches has been the investigation of 
reversible behavior in Petri nets.  Initial studies 
considered the reversal of selected transitions of conventional Petri 
nets~\cite{PetriNets,BoundedPNs} and explored decidability
problems regarding reachability and coverability in the resulting Petri net. 
Given that this approach 
to reversibility violates causality, subsequent work in~\cite{Unbounded,investigating} 
investigated whether it is possible to add a complete set of effect-reverses for a given 
transition without changing the set of reachable markings, showing that this problem is in general undecidable.
In another line of work~\cite{RPlaceTrans} proposes a causal semantics for P/T nets by identifying the 
causalities and conflicts of a P/T net through unfolding it into an equivalent occurrence 
net and subsequently introducing appropriate reverse transitions to create a coloured Petri
net (CPN) that captures a causal-consistent reversible semantics. 
On a similar note,~\cite{RON} introduces the notion of reversible occurrence nets and 
associates a reversible occurrence net to a causal reversible prime event structure, and 
vice versa. 

In this work, we focus on Reversing Petri
Nets~\cite{PP,netsWithBonds} (RPNs), a reversible model inspired by Petri nets 
that allows the modelling
of reversibility as realised by backtracking, causal-order, and out-of-causal-order
reversing.  A key challenge when reversing computations in Petri
nets is handling \emph{backward conflicts}. These conflicts arise
when tokens occur in a certain place due to different causes
making unclear which transitions ought to be reversed.
To handle this ambiguity, RPNs introduce the notion of a \emph{history}
of transitions, which records causal information of
executions. Furthermore, inspired by biochemical systems as well
as other resource-aware applications, the model 
employs named tokens that can be connected together
to form bonds, and are preserved during execution. The usefulness
of the framework and its extensions~\cite{ExpressPP,coll} was illustrated 
in a number of examples including
the modelling of long-running transactions with compensations~\cite{KP-2020}, a 
signal-passing mechanism used by the ERK pathway, and an application to Massive MIMO~\cite{RC19}, and they have been translated to 
Answer Set Programming (ASP), a declarative programming framework with 
competitive solvers~\cite{ASPtoRPNs}. One of the extensions of RPNs concerns the introduction of
multiple tokens~\cite{ExpressPP} according to the individual-token 
interpretation. The aim of this extension has been
to allow multiple tokens of the same type to exist within a net while 
distinguishing them based on their causal path. This implies that tokens 
of the same type are equally eligible to fire a transition when going forward, however, when
going backwards they are able to reverse only the transitions they have
previously fired.  

A challenge that arises is to explore the relationship between RPNs and 
classical Petri nets. Of particular interest is how the global control 
imposed via the history construct in RPNs can be captured by the strictly 
local semantics of Petri nets. To this effect, in our previous work~\cite{BGMPPP} we have 
provided a translation from a subclass of acyclic RPNs into coloured Petri 
nets (CPNs)~\cite{CPN}, an extension of traditional Petri nets where tokens can carry data values.
The study establishes that RPNs can be encoded into CPNs, demonstrating that the principles of reversible computation can be directly encoded in the traditional model. The translation involves a structural transformation from RPNs to CPNs, introducing both forward and backward instances for each transition. The translation succeeds in maintaining a local approach by carefully storing histories and causal dependencies of executed transition sequences in additional places.
Furthermore, the introduction of cycles to the RPN model and the 
consequences to the CPN translation were considered in~\cite{BG}. 
In the current work, we go a step further and consider the RPN extension 
with multiple tokens. Specifically, we propose a translation of a subclass 
of RPNs with multiple tokens to CPNs. Similarly
to previous work, the resulting translation succeeds in capturing the 
global control merely at the local level of additional places, while lifting restrictions on token uniqueness 
Furthermore, the current translation extends beyond prior 
work by presenting a refined approach that unifies the 
transformation process for the three types of reversibility, making it more
flexible and comprehensive. The correctness of the translation is formally proved establishing the correspondence between the states of an RPN
and its CPN translation as well as how a transition in a state of an
RPN can be matched by a transition of any corresponding state of the  CPN, leading to equivalent states, and vice versa.
An additional contribution of the present paper lies in the practical 
implementation of the transformation, realized through a tool. This tool 
facilitates the analysis of reversible systems using CPN Tools~\cite{CPNtools}, providing a 
bridge between theoretical models and practical applicability. We note that a previous version of this work has appeared in~\cite{RC2022}. The present paper constitutes an extension of that work with a refinement of the machinery and the complete proofs of the results.

{\bf Paper organization} Following some preliminary definitions, Sections~3 
and~4 provide an overview of reversing Petri nets and a description of coloured Petri nets. Sections 5 
and 6 define the necessary machinery for transforming RPNs to CPNs and establish 
formally the correctness of the translation. 
Section 7 presents the tool that has been developed implementing the transformation and 
providing a connection between RPNs to CPN Tools. The paper concludes by a summary of the contributions and a discussion of future work.

\section{Preliminaries}\label{sec:prel}

The set of non-negative integers is denoted by $\Nat$. 
Given a set X, the cardinality (number of elements) of $X$ is denoted by $\#X$, the powerset (set of all subsets) by~$2^X$ -- the cardinality of the powerset is $2^{\#X}$. 
Multisets over $X$ are members of $\Nat^X$, i.e., functions from $X$ into $\Nat$. 
%For convenience and readability, if the set $X$ is finite, 
%multisets in $\Nat^X$ will be represented twofold, depending on the context: 
%by vectors of $\Nat^{\#(X)}$ (assuming a fixed ordering of the set $X$) 
%or by adequate linear combinations of arguments.

In what follows every function $f:X\rightarrow Y$ might be extended in a~natural way to the domain $2^X$.

\begin{definition}\label{d:multiset}
The monoid $\Nat^X$, for a set $X$, is the set of multisets over $X$ with componentwise addition~$+$.

If $y,z\in\Nat^X$ then 
$(y+z)(x)=y(x)+z(x)$ for every $x\in X$.
For $Y,Z\subseteq\Nat^X$ we define $Y+Z=\bigcup\{y+z\mid y\in Y,z\in Z\}$.
The partial order $\leq$ is understood componentwise, while~$<$ means~$\leq$ and~$\neq$.
\end{definition} 

\section{Reversing Petri nets}
In this section we recall the model of reversing Petri nets (RPNs) defined in~\cite{netsWithBonds} and specifically their extension with multiple tokens of \cite{ExpressPP}, together with descriptions of the three reversibility semantics
which these nets support in addition to the standard forward execution, namely \emph{backtracking}, \emph{causal reversing} and   \emph{out-of-causal-order reversing}.
Due to the lack of space, we will only focus here on the most important concepts and facts. We refer the reader to \cite {netsWithBonds,ExpressPP,KP-2020}, where RPNs are presented in detail.

\comment{In this section we recall the model of reversing Petri nets (RPNs) defined in~\cite{PP}. RPNs are a mathematical and graphical model associated with formal semantics that alows the execution of Petri nets to run  in both the forward and backward directions (see~\cite{PP} for details). 

RPNs, besides the standard forward execution, allow three forms of reversible execution. 
The first approach, backtracking, is the most restrictive one, since it only  allows reversing  the last transition which was executed during the forward computation. 
The second approach, causal reversing, is  less restrictive than backtracking since it allows a transition to 
rollback if all its effects, if any, have been undone beforehand. 
Finally, out-of-causal-order reversing is the most flexible form since it allows actions to be reversed in any order as long as they have been executed. 

Undoing an arbitrary 
transition during the execution of a Petri net requires close monitoring of token manipulation within the transitions of a net. The effect on the tokens of each transition should be clearly expressed in order to be able to undo such an effect. Therefore we distinguish each token on a net along 
with its causal path, i.e. the places and transitions it has traversed before reaching its current state. 
This requirement becomes complicated when transitions consume multiple tokens but 
release only a subset of the consumed tokens or even a set of new ones.  To resolve this, a}

A \RPN is built on the basis
of a set of bases or simply tokens that correspond to the basic entities that occur in a system.
These tokens are persistent, cannot be consumed, and can be combined together as the effect 
of transitions via so-called {\em bonds}  into coalitions (also called \emph{molecules}) that record the computational history
of each token.
This approach is similar to reaction systems from biochemistry but can be applied
to a wide range of systems that feature reversible behaviour. Based on this intuition,
reversing Petri nets are defined as follows:

\begin{definition}{\rm
 A \emph{reversing Petri net} (RPN) is a tuple $(P,T, F, A, B)$ where:
\begin{enumerate}
\item $P$ is a finite set of \emph{places},
\item $T$ is a finite set of \emph{transitions},
\item $F : (P\times T  \cup T \times P)\rightarrow\Nat^{{{A}}\cup {{B}}\cup \overline{A} \cup \overline{B}}$ is 
a set of directed \emph{arcs}, 
\item $A$ is a finite set of \emph{base} or \emph{token types} ranged over
by $a$, $b$,\ldots. Instances of tokens of type $a$ are denoted by $a_i$ where $i$ is unique and
 the set of all instances of token types in $A$  is denoted by ${\cal{A}}$. Furthermore, we write  
$\overline{A} = \{\overline{a}\mid a\in A\}$ for the set containing a 
 ``negative" form for each token type\footnote{Elements of $A$ signify the presence of the base type, when elements of $\overline{A}$
the absence of it.},
\item $B\subseteq A\times A$ is a set of undirected \emph{bond types}. 
We assume that the relation $B$ is symmetric.
The two elements $(a, b), (b, a) \in B$ would be represented by one element only and
denoted by $\bondins{a}{b}$. We also use the notation $a \bond b$
for a bond type $\bondins{a}{b}\in B$\footnote{$\bondins{a}{b}$ may be also understood as two elements multiset over $A$.}.
Similarly, instances of bonds are elements of $\cal{A} \times \cal{A}$ and denoted by 
$\bondins {a_i} {b_j}$ or $a_i \bond b_j$ for $a_i,b_j\in {\cal{A}}$. 
The set of all 
instances of bond types in $B$ is denoted by~${\cal{B}}$.
$\overline{B} = \{\overline{\beta}\mid \beta\in B\}$
contains the corresponding ``negative" form for each bond type. % type and we write ${\cal{B}}=B \cup \overline{B}$.
% ({\color{red} Negative instances could be replaced by writing e.g. $(a,0)\in F(x,t)$.})
\end{enumerate}
}\end{definition}

The first two clauses of the definition state the sets of \emph{places} and \emph{transitions}, which are understood in a standard way 
(see~\cite{PN}). To preserve individuality whilst adding token multiplicity we introduce ${\cal{A}}$ and ${\cal{B}}$ which are the sets of token and bond instances,
where for each token type $a$ of a set of types $A$, instances of that type are indicated as $a_i$, $i$ being a unique index.   Intuitively, the incoming and outgoing arcs of a transition indicate the requirements of a transition firing and the effects of its execution, respectively. The negative types of tokens  $\overline{A}$ or $\overline{B}$ indicate the requirement of token absence  and  they make sense only on incoming arcs of a transition, hence $F(t,t\bullet)\cap \overline{A}=\es$ and  
$F(t,t\bullet)\cap \overline{B}=\es$, where for transition $t\in T$ we define sets $\bullet t=\{p\in P\mid F(p,t)\neq\es\}$, $t\bullet=\{p\in P\mid F(t,p)\neq\es\}$ 
(sets of input and output places of $t$), and
$\guard{t}=\bigcup_{p\in P}F(p,t)$, $\effects{t}=\bigcup_{p\in P}F(t,p)$ (unions of 
labels of the incoming/outgoing arcs of $t$),
as well as $\effect{t}=\effects{t}\setminus\guard{t}$.

As with standard Petri nets the association of tokens to places is called a \emph{marking}  such that 
$M: P\rightarrow 2^{\cal{A}\cup \cal{B}}$. %We assume that $type(M_o(p))=F(t_0, p)$ for $p\in P \setminus \{p_0\}$ and $M_0(p_0)=\emptyset$. 
In addition, we employ the notion of a \emph{history}, which assigns a memory to each
transition 
$H : T\rightarrow 2^{\Nat \times 2^{P \times \cal{A}}}$. 
Intuitively, a history of $H(t) = \emptyset$ for some $t \in T$ captures that the transition has not taken place, and a history of $(k,S)\in H(t)$,
captures that the transition was executed as the $k^{th}$ transition occurrence and it has not been reversed and $S$ contains the information 
of transferred token instances and places from where they were taken.
Note that $|H(t)|>1$ may
arise due to different token instances firing the same transition. A pair of a marking and a history, $\state{M}{H}$, describes a \emph{state} of an RPN %based on which execution is determined 
with $\state{M_0}{H_0}$ the initial state, where $H_0(t) = \emptyset $ for all $t\in T$ %and $H_0(t_0)=\{(0, \mathcal{A}  \cup \mathcal{B} \times \{ p_0  \} ) \}$ for all $t \in T \setminus \{t_0\}$. %We use the notation  to denote states.

Finally, for $a_i\in \cal{A}$ and $C\subseteq 2^{\cal{A}\cup \cal{B}}$ (usually $C$ is a marking of a given place)
we define $\connected(a_i,C)$ to be the set containing the tokens connected
to $a_i$  as well as the bonds creating these connections according to 
set $C$. In order to do that let us first define, for a given $a_i \in \cal{A}$ and $C \subseteq 2^{\cal{A}\cup \cal{B}}$ , the set $paths(a_i, C)$ as follows:
 $paths(a_i, C) = \{(c_1, c_2, \ldots, c_k) \mid c_i \in {\cal{A}} \cap C
 , i \in 1, \ldots, k ; \exists_{((c_1, c_2), (c_2, c_3), \ldots, (c_{k-1}, c_k))} c_1 = a_i  ; (c_{j-1}, c_j) \in {\cal{B}} \cap C
, j \in 2, \ldots, k\}$.

We are ready to define: 
$con(a_i, C) =\\ (\{\{a_i\} \cap C) \cup \{\bondins{b_k}{c_l}, c_l \mid \exists_{(a_i, \ldots, b_k, c_l)} (a_i, \ldots, b_k, c_l) \in paths(a_i, C)  \}\}$.

More intuitively, following graph theory, we can treat a molecule as a graph, with sets of vertices $\cal{A}$ and edges $\cal{B}$. 
Then, $paths(a_i)$ is analogous to the set of all paths from vertex $a_i$ and $con(a_i, C)$ to the connected component of $a_i$, both within the graph $C$.

Let $X \subseteq \cal{A} \cup \cal{B}$ then by $X|_x$ we denote the subset of $X$ containing all its elements of type $x$,
where $x$ can be either a base or a bond type.
Let us also define a function \emph{type} as $\type: \written{A}\rightarrow A$, which 
for a given base instance (i.e. element of $\written{A}$)
returns its type (i.e. corresponding element of $A$).

\begin{example}
Figure~\ref{fig:con} depicts a marking $M(p)$ for some place $p$. We compute the sets $paths(a_1,M(p))$ and $con(a_1,M(p))$ as follows:

$paths(a_1,M(p)) = \{(a_1,c_1),(a_1,d_2),(a_1,d_2,e_3), (a_1,b_2) \}$,  

$con(a_1,M(p)) = \{a_1,b_2,c_1,d_2,e_3,\bondins{a_1}{b_2},\bondins{a_1}{c_1},\bondins{a_1}{d_2},\bondins{d_2}{e_3}\}$. 

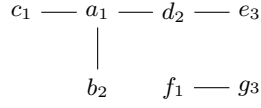
\begin{figure}[ht]
  
  \begin{center}
    \begin{tikzpicture}
		\node (n1) at (0,1) {$c_1$};
        \node (n2) at (1,1) {$a_1$};
        \node (n3) at (2,1) {$d_2$};
        \node (n4) at (3,1) {$e_3$};
        \node (n5) at (2,0) {$f_1$};
        \node (n6) at (3,0) {$g_3$};
        \node (n7) at (1,0) {$b_2$};
    
        \draw (n1) -- (n2);
        \draw (n2) -- (n3);
        \draw (n3) -- (n4);
        \draw (n5) -- (n6);
        \draw (n2) -- (n7);
\end{tikzpicture}
\end{center}
\caption{Exemplary molecules.}
\label{fig:con}
\end{figure}
\label{ex:con}
\end{example}

\begin{figure}[ht]
  
  \begin{center}
        
  \begin{tikzpicture}

    \node (type1) at (0,0) {
   \begin{tikzpicture}[scale=0.6]
        \node[circle,draw,minimum size=0.5cm] (p1) at (0,0) {};
        \node[draw,minimum size=0.4cm] (t1) at (1.75,0) {\tiny $t$};
        \node[circle,draw,minimum size=0.5cm] (p3) at (3.5,0) {};
  
        \draw[-latex] (p1) to node [above,inner sep=1.5pt] {\tiny $a$} (t1);
        \draw[-latex] (t1) to node [above,inner sep=1.5pt] {\tiny $a$} (p3);
      \end{tikzpicture}
    };

    \node(l1) [below of=type1] {\textbf{(a)} $T^{TRN}$};

    \node (type2) at (3.5,0) {
     \begin{tikzpicture}[scale=0.6]
        \node[circle,draw,minimum size=0.5cm] (p1) at (0,0) {};
        \node[draw,minimum size=0.4cm] (t1) at (1.75,0) {\tiny $t$};
        \node[circle,draw,minimum size=0.5cm] (p3) at (3.5,0) {};
  
        \draw[-latex] (p1) to node [above] {\tiny $a,b$} (t1);
        \draw[-latex] (t1) to node [above,inner sep=1.5pt] {\tiny $a$-$b\:\:$} (p3);
      \end{tikzpicture}
    };

    \node(l2) [below of=type2] {\textbf{(b)} $T^{BC1}$};

    \node (type3) at (7,0) {
     \begin{tikzpicture}[scale=0.6]
        \node[circle,draw,minimum size=0.5cm] (p1) at (0,0.75) {};
        \node[circle,draw,minimum size=0.5cm] (p2) at (0,-0.75) {};
        \node[draw,minimum size=0.4cm] (t1) at (1.75,0) {\tiny $t$};
        \node[circle,draw,minimum size=0.5cm] (p3) at (3.5,0) {};
  
        \draw[-latex] (p1) to node [above] {\tiny $a$} (t1);
        \draw[-latex] (p2) to node [below] {\tiny $b$} (t1);
        \draw[-latex] (t1) to node [above,inner sep=1.5pt] {\tiny $a$-$b\:\:$} (p3);
      \end{tikzpicture}
   
    };

    \node(l3) [below of=type3] {\textbf{(c)} $T^{BC2}$};

\end{tikzpicture}
\end{center}
\caption{The decomposition of the transitions set into the set of transitions only transferring molecules~(a), creating a bond with
a single input place (b), or two input places (c).}
\label{fig:decomposition}
\end{figure}
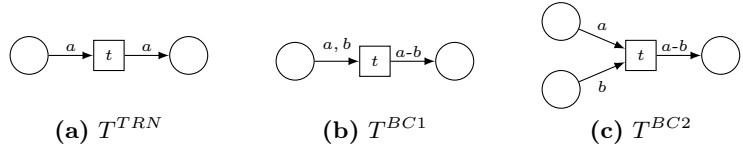

Let us define the following three sets of transitions (see Fig~\ref{fig:decomposition}):
\begin{itemize}
\item $T^{TRN} = \{t \in T \mid \#(\bullet t) = \#(t \bullet) = 1 \; ; \; \exists_{a \in A} 
(F(\bullet t, t) = \{a\} \; ; \;  F(t, t \bullet) = \{a\}  ) \}$ -- the set of transitions that transfer molecules.
\item $T^{BC1} = \{t \in T \mid \#(\bullet t) = \#(t \bullet) = 1 \; ; \; \exists_{a, b \in A} 
(F(\bullet t, t) = \{a, b\} \; ; \;  F(t, t \bullet) = \{\bondins{a}{b}\}  ) \}$ -- the set of transitions  that create a bond in a molecule.
\item $T^{BC2} = \{t \in T \mid \exists_{p_1, p_2 \in P} \bullet t = \{p_1, p_2\} \; ; \; 
 \#(t \bullet) = 1  \; ; \;   \exists_{a, b \in A} (F(p_1, t) = \{a\} \; ; \;  F(p_2, t) = \{b\}
 \; ; \;  
 F(t, t \bullet) = \{\bondins{a}{b}\}  ) \}$ -- the set of transitions with two input places that create a bond in a~molecule.
\end{itemize}
Note that a transition $t \in T^{BC1}$ may create a bond even
between bases which are already parts of the same molecule. 
For a transition $t\in T^{TRN}$, we say that $t$ is \emph{of type $TRN$}. Analogously, $t\in T^{BC1}$ 
is \emph{of type $BC1$} and $t\in T^{BC2}$ is \emph{of type $BC2$}.

\begin{definition}\label{def:lowLevel}{\rm
A \RPN $(P,T,F,A,B)$ is called \emph{low-level}, if it satisfies the following conditions for all $t\in T$:
\begin{enumerate}
\item $A\cap\guard{t}=A\cap\effects{t}$,
\item if $a\bond b\in \guard{t}$ then $a\bond b \in \effects{t}$,
\item $T$ can be decomposed into three sets: $T=T^{BC1}\cup T^{BC2}\cup T^{TRN}$, where $T^{TRN}, T^{BC1}, T^{BC2}$ are defined above,
\item if $a,b\in F(p,t)$ and $\beta=a\bond b\in F(t,q)$ then $\overline{\beta}\in F(p,t)$.
\end{enumerate}}
\end{definition}

We say that a transition is forward enabled when the following conditions are satisfied: 

\begin{definition}\label{forward}{\rm
Consider an RPN $(P,T, F, A, B)$, a transition $t\in T$, and a state $\state{M}{H}$. We say that
 $t$ is \emph{forward enabled} in $\state{M}{H}$ if there exist $S:P\rightarrow 2^{\cal{A}}$, such that 
  following hold:
 \begin{enumerate}
 \item $ \forall_{p \in \bullet t, a \in A} F(p,t)(a) = \#(S(p) \cap M(p))|_a$,
% \item $ \forall_{p \in \bullet t, a_1\bond a_2 \in B} F(p,t)(a_1\bond a_2) \leq
%  \#\bigg(\bigg(\bigcup_{\alpha \in S \cap M(p) } con(\alpha, M(p) \bigg) \bigg|_{a_1\bond a_2}\bigg)$
 \item $ \forall_{p \in \bullet t} \overline{a} \in F(p, t) \Longrightarrow 
 \bigg( \bigcup_{\alpha \in S(p) \cap M(p) } con(\alpha, M(p) \bigg) \bigg|_a = \es, $
 \item $ \forall_{p \in \bullet t} \overline{a_1 \bond a_2} \in F(p, t) \Longrightarrow 
 \bigg( \bigcup_{\alpha \in S(p) \cap M(p) } con(\alpha, M(p) \bigg) \bigg|_{a_1\bond a_2}= \es, $
 \item $\forall_{a \in A} F(t, t\bullet)(a) = F(\bullet t, t)(a)$,\\
 \item $F(t, t \bullet)(a_1 \bond a_2) = 1 \Longrightarrow $\\
 $\exists_{\alpha_1, \alpha_2 \in S(\bullet t)}, \alpha_1\neq \alpha_2,  
 \type(\alpha_1) = a_1 ; \type(\alpha_2) = a_2 ; (\alpha_1 \bond \alpha_2) 
 \notin M(\bullet t).$
 \end{enumerate}
}\end{definition}
In other words, a transition $t$ is forward enabled in a state $\state{M}{H}$ if (1) all token instances selected in the set $S(p)$ exist in the respective marking $M(p)$ and correspond to the requirements of the incoming arc $F(p,t)$ (i.e. tokens required by the action for execution), (2),(3) none 
of the tokens or bonds whose absence is required exist in the connected components that have been selected in $S$, (4) tokens are preserved according to the labels of the arcs, and (5) indicates that a transition can create a bond only if its input contains token instances of suitable types and
those instances are not bonded yet. 

We may now define the firing rule of forward execution in RPNs: 

\begin{definition}{\rm \label{forw}
Given a \RPN $(P,T, F, A, B)$, a state $\state{M}{H}$, and a transition $t$ enabled in $\state{M}{H}$ with $S:P\rightarrow 2^{\cal{A}} $ selected in the definition of forward enabledness, we write $\state{M}{H}
\trans{t} \state{M'}{H'}$:
\[
\begin{array}{rcl}
	M'(p) & = & \left\{
	\begin{array}{ll}
		M(p)\setminus \bigcup_{\alpha_i \in S(p)}\connected(\alpha_i,M(p))  & \textrm{if } p\in \bullet t \\
		M(p) \cup \bigcup_{ \alpha_i \in S(\bullet t)}\connected(\alpha_i,M(\bullet t))\cup \\ 
		\hspace{1cm}\bigcup_{\type(\alpha_1) \bond \type(\alpha_2) \in F(t,p)} \{\bondins{\alpha_1}{\alpha_2} \mid \\
		\hspace{1.5cm}\alpha_1 \neq \alpha_2 \in S(\bullet t) ;
		(\alpha_1 \bond \alpha_2) \notin M(\bullet t) \} 
		 & \textrm{if }  p\in t \bullet\\
        	M(p), &\textrm{otherwise}
	\end{array}
	\right.
\end{array}
\]
and
\[
\begin{array}{rcl}
H'(t') & = & \left\{
	\begin{array}{lll}
		H(t') \cup \{ (max\{k'\mid (k', S') \in H(t'),t'\in T\}+1, S )\}%\bigcup_{\alpha_i\in S\cap M(x)}(\alpha_i,x)] 
		\hspace{0.2in} & \textrm{if } t'= t\\
        	H(t'), & \textrm{ otherwise}
	\end{array}
	\right.
	\end{array}
\]
}\end{definition}

After the forward execution of the transition $t$, all tokens and bonds selected in $S$ and occurring in its incoming arcs are 
transferred from the input places to the output place of $t$. Any newly created bonds will also be added to the output places. Moreover, the history function 
$H$ is changed by assigning the next available integer number to the transition along with the selected set of token instances.

Since we have added token multiplicity we need to dynamically define the \emph{bonding effect} $\beffect{t,S}$ of transition $t$ concerning the particular set $S$ of token instances, that are involved in the occurrence of the transition. 

\[\beffect{t,S} = \bigcup_{a_1\bond a_2 \in F(t,t\bullet)}\{\bondins{\alpha_1}{\alpha_2}\mid \alpha_1\neq\alpha_2 \in S(\bullet t), \type(\alpha_1)=a_1, \type(\alpha_2)=a_2 \} \]

\begin{figure}[ht]
\begin{center}
\begin{tikzpicture}[scale=0.9]
%\node[circle,draw,minimum size=1cm, blue,dotted] (p0) at (-1,4) {};
\node[circle,draw,minimum size=1cm,label=left:{$p_1$}] (p1) at (1.5,4) {};
\node at (1.55,4.2) {$\bullet a_1$};
\node at (1.55,3.8) {$\bullet a_2$};
\node[circle,draw,minimum size=1cm,label=left:{$p_2$}] (p2) at (1.5,2) {};
\node at (1.55,2.2) {$\bullet b_1$};
\node at (1.55,1.8) {$\bullet b_2$};
\node[circle,draw,minimum size=1cm,label=left:{$p_3$}] (p3) at (1.5,0) {};
\node at (1.55,0.2) {$\bullet c_1$};
\node at (1.55,-0.2) {$\bullet c_2$};
\node[circle,draw,minimum size=1cm] (p4) at (6,2) {};
\node[circle,draw,minimum size=1cm] (p5) at (10,2) {};

%\node[draw,minimum size=1cm, blue, dotted] (t0) at (-1,2) {$t_0$};
\node[draw,minimum size=1cm] (t1) at (4,4) {$t_1$};
\node at (4,4.8) {\textcolor{red}{TRN}};
\node[draw,minimum size=1cm] (t2) at (4,1) {$t_2$};
\node at (4,1.8) {\textcolor{red}{BC2}};
\node[draw,minimum size=1cm] (t3) at (8,2) {$t_3$};
\node at (8,2.8) {\textcolor{red}{BC1}};
%\draw[-latex,blue, dotted] (p0) to node [left] {$a,b,c$} (t0) ;
%\draw[-latex,blue, dotted] (t0) to node [above] {$a$} (p1) ;
%\draw[-latex,blue, dotted] (t0) to node [above] {$b$} (p2) ;
%\draw[-latex,blue, dotted] (t0) to node [above] {$c$} (p3) ;
\draw[-latex] (p1) to node [above] {$a$} (t1) ;

\draw[-latex] (p1) to node [above] {$a$} (t1) ;
\draw[-latex] (p2) to node [above] {$b$} (t2) ;
\draw[-latex] (p3) to node [above] {$c$} (t2) ;
\draw[-latex] (t1) to node [above] {$a$} (p4) ;
\draw[-latex] (t2) to node [below,inner sep=6pt] {$b-c$} (p4) ;
\draw[-latex] (p4) to node [above] {$a,b$} (t3) ;
\draw[-latex] (t3) to node [above] {$a-b$} (p5) ;

\node at (9,4) {\textbf{(a)}};
\end{tikzpicture}
\vfill
\begin{tikzpicture}[scale=0.9]
%\node[circle,draw,minimum size=1cm, blue,dotted] (p0) at (-1,4) {};
\node[circle,draw,minimum size=1cm,label=left:{$p_1$}] (p1) at (1.5,4) {};
\node at (1.55,4) {$\bullet a_2$};
\node[circle,draw,minimum size=1cm,label=left:{$p_2$}] (p2) at (1.5,2) {};
\node at (1.55,2) {$\bullet b_2$};
\node[circle,draw,minimum size=1cm,label=left:{$p_3$}] (p3) at (1.5,0) {};
\node at (1.55,0) {$\bullet c_2$};
\node[circle,draw,minimum size=1cm] (p4) at (6,2) {};
\node[circle,draw,minimum size=1cm] (p5) at (10,2) {};

\node at (6.05,2.3) {$\bullet a_1$};
\node (tok1) at (6.05,2) {$\bullet {b_1}$};
\node (tok2) at (6.05,1.6) {$\bullet {c_1}$};
\draw[thick] (5.9,2) -- (5.9,1.6) ;

%\node[draw,minimum size=1cm, blue, dotted] (t0) at (-1,2) {$t_0$};
\node[draw,minimum size=1cm] (t1) at (4,4) {$t_1$};
\node[draw,minimum size=1cm] (t2) at (4,1) {$t_2$};
\node[draw,minimum size=1cm] (t3) at (8,2) {$t_3$};

%\draw[-latex,blue, dotted] (p0) to node [left] {$a,b,c$} (t0) ;
%\draw[-latex,blue, dotted] (t0) to node [above] {$a$} (p1) ;
%\draw[-latex,blue, dotted] (t0) to node [above] {$b$} (p2) ;
%\draw[-latex,blue, dotted] (t0) to node [above] {$c$} (p3) ;
\draw[-latex] (p1) to node [above] {$a$} (t1) ;

\draw[-latex] (p1) to node [above] {$a$} (t1) ;
\draw[-latex] (p2) to node [above] {$b$} (t2) ;
\draw[-latex] (p3) to node [above] {$c$} (t2) ;
\draw[-latex] (t1) to node [above] {$a$} (p4) ;
\draw[-latex] (t2) to node [below,inner sep=6pt] {$b-c$} (p4) ;
\draw[-latex] (p4) to node [above] {$a,b$} (t3) ;
\draw[-latex] (t3) to node [above] {$a-b$} (p5) ;
\node at (2,5) {};
\node at (9,4) {{\textbf{(b)}}};
\end{tikzpicture}
\end{center}
\caption{
(a) Reversing Petri Net $N$ in its initial marking.
(b) Reversing Petri net $N$ from part (a) after the execution of sequence $t_1t_2$, or, equivalently, after the execution of sequence $t_2t_1$.
}
\label{f:rpn-ex}
\end{figure}
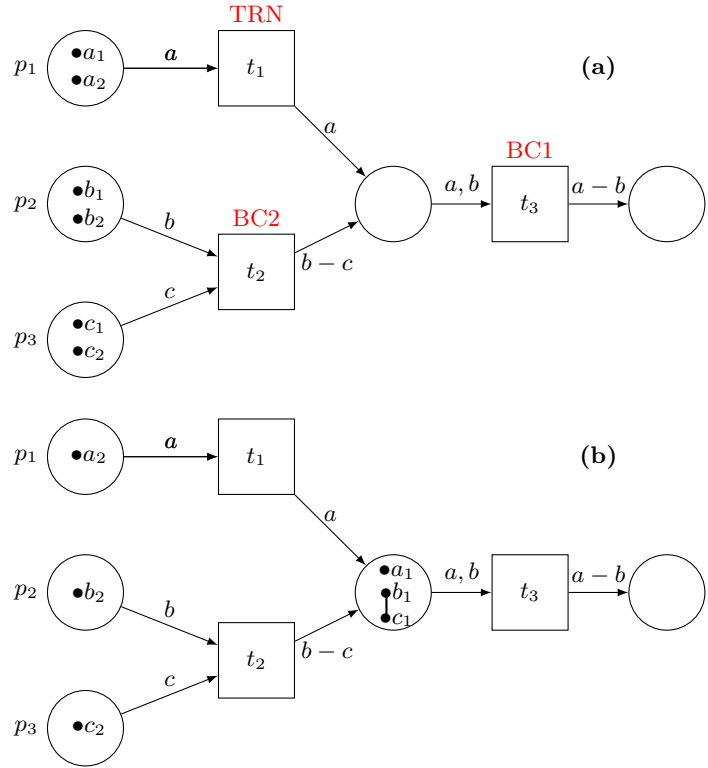

The example depicted in Figure~\ref{f:rpn-ex}a shows a reversible Petri net in its initial marking. 
%The distinguished transition $t_0$ and place $p_0$ together with arcs connected to them are depicted dashed blue. 

Note that all kinds of transitions occurs in the net: transition $t_1$ is a transporting one, hence $t_1\in T^{TRN}$ (as indicated in red in the figure), while $t_2$ and $t_3$ create bonds; moreover $t_2\in T^{BC2}$ and $t_3\in T^{BC1}$. In addition, note that,
for simplicity, we only indicate the positive token and bond types that are necessary for the transition to fire.

Initially, the history of all transitions equals $\emptyset$.
We can see that transitions $t_1$ and $t_2$ are enabled and might be executed in arbitrary order. After the executions of sequence $t_1t_2$ or $t_2t_1$ we obtain the marking presented in Figure~\ref{f:rpn-ex}b. But the history function varies depending on the order of the executed transitions. Namely, after $t_1t_2$
we have 
$H(t_1)=\{(1,\{(p_1,a_1)\})\}$, 
$H(t_2)=\{(2,\{(p_2,b_1),(p_3,c_1)\})\}$, 
$H(t)=\emptyset$ 
for $t\in\{t_0,t_3\}$, while after $t_2t_1$: 
$H(t_1)=\{(2,\{(p_1,a_1)\})\}$, 
$H(t_2)=\{(1,\{(p_2,b_1),(p_3,c_1)\})\}$, 
$H(t)=\emptyset$ for $t\in\{t_0,t_3\}$.

\subsection{Backtracking}\label{br-enb}
We now present the semantics for three
forms of reversibility as proposed in~\cite{netsWithBonds}, starting with the simplest form of reversibility namely, backtracking. 
\begin{definition}{\rm
		
		Consider an RPN $(P,T, F, A, B)$ a state $\state{M'}{H'}$ and a transition $t$ where $t\in T$. We say that $t$ is \emph{$bt$-enabled} in
		$\state{M'}{H'}$ if $(k,S)\in H(t)$,
		with $k=max\{k'\mid(k', S') \in H(t'),t'\in T\}$. %\geq k'$ for all $[k':S'] \in H(t')$, $t'\in T$.
		}\end{definition}

We say that a transition is backward enabled  only if its the last transition executed during a forward computation, i.e.  it has the highest $H$ value. 
The effect of backtracking a transition in a \RPN is defined as follows:

\begin{definition}\label{br-def}{\rm
		Given an RPN $(P,T, F, A, B)$, a state $\state{M'}{H'}$, and a transition $t$ with $(k,S) \in H'(t)$ $bt$-enabled in $\state{M'}{H'}$, we write $ \state{M'}{H'}
		\btrans{t} \state{M}{H}$:
		\[
		\begin{array}{rcl}
		M(p) & = & \left\{
		\begin{array}{ll}
		M'(p)\cup\bigcup_{ q \in t\bullet,\alpha_i \in S(p)}\connected(\alpha_i,M'(q)\setminus\beffect{t,S}), \hspace{0.3in} & \textrm{if } p\in \bullet {t} \\
		M'(p)\setminus \bigcup_{\alpha_i\in S(\bullet t)}\connected(\alpha_i,M'(p)) , & \textrm{if }  p\in t\bullet\\
		M'(p), &\textrm{otherwise}
		\end{array}
		\right.
		\end{array}
		\]
		and
			\[
	\begin{array}{rcl}
	H(t') & = & \left\{
	\begin{array}{ll}
	H'(t')\setminus\{(k,S)\}, \hspace{0.3in} & \textrm{if }t'=t \\
	H'(t'), &\textrm{otherwise}
	\end{array}
	\right.
	\end{array}
	\]
}\end{definition}

Thus, when a transition $t$ is reversed in a backtracking fashion  all tokens and bonds in the output place of the transition, as well as their connected components, are transferred to the incoming places of the transition and any newly-created bonds are broken.

\subsection{Causal Reversing}

The ability to reverse in a causal manner is defined as follows.

\begin{definition}\label{co-enabled}{\rm
		Consider an RPN $(P,T, F, A, B)$, a state $\state{M'}{H'}$, and a transition $t\in T$. Then $t$ is
		$co$-enabled in  $\state{M'}{H'}$ if
		$(k,S) \in H'(t)$ and, for all $\alpha_i\in S$, if $\connected(\alpha_i,M(q))\cap S' \neq \emptyset$
		for some $q$ and some $t'$ where $(k',S')\in H'(t')$  then $k'\leq k$. 
}\end{definition}

Note that the causal reversing of a transition $t \in T$ is allowed if all transitions executed causally after 
$t$ have been reversed.
The effect of causally reversing a transition in an \RPN is as follows:
 
\begin{definition}\label{co-def}{\rm
		Given an \RPN $(P,T, F, A, B)$, a state $\state{M'}{H'}$, and a transition $t$ with $(k,S) \in H'(t)$ $co$-enabled in $\state{M'}{H'}$, we write $ \state{M'}{H'}
		\ctrans{t} \state{M}{H}$ where $M$ is defined as in Definition~\ref{br-def} and $H$ as:
			\begin{eqnarray*}
			H(t') & = & \;\{(k',S')\mid  t'\in T, (k',S')\in H'(t'), k'<k \} \\ 	 
			&& \cup \{(k'-1,S')\mid  t'\in T, (k',S')\in H'(t'), k'>k \}  
	\end{eqnarray*}
	}
\end{definition}

Reversing a transition in a causally-respecting order is implemented in exactly the same way as in
backtracking (Definition~\ref{br-def}), i.e., the tokens are moved from the out-places to the in-places of the transition, all bonds
created by the transition are broken, and the reversal adjusts the history function. The history function is updated by removing history $(k,S)$ of the reversed transition and shifting down all identifiers that are greater than $k$ by one.

\subsection{Out-of-causal-order Reversibility}
We are now ready to define out-of-causal-order reversing enabledness.

We begin by noting that in out-of-causal-order reversibility any executed transition can be reversed at any time.
\begin{definition}\label{o-enb}{\rm
		Consider an RPN $(P,T, F, A, B)$, a state $\state{M'}{H'}$, and a transition $t\in T$. We say that $t$ is \emph{$o$-enabled} in $\state{M'}{H'}$, if $ H'(t)\neq\emptyset$.
}\end{definition}

According to the above definition, in this setting, every executed
transition can be reversed. We can use the history function to identify which token instances have been used for the firing of a particular transition. 
The following notion helps to define the last executed transition manipulating a given set of tokens, where we write $\bot$ to express
that the value is undefined.

\begin{definition}\label{last}{\rm
		Given a \RPN $(P,T, F, A, B)$, 
		a history $H$, and 
		a set of bases and bonds instances $C \subseteq \cal{A} \cup \cal{B}$ we write:
		\[
		\begin{array}{rcl}
		\first{C,H} &=& \left\{
		\begin{array}{ll}
		t , \;\;\textrm{ if }\exists t,\;(k,S)\in H(t), \; S\cap C\neq \emptyset, \mbox{ and }\\
		\hspace{0.2in} \forall t', \;(k',S') \in H(t'),  S'\cap C\neq \emptyset, \; k'\leq k  \\
		\bot,  \;\textrm{ otherwise }
		\end{array}
		\right.
		\end{array}
		\]
}\end{definition}

The effect of reversing a transition in out-of-causal order is as follows:

\begin{definition}\label{oco-def}{\rm
		Given a \RPN $(P,T, F, A, B)$ with initial marking $M_0$, a state $\state{M'}{H'}$ and a transition $t$ with $(k,S)$ that is o-enabled in $\state{M'}{H'}$, we write $\state{M'}{H'}\otrans{t} \state{M}{H}$ where $H$ is adjusted as in Definition~\ref{co-def} and $M$ as follows for all $p \in P$:
		\begin{eqnarray*}
			M(p) & = & M'(p)\setminus \beffect{t,S} \setminus \;\{C_{\alpha,p}\mid (\exists_{t' \in T}) 
			p\in t'\bullet, \alpha\in M'(p), t'\neq{\first{C_{\alpha,p},  H}}\} \\ 	 
			&& \cup \;\{C_{\alpha,q} \mid (\exists_{t' \in T})p\in t'\bullet,\; \alpha\in M'(q), \first{C_{\alpha,q},H} =t'\} \\%\mbox{ for some } y\in t\circ\}  
			& & 			 \cup \;\{C_{\alpha,q}\mid \alpha\in M'(q), \first{C_{\alpha,q},H} =\bot,C_{\alpha,q}\subseteq  M_0(p) \}
		\end{eqnarray*}
		where we use the abbreviation $C_{\alpha,z} = \connected(\alpha,M'(z)\setminus \beffect{t,S})$ for $\alpha\in S$, $z\in P$.
}
\end{definition}

In the above formula, $C_{\alpha,z}$ indicates the part of a molecule containing instance $\alpha$ obtained from place $z$ in marking $M'$ by removing the bond created by transition $t$. To compute marking $M(p)$ we have to check whether a place $p$ is an exit from the last executed transition manipulating an instance $\alpha$ in the current history $H$ or, in case the last such place is undefined,
whether $p$ was the place in which the component of $\alpha$ existed in the initial marking. 
%Note that the formula described in Definition~\ref{oco-def} consists of the sum of the separate components. 
We illustrate the definition with the following example. 

\begin{figure}[!h]
  
  \begin{center}  

  \begin{tikzpicture}[scale=1]

        \node[circle,draw,minimum size=0.8cm,label=left:{\tiny{$p_1$}}] (p1) at (0,4.5) {$\,\bullet\, a_1$};
        \node[circle,draw,minimum size=0.8cm,label=left:{\tiny{$p_2$}}] (p2) at (0,3) {$\,\bullet\, b_1$};
        \node[circle,draw,minimum size=0.8cm,label=left:{\tiny{$p_3$}}] (p3) at (0,1.5) {$\,\bullet\, c_1$};
        \node[circle,draw,minimum size=0.8cm,label=left:{\tiny{$p_4$}}] (p4) at (0,0) {$\,\bullet\, d_1$};
        \node[circle,draw,minimum size=0.8cm,label=above:{\tiny{$p_5$}}] (p5) at (3,3) {};
 		\node[circle,draw,minimum size=0.8cm,label=below:{\tiny{$p_6$}}] (p6) at (3,1.5) {};
        \node[circle,draw,minimum size=0.8cm,label=above:{\tiny{$p_7$}}] (p7) at (6,1.5) {};
        \node[circle,draw,minimum size=0.8cm,label=left:{\tiny{$p_8$}}] (p8) at (6,0) {$\,\bullet\, e_1$};
        \node[circle,draw,minimum size=0.8cm,label=right:{\tiny{$p_9$}}] (p9) at (9,0) {};

        \node[draw,minimum size=0.6cm] (t1) at (1.5,3.75) {$t_1$};
        \node[draw,minimum size=0.6cm] (t2) at (1.5,0.75) {$t_2$};
        \node[draw,minimum size=0.6cm] (t3) at (4.5,2.25) {$t_3$};
        \node[draw,minimum size=0.6cm] (t4) at (7.5,0.75) {$t_4$};

        \draw[-latex] (p1) to node [above,inner sep=2pt] {\footnotesize{$a$}} (t1);
        \draw[-latex] (p2) to node [below,inner sep=2pt] {\footnotesize{$b$}} (t1);
        \draw[-latex] (t1) to node [above,inner sep=5pt] {\footnotesize{$a-b$}} (p5);
        \draw[-latex] (p3) to node [above,inner sep=2pt] {\footnotesize{$c$}} (t2);
        \draw[-latex] (p4) to node [below,inner sep=2pt] {\footnotesize{$d$}} (t2);
        \draw[-latex] (t2) to node [below,inner sep=5pt] {\footnotesize{$c-d$}} (p6);
        \draw[-latex] (p5) to node [above,inner sep=2pt] {\footnotesize{$a$}} (t3);
        \draw[-latex] (p6) to node [below,inner sep=2pt] {\footnotesize{$c$}} (t3);
        \draw[-latex] (t3) to node [above,inner sep=5pt] {\footnotesize{$a-c$}} (p7);
        \draw[-latex] (p7) to node [above,inner sep=2pt] {\footnotesize{$a$}} (t4);
        \draw[-latex] (p8) to node [below,inner sep=2pt] {\footnotesize{$e$}} (t4);
        \draw[-latex] (t4) to node [above,inner sep=5pt] {\footnotesize{$a-e$}} (p9);

		\node at (3,4.7) {(a)};
		\node at (6.2,4.7) {(b)};
        \node[label=above:{$d_1$}] (v1) at (7,3.5){$\bullet$};
        \node[label=above:{$c_1$}] (v2) at (7.5,3.5) {$\bullet$};
        \node[label=above:{$a_1$}] (v3) at (8,3.5) {$\bullet$};
        \node[label=above:{$b_1$}] (v4) at (8.5,3.5) {$\bullet$};
        \node[label=below:{$e_1$}] (v5) at (8,3) {$\bullet$};
%        \draw (v1) -- (v2);
%        \draw (v2) -- (v3);
%        \draw (v3) -- (v4);
%        \draw (v3) -- (v5);
		\draw[thick] (7,3.51) -- (8.5,3.51) ;
		\draw[thick] (8,3.5) -- (8,3) ;

        \node[circle,draw,minimum size=2.25cm,label=right:{\tiny{$p_9$}}] (p1) at (7.75,3.5) {};
        \node at (0,-1) {};

%\draw[-latex,ultra thick,gray,snake=snake,segment amplitude=.4mm,segment length=2mm,line after snake=1mm] (0.75,0) -- (-0.5,0);

\end{tikzpicture}

\end{center}
\caption{Example of a reversing Petri net working according to the out-of-causal-order semantics. (a) Initial marking $M_0$. (b) The content of place $p_9$ in marking $M'$ obtained from $M_0$ after the execution of the sequence $t_1t_2t_3t_4$, other places are empty. }
\label{fig:ex1-ooc}
\end{figure}
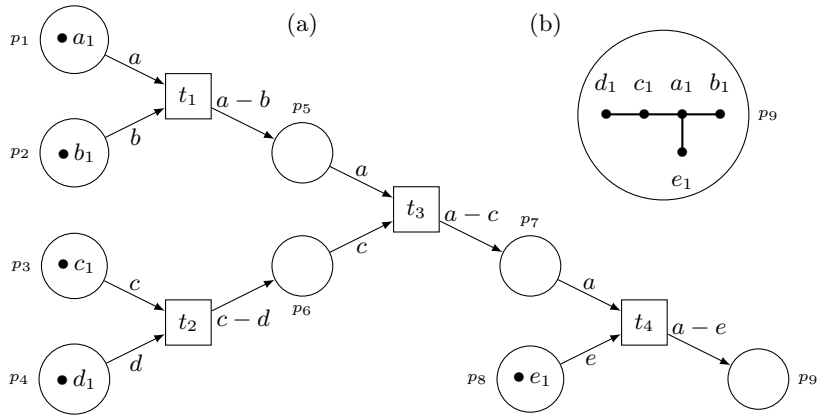

\begin{figure}[!ht]
  
  \begin{center}

\begin{tikzpicture}

        \node[circle,draw,minimum size=0.8cm,label=left:{\tiny{$p_1$}}] (p1) at (0,4.5) {};
        \node[circle,draw,minimum size=0.8cm,label=left:{\tiny{$p_2$}}] (p2) at (0,3) {};
        \node[circle,draw,minimum size=0.8cm,label=left:{\tiny{$p_3$}}] (p3) at (0,1.5) {};
        \node[circle,draw,minimum size=0.8cm,label=left:{\tiny{$p_4$}}] (p4) at (0,0) {};
        \node[circle,draw,minimum size=0.8cm,label=above:{\tiny{$p_5$}}] (p5) at (3,3) {};
 		\node[circle,draw,minimum size=0.8cm,label=below:{\tiny{$p_6$}}] (p6) at (3,1.5) {};
 		\node at (3.05,1.33) {$\bullet d_1$};
 		\node at (3.05,1.65) {$\bullet c_1$};
		\draw[thick] (2.88,1.3) -- (2.88,1.64) ;

        \node[circle,draw,minimum size=0.8cm,label=above:{\tiny{$p_7$}}] (p7) at (6,1.5) {};
        \node[circle,draw,minimum size=0.8cm,label=left:{\tiny{$p_8$}}] (p8) at (6,0) {};
        \node[circle,draw,minimum size=1.3cm,label=right:{\tiny{$p_9$}}] (p9) at (9,0) {};
 		\node at (8.75,0.20) {$a_1\bullet $};
 		\node at (9.35,0.25) {$\bullet b_1$};
 		\node at (8.75,-0.2) {$e_1\bullet $};

		\draw[thick] (8.95,0.22) -- (9.2,0.22) ;
		\draw[thick] (8.9,0.22) -- (8.9,-0.1) ;

        \node[draw,minimum size=0.6cm] (t1) at (1.5,3.75) {$t_1$};
        \node[draw,minimum size=0.6cm] (t2) at (1.5,0.75) {$t_2$};
        \node[draw,minimum size=0.6cm,fill=blue!20] (t3) at (4.5,2.25) {$t_3$};
        \node[draw,minimum size=0.6cm] (t4) at (7.5,0.75) {$t_4$};

        \draw[-latex] (p1) to node [above,inner sep=2pt] {\footnotesize{$a$}} (t1);
        \draw[-latex] (p2) to node [below,inner sep=2pt] {\footnotesize{$b$}} (t1);
        \draw[-latex] (t1) to node [above,inner sep=5pt] {\footnotesize{$a-b$}} (p5);
        \draw[-latex] (p3) to node [above,inner sep=2pt] {\footnotesize{$c$}} (t2);
        \draw[-latex] (p4) to node [below,inner sep=2pt] {\footnotesize{$d$}} (t2);
        \draw[-latex] (t2) to node [below,inner sep=5pt] {\footnotesize{$c-d$}} (p6);
        \draw[-latex] (p5) to node [above,inner sep=2pt] {\footnotesize{$a$}} (t3);
        \draw[-latex] (p6) to node [below,inner sep=2pt] {\footnotesize{$c$}} (t3);
        \draw[-latex] (t3) to node [above,inner sep=5pt] {\footnotesize{$a-c$}} (p7);
        \draw[-latex] (p7) to node [above,inner sep=2pt] {\footnotesize{$a$}} (t4);
        \draw[-latex] (p8) to node [below,inner sep=2pt] {\footnotesize{$e$}} (t4);
        \draw[-latex] (t4) to node [above,inner sep=5pt] {\footnotesize{$\,\,\,a-e$}} (p9);

		\node at (3,4.7) {(a)};
        \node at (0,-1) {};
		
%\draw[-latex,ultra thick,gray,snake=snake,segment amplitude=.4mm,segment length=2mm,line after snake=1mm] (0.75,0) -- (-0.5,0);

\end{tikzpicture}

\begin{tikzpicture}

        \node[circle,draw,minimum size=0.8cm,label=left:{\tiny{$p_1$}}] (p1) at (0,4.5) {};
        \node[circle,draw,minimum size=0.8cm,label=left:{\tiny{$p_2$}}] (p2) at (0,3) {$\,\bullet\, b_1$};
        \node[circle,draw,minimum size=0.8cm,label=left:{\tiny{$p_3$}}] (p3) at (0,1.5) {};
        \node[circle,draw,minimum size=0.8cm,label=left:{\tiny{$p_4$}}] (p4) at (0,0) {};
        \node[circle,draw,minimum size=0.8cm,label=above:{\tiny{$p_5$}}] (p5) at (3,3) {};
 		\node[circle,draw,minimum size=0.8cm,label=below:{\tiny{$p_6$}}] (p6) at (3,1.5) {};
        \node[circle,draw,minimum size=0.8cm,label=above:{\tiny{$p_7$}}] (p7) at (6,1.5) {};
        \node[circle,draw,minimum size=0.8cm,label=left:{\tiny{$p_8$}}] (p8) at (6,0) {};
        \node[circle,draw,minimum size=1.3cm,label=right:{\tiny{$p_9$}}] (p9) at (9,0) {};

        \node[draw,minimum size=0.6cm,fill=blue!20] (t1) at (1.5,3.75) {$t_1$};
        \node[draw,minimum size=0.6cm] (t2) at (1.5,0.75) {$t_2$};
        \node[draw,minimum size=0.6cm] (t3) at (4.5,2.25) {$t_3$};
        \node[draw,minimum size=0.6cm] (t4) at (7.5,0.75) {$t_4$};

        \draw[-latex] (p1) to node [above,inner sep=2pt] {\footnotesize{$a$}} (t1);
        \draw[-latex] (p2) to node [below,inner sep=2pt] {\footnotesize{$b$}} (t1);
        \draw[-latex] (t1) to node [above,inner sep=5pt] {\footnotesize{$a-b$}} (p5);
        \draw[-latex] (p3) to node [above,inner sep=2pt] {\footnotesize{$c$}} (t2);
        \draw[-latex] (p4) to node [below,inner sep=2pt] {\footnotesize{$d$}} (t2);
        \draw[-latex] (t2) to node [below,inner sep=5pt] {\footnotesize{$c-d$}} (p6);
        \draw[-latex] (p5) to node [above,inner sep=2pt] {\footnotesize{$a$}} (t3);
        \draw[-latex] (p6) to node [below,inner sep=2pt] {\footnotesize{$c$}} (t3);
        \draw[-latex] (t3) to node [above,inner sep=5pt] {\footnotesize{$a-c$}} (p7);
        \draw[-latex] (p7) to node [above,inner sep=2pt] {\footnotesize{$a$}} (t4);
        \draw[-latex] (p8) to node [below,inner sep=2pt] {\footnotesize{$e$}} (t4);
        \draw[-latex] (t4) to node [above,inner sep=5pt] {\footnotesize{$\,\,\,a-e$}} (p9);

 		\node at (8.75,0.20) {$c_1\bullet $};
 		\node at (9.35,0.2) {$\bullet a_1$};
 		\node at (8.75,-0.15) {$d_1\bullet $};
 		\node at (9.35,-0.2) {$\bullet e_1$};

		\draw[thick] (8.95,0.22) -- (9.2,0.22) ;
		\draw[thick] (8.9,0.22) -- (8.9,-0.1) ;
		\draw[thick] (9.19,0.22) -- (9.19,-0.1) ;

		\node at (3,4.7) {(b)};
%        \node at (0,-1) {};
		
%\draw[-latex,ultra thick,gray,snake=snake,segment amplitude=.4mm,segment length=2mm,line after snake=1mm] (0.75,0) -- (-0.5,0);

\end{tikzpicture}
\end{center}
\caption{(a) Marking obtained after reversing transition $t_3$ at marking $M'$ described in Figure~\ref{fig:ex1-ooc}b. (b) Marking obtained after reversing transition $t_1$ at marking $M'$ described in Figure~\ref{fig:ex1-ooc}b.}
\label{fig:ex2-ooc}
\end{figure}
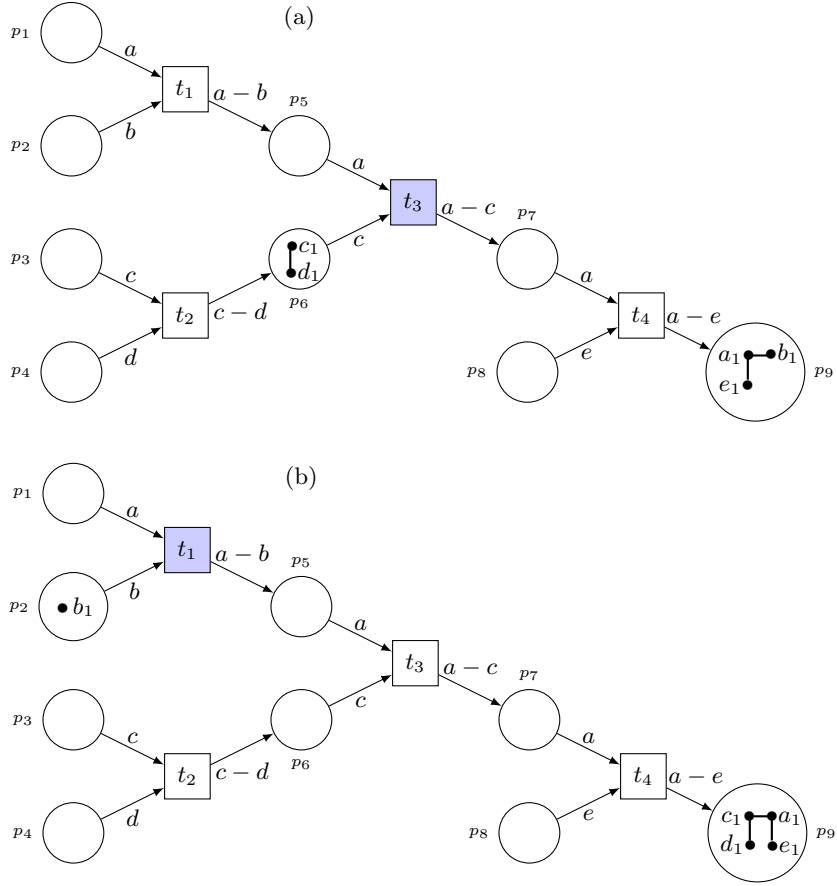

\begin{example}
Let us look at Figure~\ref{fig:ex1-ooc}a.
We can execute the sequence $t_1t_2t_3t_4$ at the initial marking $M_0$, obtaining marking $M'$ as follows:
place $p_9$ contains a molecule depicted in the circle (Figure~\ref{fig:ex1-ooc}b), other places are empty. 
%The marking obtained after reversing a transition in out of causal manner is called $M$, according to Definition~\ref{oco-def}. 

Assume now that $t_3$ is the transition to be reversed. (Note that this reversing transition is described as $t$ in the formula of Definition~\ref{oco-def}.)
Let us focus on place $p_9$ first.
We~want to determine the marking $M(p_9)$ obtained after reversing $t_3$, hence for $p$ from the definition we take $p_9$. $M'(p)$ contains the whole molecule from Figure~\ref{fig:ex1-ooc}b. 
According to the formula, we have to remove the bond created by the forward execution of transition  $t_3$, i.e. $\beffect{t,S}$, namely: the bond $a_1-c_1$. 
This results in partitioning the whole molecule into two parts: $e_1-a_1-b_1$ and $c_1-d_1$. For each base
 instance $\alpha$ we have to analyse its connected component $C_{\alpha,p}$ from the formula. Base instance $a_1$ (and also 
$e_1$) has been used by $t_4$, hence the instance together with its whole connected component $C_{a_1,p}=\{e_1,a_1,b_1,\bondins{e_1}{a_1},\bondins{a_1}{b_1}\}$ stays a part of $M(p)$. 
For $C_{c_1,p}$ we have to analyse the next part of the formula from Definition~\ref{oco-def}. 
%First, transition $t'$, for 
%which place $p_9$ is its output place, have to be determined - is it $t_4$. 
Base instance $c_1$ was present in $p_9$ at $M'$. 
However, $t_4$ has not been the last transition (according to history $H$) using $c_1$ (and $d_1$ from the same connected 
component). That is why molecule $c_1-d_1$ should be removed from $p_9$ at $M$. Other parts of the formula do not apply in 
this case. 
Assume next, that the reversed transition $t$ is still $t_3$, but we focus on place $p_6$  (i.e. $p$ from the formula equals $p_6$ now), and we want to determine $M(p_6)$. At $M'$ (Figure~\ref{fig:ex1-ooc}b) place $p_6$ is empty and we cannot subtract anything from its marking. Base instances $a_1$, $c_1$\footnote{
We mention only those two because after reversing $t_3$ we have two molecules in $p_9$, the first one can be described as $con(a_1, M(p_9))$, the second $con(c_1, M(p_9))$. Any other base instance from those two connected components can be taken instead.
}
and their connected components at $M'$ are located in place $p_9$, hence for $q$ we take $p_9$. (As said before, $C_{a_1, q}$ stays in place $p_9$ at $M$.) For $C_{c_1, q}=\{c_1,d_1,\bondins{c_1}{d_1}\}$ we have to determine transition $t'$, for which place $p$ is the output place - it is transition $t_2$. This transition has been the last one (according to history $H$) using $c_1$, hence $C_{c_1, q}$ should be added to $M(p)$ (which is in this case  $M(p_6)$). The last part of the formula from Definition~\ref{oco-def} does not apply here - $p_6$ is not an initial place for any base instance. The  marking obtained from $M'$ (depicted in Figure~\ref{fig:ex1-ooc}a) by reversing transition $t_3$ is presented in Figure~\ref{fig:ex2-ooc}a.

In the last case, we want to reverse transition $t_1$ (not $t_3$) starting from the marking $M'$ from Figure~\ref{fig:ex1-ooc}b and determine the obtained marking $M$ -- for $t$ from the formula from Definition~\ref{oco-def} we take $t_1$. Transition $t_1$ has created bond $a_1-b_1$. Let us now take for $p$ (from the definition) the place $p_2$, which is an initial place for $b_1$ and it is not an output place for any transition $t'\in T$.
All base instances from the analysed RPN in $M'$ were located in $p_9$, hence for all of them $q$ (from the Definition~\ref{oco-def} formula) equals $p_9$.  
According to the formula, we have to remove the bond created by the forward execution of transition  $t_1$, i.e. $\beffect{t,S}$, namely: the bond $a_1-b_1$. 
The connected component for $a_1$ obtained from $p_9$ in $M'$ after removing the bond created by $t_1$ is as follows: $C_{a_1, q}=\{a_1, e_1, c_1, d_1,\bondins{d_1}{c_1},\bondins{c_1}{a_1},\bondins{a_1}{e_1}\}$. The last transition that used those instances exists. However, for $C_{b_1, p_9}=\{b_1\}$ we cannot determine the last transition that used $b_1$ in accordance with history $H$ - the last one was $t_1$ but now is being reversed. $b_1$ in the initial marking $M_0$ was located in $p_2$, hence it goes back to that place - it is described by the last part of the formula. The  marking obtained from $M'$ (depicted in Figure~\ref{fig:ex1-ooc}b) by reversing transition $t_1$ is presented in Figure~\ref{fig:ex2-ooc}b.

\end{example}
\label{ex:ooc}

\section{Coloured Petri Nets}
\label{sec:CPN}

Recall that RPNs constitute a model in which transitions can be reversed 
according to three semantics: backtracking, causal, and out-of-causal-order 
reversing. A main characteristic of RPNs is the concept of a \emph{history}, which 
assigns 
a memory to each transition.
While this construct enables transition reversal, it imposes
the need of a global control during computation. Our goal is 
to recast the model of RPNs into one without any form of global control
while establishing the expressiveness relation
between RPNs and the model of bounded coloured Petri nets. 
In this section we recall the notion of coloured Petri nets.

\begin{definition}[\cite{CPN}]{\rm
\label{def:CPN}
A (non-hierarchical) \emph{coloured Petri net} is a nine-tuple
$CPN=(P,T,D,\Sigma,V,C,G,E,I)$, where:
\begin{itemize}
\item $P$ and $T$ are finite, disjoint sets of \emph{places} and \emph{transitions};
\item $D\subseteq P\times T\cup T\times P$ is a set of \emph{directed arcs};
\item $\Sigma$ is a finite set of non-empty \emph{colour sets};
\item $V$ is a finite set of \emph{typed variables} such that $Type[v]\in\Sigma$ for all $v\in V$,
where $Type$ is a function returning a colour of variable;
\item $C:P\rightarrow\Sigma$ is a \emph{colour set function} that assigns colour sets to places;
\item $G:T\rightarrow \mathit{EXPR}_V$ is a \emph{guard function} that assigns a guard to each transition $t$ such that $\mathit{Type}[G(t)]=Bool$;
\item $E:D\rightarrow \mathit{EXPR}_V$ is an \emph{arc expression function} that assigns an arc expression to each arc $d\in D$ such that $\mathit{Type}[E(d)]=\Nat^{C(p)}$, where $p$ is the place connected with the arc $d$;
\item $I:P\rightarrow \mathit{EXPR}_\emptyset$ is an \emph{initialisation function} that assigns an initialisation expression to take each place $p$ such that $\mathit{Type}[I(p)]=\Nat^{C(p)}$. 
\end{itemize}}
\end{definition} 

Note that, according to the utilised CPN-Tools \cite{cpn-tools}, 
$\mathit{EXPR}_V$ is the set of \emph{net inscriptions} (over a set of variables $V$, possibly empty,
i.e., using only constant values) provided by CPN ML.
Moreover, by $\mathit{Type}[e]$ we denote the type of values obtained by the evaluation
of expression $e$.
The set of \emph{free variables} in an expression $e$ is denoted by $\mathit{Var}[e]$.
The setting of a particular value to free variable $v$ is called~a \emph{binding} 
$b(v)$. We~require that $b(v)\in \mathit{Type}[v]$ 
and denote with the use of $\langle\rangle$ filled by the list of valuations and
written next to the element to whom it relates. 
The set of bindings of $t$ is denoted by $B(t)$.
The \emph{binding element} is a~transition $t$ together with a valuation $b(t)$
of all the free variables related to $t$. 
We~denote it by $(t,b)$, for $t\in T$ and $b\in B(t)$. 

A \emph{marking} $M$ in coloured Petri nets is a function which assigns  
a set of tokens $M:P \rightarrow \Nat^\Sigma$ consistently with $C$.
An initial marking is denoted by $M_0$ and defined for each $p\in P$ as follows: 
$M_0(p) = I(p)\langle\rangle$.

A binding element $(t,b)$ is \emph{enabled} at a marking $M$ if $G(t)\langle b\rangle$ is true and
at each place $p\in P$ there are enough tokens in $M$ to fulfil 
the evaluation of the arc expression function $E(p,t)\langle b\rangle$. 
The resulting marking is obtained by removing the tokens given by $E(p,t)\langle b\rangle$ from $M(p)$ 
and adding those given by $E(t,p)\langle b\rangle$ for each $p\in P$.

We define the \emph{enabledness} of transitions in CPN as follows:
a transition $t \in T$ is \emph{enabled} in $M$ and its execution 
leads to marking $M'$ (denoted $M[t\rangle M'$) if there exists a binding $b \in B(t)$,
such that the binding element $(t, b)$ is enabled at $M$.

\section{Transformation}
\label{sec-transformation}

Let us recall that for a given low-level RPN  $N_R=(P_R,T_R,F_R,A_R,B_R)$ 
with {$\written{A}$} and {$\written{B}$} for bases and bonds instances, we have the following decomposition: 
$T_R=T_R^{BC1}\cup T_R^{BC2}\cup T_R^{TRN}$. 
We assume that we number transitions and the enumeration
is consistent with $\prec$ (i.e. if $t_i \prec t_j$ then $i < j$).

For such an RPN we define the following:

\begin{itemize} 
\item a relation $\rightarrow$ on $P_R\cup T_R$ as follows,
$x\rightarrow y$ if $F(x,y)$ is not empty
and we call it \emph{direct order};  

\item a relation $\prec$ on $P_R\cup T_R$ as a transitive
(but irreflexive) closure of $\rightarrow$;

\item a bounded set of integers $\Nat_b=\{0,...,nb\}$, where $nb$ is 
an integer number which is larger
than the number of transitions and the number of instances of bases and bonds;
\item $ConCom(X)$ - a connected component of the set $X \subseteq \written{A} \cup \written{B}$; having a set $X$ consisting of bases and bonds instances and treating $X$ as an undirected graph, the function $ConCom(X)$ returns a set of connected components of~$X$;

\item \emph{mol} is a pair $(V, E)$ called a molecule, where $V \subseteq \written{A}$ and 
$E \subseteq \written{B}$, such that
$(V, E)$ is a connected graph, by $mol_\alpha$ we understand a molecule $(V, E)$ such 
that $\alpha \in V$;

\item if a multiset $X \in \Nat^Y$ contains only one element (is a singleton)
we denote it by this only element $y \in Y$, such that $X(y) = 1$.

\end{itemize}

\begin{remark}
Note that the relation $\rightarrow$ could be defined only between places and transitions.
On the other hand, such a~restriction does not hold  for $\trord$.
\end{remark}

Furthermore, for any element $x\in P_R\cup T_R$ or $t\in T_R$, we consider the following five sets:

\begin{itemize}

\item a set of \emph{neighborhood} of an element $x\in P_R\cup T_R$ as 
$nei(x)$; 
%\todo{Bledy w def, i rozdzielic na dwie, i wyjanic.}
%\item a set \emph{out of causal dependencies counters} of a transition $t\in T_R$ as 
%$dpc_{OOC}(t)=nei_{OOC}\cap T_R$ -- a set of transitions
%used to decide whether a transition is enabled and
%compute the effect of its reversing;
%
%\item a set \emph{out of causal dependencies histories} of a transition $t\in T_R$ as 
%$dph_{OOC}(t)=nei_{OOC}\cap T_R$ -- 
%a set of transitions, which have a common place $h_{ij}$ with $t$ -- 
%the set of transitions, which are taken into consideration paired with $t$ as an elements
%of quadruples in $h_i$, where $h_i$ is the transition history place of $t$; 
% -- a set of transitions, which affect reversing of a given transition together, with those
%affected by the reverse -- related to the effect of forward execution of a transition;
\item a set of \emph{dependency counters} of a transition $t\in T_R$ as 
$dpc(t)$ -- a set of transitions
used to decide whether a reversing transition is enabled;

\item a set \emph{dependency histories} of a transition $t\in T_R$ as 
$dph(t)$ -- 
a set of transitions which would be affected by the execution of the given transition or 
its reverse;
\item a set of \emph{reversing input places} of a transition $t\in T_R$ as 
$rin(t)$ -- the set of places in which one needs to search for molecules, while reversing transition $t$; 

\item a set \emph{reversing output places} of a transition $t\in T_R$ as 
$rout(t)$ -- the set of places where a molecule might be placed after reversing transition $t$.

%\item We say that transition \emph{operates} on a molecule when it puts
%it to its output place.
%\item \emph{local history} for a given transition $t_i \in T_R$
%is a set of transition history places of $t_j \in dph(t_i)$ together with transition history place 
%of $t_i$.
%\item $blockT$ is a function which for a given local history $LH$ of transition $t$, 
%a base instance $\alpha$ and transition $t$
%returns
%the last transition from $T_R \setminus \{t\} $ which operates on a molecule containing $\alpha$
%or $\lambda$ when such a transition does not exist. 

\end{itemize}

During the transformation from an RPN to a CPN two types of new places are added to the CPN: \emph{transition history places}
 and \emph{connection history places}. A transition history place is created for every transition and it contains information about history of executions of that transition. Histories are important during 
 reversing and to reverse a transition $t$ sometimes it is necessary to check and modify content of
 history places of other transitions - the set of those transitions is denoted as $dph(t)$. 
 A connection history place is created for a pair of transitions and it contains a number (counter) which describes how many times transitions from the pair were executed. Those places are not created for every 
 pair of transitions, but for a given transition $t$ they are added to the net only for transition $t$ and 
 transitions from $dpc(t)$.

The above sets differ depending on the assumed operational semantics of reversing. We use subscripts $BT$, $C$ and $OCC$ to clearly indicate that we operate according to backtracking, causal-order reversing and out-of-causal-order reversing, respectively.

In Table~\ref{table:rel.sem} we present how these sets are defined depending on the relative semantics.

\begin{table}[ht]
\begin{center}

\begin{tabular}{|c|c|}
\hline
  & \bf{backtracking} \\
\hline\hline
$nei$ &$nei_{BT}(x)=\{y\in P_R\cup T_R \mid (x\rightarrow y \lor y\rightarrow x) \lor (\exists_{z\in P_R\cup T_R} \;x\rightarrow z\rightarrow y \lor y\rightarrow z \rightarrow x)\}$ \\\hline
$dpc$&$dpc_{BT}(t)=T_R \setminus \{t\}$ \\\hline
$dph$&$dph_{BT}(t)=nei_{BT}(t)\cap T_R$\\\hline
$rin$&$rin_{BT}(t)=nei_{BT}(t)\cap P_R$ \\\hline
$rout$&$rout_{BT}(t)=nei_{BT}(t)\cap P_R$ \\\hline
\hline
\hline
 & \bf{causal-order reversing}\\
\hline\hline
$nei$ &$nei_{C}(x)=\{y\in P_R\cup T_R \mid (x\rightarrow y \lor y\rightarrow x) \lor (\exists_{z\in P_R\cup T_R} \;x\rightarrow z\rightarrow y \lor y\rightarrow z \rightarrow x)\}$ \\\hline
$dpc$&$dpc_{C}(t)=nei_{C}(t)\cap T_R$ \\\hline
$dph$&$dph_{C}(t)=nei_{C}(t)\cap T_R$\\\hline
$rin$&$rin_{C}(t)=nei_{C}(t)\cap P_R$ \\\hline
$rout$&$rout_{C}(t)=nei_{C}(t)\cap P_R$ \\\hline
\hline
 & \bf{out-of-causal-order reversing}\\
\hline\hline
$nei$ &$nei_{OOC}(x)=\{y\in P_R\cup T_R \mid  x\trord y \lor y\trord x\}$ \\\hline
$dpc$&$dpc_{OOC}(t)=nei_{OOC}(t)\cap T_R$ \\\hline
$dph$&$dph_{OOC}(t)=nei_{OOC}(t)\cap T_R$\\\hline
$rin$&$rin_{OOC}(t)=nei_{OOC}(t)\cap P_R$ \\\hline
$rout$&$rout_{OOC}(t)=nei_{OOC}(t)\cap P_R$ \\\hline
\end{tabular}

\end{center}
\caption{Sets $nei$, $dpc$, $dph$, $rin$, and $rout$ for the three operational semantics of reversing. }
\label{table:rel.sem}
\end{table}

%\begin{table}[ht]
%\begin{center}
%\begin{tabular}{|c|c|c|c|}
%\hline
%relation\textbackslash semantics & backtracking & causal-order reversing& out-of-causal-order reversing \\
%\hline\hline
%$dpc$ &$all=Nei_{BT}$&$Nei_{C}$&$Nei_{OCC}$ \\
%$dph$ &$h_i$ or $Nei_{BT}$ &$Nei_{C}$ &$Nei_{OCC}$\\
%$rin$ &$Nei_{BT}$&$Nei_{C}$&$Nei_{OCC}$ \\
%$rout$ &$Nei_{BT}$&$Nei_{C}$&$Nei_{OCC}$ \\
%
%\hline
%
%\end{tabular}
%\end{center}
%\caption{add caption \textcolor{blue}{do we really need this table???}}
%\label{table:rel.sem}
%\end{table}

Note that in what follows, we use places that contain exactly one token and we
denote the marking by the value of this token instead of a multiset containing only this value.

Now, we are ready to define the CPN $N_C(N_R)=(P_C,T_C,D_C,\Sigma_C,V_C,$ $C_C,G_C,E_C,I_C)$ corresponding to $N_R$. According to the definition of CPNs, in the following transformation we use notations: $P_C$ - set of places,
$T_C$ - set of transitions, $D_C$ - set of arcs, $\Sigma_C$ - set of colours, $V_C$ - set of variables,
$C_C$ - function that assigns colours to places, $G_C$ - function that assigns guards 
to transitions, $E_C$ - function that assigns arc expressions to arcs, $I_C$ - function that assigns
initial expressions to places. \\

Before we go into the details of the transformation, let us explain the 
meaning of the $max$ function we use.
\begin{remark}\label{prec}
The function $max$ operates over the set of transitions and
returns the maximal one according to the order determined by the sequence of firings - the last, executed  
transition is the largest. The sequence of firings is stored by the history. In RPNs it is conducted explicitly, with the use of the history function.
On the other hand, in CPNs, the history is scattered among transitions history places. However, it is possible to reproduce the whole sequence
by calculations. Appropriate formulas are presented in Section~\ref{equivalence}.  
In function $max$ the information about the history of executions is obtained from places $h_j$, where 
$t_j \in (dph(t_i) \cup t_i)$ and $t_i$ is a transition to be reversed. 
Moreover, not the whole sequence of firings is considered during $max$ calculations. Only those transitions,
which during their execution have used any token instance from $con(\alpha, C_{\alpha})$ for given $\alpha$ and $C$, are considered. 
Notice, that $max$ in CPNs is very similar to $\first{C,H}$ defined in Definition~\ref{last}, where $C$ is $con(\alpha, C_{\alpha})$
and $H$ is stored in places $h_j$. When $\first{C,H}$ exists in an RPN, it is equal to the value of $max$ in the corresponding CPN - for
the same $C$ and $H$. When $\first{C,H}$ equals $\bot$, then function $max$ would return $t_0$ (the initialization transition which
is added to the CPN during the transformation to generate the initial marking). 
Furthermore, the order of transitions determined by the history, limited to a given set of token instances $con(\alpha, C_{\alpha})$ 
is compatible to the order determined by relation $\prec$ over the same set of transitions. This holds because the 
considered PNs are acyclic, hence token instances can be transported only along the arcs, so according to $\prec$.
Because of that, considering only executions related to some set of token instances $con(\alpha, C_{\alpha})$, it is not possible for $t_i$ 
to be executed after $t_j$ when $t_i \prec t_j$.
\end{remark}

First, let us specify precisely the set $\Nat_b=\{0,...,nb\}$
previously introduced. 
Let $\bound$ be the number of different instances occurring in \RPN $N_R$ in the initial marking increased by 2. 
Then $nb = 2\bound$ and hence $\Nat_b=\{0,...,2\bound\}$. 

\begin{remark}
Note that the numbers of tokens in each place is $\bound$ strong safe, i.e. at any marking in every possible situation $\bound$ tokens are placed, hence we deal with
 multisets over $\Sigma_C$. This approach has been used here for technical reasons - we consider special kind of tokens - idle ones denoted as $(\es, \es)$. However, wherever this does not lead to misunderstandings, we use the denotation for a single token. 

\end{remark}

In the following transformation, in some places, the CPN-Tools semantics is used. 
The transformation is prepared with a view to putting it into practice - 
implementation in the program.
The most frequently used CPN-Tools semantics elements are: \emph{++} which means concatenation, and 
\emph{n`} which describes quantity of elements.
\begin{itemize}
\item[{$\bm{ P_{C}}$=}]$P_R\cup \{h_i \mid t_i\in T_R\} \cup \{h_{ij} \mid  t_i,t_j\in T_R; i < j ; t_j\in dpc(t_i)\}$ 

Set of places contains places from the original RPN net, transition history places for each transition,
and connection history places for pairs of transitions (set $dpc$ determines for which
transitions connection history place is added). Notice that indexes of transition history and connection history places are very important. For transitions $t_i$ and $t_j$ transition history places
are denoted as $h_i$ and $h_j$ (respectively), and the connection history place is denoted as $t_{ij}$ (for $i<j$) or $t_{ji}$ (for $i>j$).
\item[{$ \bm {T_{C}}$=}]$T_R\cup\{tr_i \mid  t_i\in T_R\}\cup \{t_{0}\}$

Set of transitions contains transitions from the original RPN and reversing transitions - one for each original transition. A reversing transition for $t_i$ is denoted as $tr_i$.
Transition $t_{0}$ is added for technical reasons - more about this transition can be
found at the end of this section.
\item[{$ \bm {D_{C}}$=}]$Domain(F_R)\cup(Domain(F_R))^{-1}\cup$ \\
                   $\{(t_i,h_i),(h_i,t_i),(tr_i,h_i),(h_i,tr_i)\mid t_i\in T_R\}\cup$\\
                   $\{(tr_i,h_j),(h_j,tr_i)\mid t_i\in T_R ; t_j\in dph(t_i)\}\cup$\\
                   $\{(t_i,h_{jk}),(h_{jk},t_i),(tr_i,h_{jk}),(h_{jk},tr_i)\mid t_i\in T_R ; \{i,l\}=\{j,k\} 
                   ; t_l\in dpc(t_i)\}$

This set contains all arcs: arcs from RPN $N_R$, arcs opposite to those from $N_R$,
arcs between every transition $t_i$ and its history places (in both directions), 
arcs between every reversing transition $tr_i$ and the history place of $t_i$ (in both directions),
arcs between every reversing transition $tr_i$ and history places of transitions from $dph(t_i)$ 
(in both directions),
arcs between every transition $t_i$ and all its connection history places (in both directions),
arcs between every reversing transition $tr_i$ and all connection history places of $t_i$ (in both directions).
\item[{$ \bm{ \Sigma_{C}}$=}]$\Nat_b\;\cup\;$
						$A_R\;\cup\;$
						$B_R\;\cup$
						$\overline{A_R}\;\cup$
						$\overline{B_R}\;\cup$
					    $\written{A}\;\cup$
					    $\written{B}\;\cup$
					    $(2^{\written{A}}\times 2^{\written{B}})\;\cup$
					    $2^{(\Nat_b\times T_R \times T_R \times 2^{\written{A}})}\;$
					    
We define the following colours: a bounded set of natural numbers, base types, bond types, negative base types,
negative bond types, instances of bases, instances of bonds, Cartesian product of subsets of token instances and subsets of bond instances  -- \textbf{molecules},
subsets of 4-tuples (one 4-tuple contains the following information:
the second transition in the tuple, in the context of the first one in the tuple, was
$n$-th in the sequence of executions and has used the given base instances).
					    
\item[{$\bm{V_{C}}$=}]$\{( {\cal{X}}_i , {\cal{Y}}_i )\in (2^{\written{A}}\times 2^{\written{B}}) \mid i\in\Nat_b\}\;\cup$
					$({\cal{X}},{\cal{Y}}) \in (2^{A}\times 2^{B}) \;\cup $\\
                   $\{(\alpha_i\in {\written{A}} \mid i\in\Nat_b\}\;\cup$
                   $\{cnt_i \in \Nat_b \mid  i\in \{1,...,|T_R|\}\}\;\cup$ 
%\textcolor{blue}{ - counters, numbers, which occurs in places $h_{ij}$}\\
                   $\{H_i \in 2^{(\Nat_b\times T_R \times T_R \times 2^{\written{A}})} \mid  i\in \{1,...,|T_R|\}\}$
\item[{ $\bm{C_{C}}$=}]$\{\bm{p}\mapsto (2^{\written{A}}\times 2^{\written{B}})\mid p\in P_R\}\;\cup$\\
                   %$\{h_i\mapsto (\Nat_b\times (dpc(t_i)\cap T_R) \times t_i \times 2^{\written{A}}) \mid  t_i \in T_R\}\;\cup$\\
                   $\{\bm{h_{i}}\mapsto 2^{(\Nat_b\times T_R \times T_R \times 2^{\written{A}})} \mid  t_i \in T_R\}\;\cup$\\
                   $\{\bm{h_{ij}}\mapsto \Nat_b \mid  t_i,t_j \in T_R ; i < j ; t_j\in dpc(t_i)\}$  \\
This set describes which colours are assigned to which places (respectively):
colour \textit{molecule} to places from the RPN, set of 4-tuples to history places, and 
set of bounded natural numbers to connection history places.
\item[{$\bm{G_{C}}$=}] $G_C^{TRN} \cup G_C^{BC1} \cup G_C^{BC2} \cup  G_C^{t_0} \cup G_C^{\overline{t_0}} \cup G_C^{\overline{TRN}} \cup  G_C^{\overline{BC1 \cup BC2}}$ \\
where\\
%$G_C^{BC1} = \{t_i\mapsto (\{\alpha_1,\alpha_2\} \in {\cal{X}} )\mid $\\
%                      $t_i\in T_R^{BC1} ; 
%                      \{\type(\alpha_1),\type(\alpha_2)\}\subseteq F_R(\bullet t_i,t_i) ;$\\
%                      $\alpha_1\neq \alpha_2 ;$
%                      $\{({\cal{X}}, {\cal{Y}})\} \subseteq Var[E_C(\bullet t_i, t_i)] ;$\\
%                      $F_R(\bullet t_i,t_i)\cap(\overline{A}\cup\overline{B})\cap\type({\cal{X}}\cup {\cal{Y}})=\emptyset
%                      \}$ \\
%                      
%                      
%                      
%                      
                      ${\bm{G_{C}^{BC1}}} = \{\bm{t_{i}}\mapsto 
                      (\alpha_1\in {\cal{X}}_1 \land \alpha_2\in {\cal{X}}_2) 
                      \lor (\alpha_1, \alpha_2 \in {\cal{X}}_1 \land \bondins {\alpha_1} {\alpha_2} \not\in {\cal{Y}}_1 $
                      \\
                      $ \land \; ({\cal{X}}_2,{\cal{Y}}_2)  = (\es, \es)
                      \mid $\\
                      $t_i\in T_R^{BC1} ; 
                      \{\type(\alpha_1),\type(\alpha_2)\}\subset {F_R(\bullet t_i,t_i)} 							;$\\
                      $\alpha_1\neq \alpha_2 ;$
                      $\{({\cal{X}}_1, {\cal{Y}}_1), ({\cal{X}}_2, {\cal{Y}}_2)\} \subseteq \mathit{Var}[E_C(\bullet t_i, t_i)] ;$\\
                     $ F_R(\bullet t_i,t_i) \cap(\overline{A}\cup\overline{B})\cap\type({\cal{X}}_1\cup {\cal{X}}_2\cup {\cal{Y}}_1\cup {\cal{Y}}_2)=\emptyset
                      \}$\\
-- Guard of BC1 transition evaluates whether a set of base instances ${\cal{X}}_1$ of a molecule $({\cal{X}}_1, {\cal{Y}}_1)$ contains an instance $\alpha_1$ and a set of instances ${\cal{X}}_2$ of a molecule 
$({\cal{X}}_2, {\cal{Y}}_2)$ contains an instance $\alpha_2$, both sets are obtained for the only input place.
The types of those instances form a label of an arc between the input place and the transition 
in the original RPN,
$\alpha_1$ and $\alpha_2$ differs, and the molecules do not contain negative base nor bond types. 
It might also happen that a new bond is created within the already existing molecule
- in that case both instances $\alpha_1, \alpha_2$ are unbonded and contained in molecule $({\cal{X}}_1, {\cal{Y}}_1)$, 
and the second molecule $({\cal{X}}_2,{\cal{Y}}_2)$ is empty (an idle token).
        
                   ${\bm{G_{C}^{BC2}}} = \{\bm{t_{i}}\mapsto (\alpha_1\in {\cal{X}}_1 \land \alpha_2\in {\cal{X}}_2) \mid $\\
                      $t_i\in T_R^{BC2} ; 
                      \{\type(\alpha_1),\type(\alpha_2)\}\subset \bigcup_{X\in{F_R(\bullet t_i,t_i)}}X 							;$\\
                      $\alpha_1\neq \alpha_2 ;
                      p_1 \neq p_2 \in \bullet t_i ; $\\
                      $ ({\cal{X}}_1, {\cal{Y}}_1)\in Var[E_C(p_1, t_i)] ;$
                      $ ({\cal{X}}_2, {\cal{Y}}_2)\in Var[E_C(p_2, t_i)] ;$\\
                     $ \bigcup_{X\in{F_R(\bullet t_i,t_i)}}X\cap(\overline{A}\cup\overline{B})\cap\type({\cal{X}}_1\cup {\cal{X}}_2\cup {\cal{Y}}_1\cup {\cal{Y}}_2)=\emptyset
                      \}$\\
-- Guard of BC2 transition evaluates whether a set of base instances ${\cal{X}}_1$ of a molecule $({\cal{X}}_1, {\cal{Y}}_1)$
obtained from the first input place contains an instance $\alpha_1$, set of base instances ${\cal{X}}_2$ of a molecule 
$({\cal{X}}_2, {\cal{Y}}_2)$
obtained from the second input place contains an instance $\alpha_2$, 
types of those instances form labels of arcs between the input places and the transition 
in the original RPN,
$\alpha_1$ and $\alpha_2$ differs, there are two different input places,
and the molecules obtained from input places do not contain negative base nor bond types.\\
                   ${\bm{G_{C}^{TRN}}} = \{\bm{t_{i}}\mapsto (\alpha\in {\cal{X}}) \mid $\\
                      $t_i\in T_R^{TRN} ; $
                      $\type(\alpha)\in F_R(\bullet t_i,t_i) ; $
                      $E_C(\bullet t_i, t_i)=\{({\cal{X}},{\cal{Y}})\} ;$\\
                      $F_R(\bullet t_i,t_i)\cap(\overline{A}\cup\overline{B})\cap\type({\cal{X}}\cup {\cal{Y}})=\emptyset
                      \}$\\
-- Guard of TRN transition evaluates whenever a set of base instances ${\cal{X}}$ of a molecule $({\cal{X}}, {\cal{Y}})$
obtained from the input place contains an instance $\alpha$,
type of $\alpha$ form a label of an arc between the input place and the transition,
and the molecule does not contain negative base nor bond types.
 %-----------
 % GUARDS FOR REVERSE TRANSITIONS
 %------------
 
% $\{tr_i\mapsto ( 
% ( \bigwedge_{t_l\in dph(t_i)} (n_l,t_l,t_i,\{a_1,a_2\})\in H_i ) \wedge
% (  \{a_1,a_2\}\in \bigcup_{p_l\in rin(t_i)}{y_l} ; \\
% (x_j \neq \emptyset \Rightarrow \bigwedge_{p_l\in rin(t_i), l \neq j} (x_l = \emptyset ; y_l = \emptyset) ))
% \mid \\
% t_i\in T_R^{BC1} \cup T_R^{BC2},
% n_l\in \Nat_b, \type(a_1)=\alpha_1, \type(a_2)=\alpha_2
% \wedge 
% H_i=E_C(h_i,tr_i)  
% \wedge \\
% (x_l,y_l)=E_C(p_l,tr_i )\ \mathrm{for}\ {p_l\in rin(t_i)} 
% )
% $
%   
% $\{tr_i\mapsto ( 
% ( \bigwedge_{t_l\in dph(t_i)} (n_l,t_l,t_i,\{a_1,a_2\})\in H_i ) \wedge
% (  \{a_1,a_2\}\in \bigcup_{p_l\in rin(t_i)}{x_l} ; \\
% (x_j \neq \emptyset \Rightarrow \bigwedge_{p_l\in rin(t_i), l \neq j} (x_l = \emptyset ; y_l = \emptyset) ))
% \mid \\
% t_i\in T_R^{TRN},
% n_l\in \Nat_b, \type(a_1)=\alpha_1, \type(a_2)=\alpha_2
% \wedge 
% H_i=E_C(h_i,tr_i)  
% \wedge
% (x_l,y_l)=E_C(p_l,tr_i )\ \mathrm{for}\ {p_l\in rin(t_i)} 
% )
% $
%=====================================================================================

$\bm G_{C}^{t_0} = \{\bm{t_{0}}\mapsto false\}$ \\
-- Guard of the initial transition $t_0$ - always returns \textit{false},
hence the transition cannot be executed.\\
$\bm G_{C}^{\overline{t_0}} = \{\bm{tr_{0}}\mapsto false\}$ \\
-- Guard of reversal transition for the initial transition $t_0$ - always returns \textit{false},
hence the transition cannot be executed.\\
The last two guards are a little bit more complex, that is why we include functions in their descriptions.
Those functions are: \emph{isElement} and \emph{numOfnonEmpty}\footnote{ 
Exemplary implementations of those functions are included in colour Petri nets 
generated by our application. Their formal definitions are included 
in descriptions of guards.}. The function \emph{isElement} returns \textit{true} if its first argument
is an element of the set given as the second argument. In the opposite case the function returns \textit{false}.
The function \emph{numOfnonEmpty} counts how many of its arguments are equal to $(\es, \es)$ and 
returns that number. 

Both following guards have the same construction. The first element of a guard
is a logical conjunction of $\#dpc(t_i)$ conditions, each of them ensures that
the token describing the history of $t_i$ and obtained from the place $h_i$ (which is 
represented as $E_C(h_i,tr_i)$ in the guards) contains 4-tuple related to the execution
which is reversed. In the forward execution of the transition (to be reversed) the base $\alpha$ was transported (for transporting transition) 
or a bond between instances $\alpha_1$ and $\alpha_2$ was created (for BC1 or BC2 transition), which
from now on is denoted as $\bondins{\alpha_1}{\alpha_2}$.
This part of the guards is very important because exactly here the choice: \emph{which execution would be reversed?}
(which is equivalent to the choice: \emph{execution related to which instances would be reversed?})
is made. In CPN examples, prepared using CPN-Tools, this choice can be made by the user or randomly.
The next part of the guards checks whether the set consisting of instances of bases (for TRN transition)
or bonds (for BC1 and BC2 transition) obtained from all input places
of the transition (to be reversed), contains instances related to its forward execution.
The last part of the guards assures that from all tokens obtained from input places only one
describes a molecule, the remaining ones should be idle tokens.

$\bm{G_{C}^{\overline{TRN}}} = \{\bm{tr_{i}}\mapsto$ 
$(\bigwedge_{t_j\in dpc(t_i)}$ isElement($(k_j,t_j,t_i,\{\alpha\}), E_C(h_i,tr_i))$;
isElement($\alpha, \bigcup_{p_g \in rin(t_i)} {\cal{X}}_g)$; 
\\
numOfnonEmpty($\{ ({\cal{X}}_g, {\cal{Y}}_g) \mid p_g \in rin(t_i)\})$)$=1$\\
where\\
$t_i\in T_R^{TRN}$
$;$
$\type(\alpha)\in F_R(t_i,t_i \bullet);$ 
$\forall_{p_g \in rin(t_i)}({\cal{X}}_g,{\cal{Y}}_g)=b(E_C(p_g,tr_i))$
isElement($q, Q$)$=true$ if and only if $q \in Q ;$\\
numOfnonEmpty($Q$)$=\#\{(q_1, q_2) \in Q \mid (q_1, q_2) \neq (\es, \es)\} $
and for backtracking $k_j$ is fixed as follows: $k_j = b(E_C(h_{ij}, tr_i))$ for $i < j$ or $k_j = b(E_C(h_{ji}, tr_i))$
for $i > j$
$\}$

$\bm{G_{C}^{\overline{BC1 \cup BC2}}} = \{\bm {tr_{i}}\mapsto$ 
$(\bigwedge_{t_j\in dpc(t_i)}$ isElement($(k_j,t_j,t_i,\{\bondins{\alpha_1}{\alpha_2}\}),$ $ E_C(h_i,tr_i)$); 
isElement($\bondins{\alpha_1}{\alpha_2}, \bigcup_{p_g \in rin(t_i)} {\cal{Y}}_g$);\\
numOfnonEmpty($\{ ({\cal{X}}_g, {\cal{Y}}_g) \mid p_g \in rin(t_i)\})$)$=1$\\
where\\
$t_i\in (T_R^{BC1} \cup T_R^{BC2});$
$\{\type(\alpha_1), \type(\alpha_2)\} \in F_R(t_i,t_i\bullet )$ 
$;$\\
$\forall_{p_g \in rin(t_i)}({\cal{X}}_g,{\cal{Y}}_g)=b(E_C(p_g,tr_i)) ;$\\
isElement($q, Q$)$=true$ if and only if $q \in Q ;$\\
numOfnonEmpty($Q$)$=\#\{(q_1, q_2) \in Q \mid (q_1, q_2) \neq (\es, \es)\} $
and for backtracking $k_j$ is fixed as follows: $k_j = b(E_C(h_{ij}, tr_i))$ for $i < j$ or $k_j = b(E_C(h_{ji}, tr_i))$
for $i > j$
$\}$\\

%=====================================================================================
\item[{\bm{ $E_{C}$}=}]$\{\bm{(p,t)}\mapsto ({\cal{X}},{\cal{Y}})  \mid  (p,t)\in Domain(F_R);
t \in T^{TRN}_R \cup T^{BC2}_R\} $\\
Description of input arcs from the original RPN for $TRN$ and $BC2$ transitions 
 - the transfer of one molecule
$({\cal{X}},{\cal{Y}})$ obtained from the place $p$. \\
 $\;\cup$ \\
$\{\bm{(p,t)}\mapsto (1`({\cal{X}}_1,{\cal{Y}}_1) +\!+ 1`({\cal{X}}_2,{\cal{Y}}_2)) \mid  (p,t)\in Domain(F_R);
t \in T^{BC1}_R\} $\\
Description of input arcs from the original RPN for $BC1$ transition 
 - the transfer of two molecules from place $p$ (in the guard it is assumed that one of those
 molecules may be empty).
				\\$\;\cup$\\
                   $\{\bm{(t,p)}\mapsto \idle  \mid  (p,t)\in Domain(F_R)\}; t \in T^{TRN}_R \cup T^{BC2}_R\}$\\
Description of arcs opposite to input arcs from the original RPN. An~idle token is transferred for $TRN$ or $BC2$ transition.
\\$\;\cup$\\
                   $\{\bm{(t,p)}\mapsto 2`\idle  \mid  (p,t)\in Domain(F_R)\}; t \in T^{BC1}_R\}$\\
Description of arcs opposite to input arcs from the original RPN. An~idle token is transferred for $BC1$ transition.\\
%                   $\;\cup$\\
%                   $\{(t,p)\mapsto ({\cal{X}},{\cal{Y}}\cup \{\bondins{\alpha_1}{\alpha_2}\} )\mid  
%                           E_C(\bullet t,t)=\{({\cal{X}},{\cal{Y}})\} ; 
%                           \{\type(\alpha_1),\type(\alpha_2)\}\in F_R(t,p) ;
%                           t\in T_R^{BC1} ; (t,p)\in Domain(F_R)\}$ \\
%\todo{w CPN tool zobaczyc czy to jest wykonalne czy da sie z jednej molukuly albo z dwoch
%i wtedy zmienic dopiero tu i guarda}
%Description of output arcs for BC1 transitions, similar to the ones from the RPN. It contains the transfer 
%of instances of bases and bonds 
%obtained from the input place and the new bond. Types of the instances in the new bond should be
%consistent with the label of the arc in the RPN. \\
                   $\;\cup$\\
                   $\{\bm{(t,p)}\mapsto ({\cal{X}},{\cal{Y}})\mid  
                           E_C(\bullet t,t)=\{({\cal{X}},{\cal{Y}})\} ; 
                           t\in T_R^{TRN} ; (t,p)\in Domain(F_R)\}$ \\
Description of output arcs for TRN transitions, similar to the ones from the RPN. 
They contain transfer of the molecule obtained from the input place.\\
                   $\;\cup$\\
                   $\{\bm{(t,p)}\mapsto ({\cal{X}}_1\cup {\cal{X}}_2,{\cal{Y}}_1\cup {\cal{Y}}_2\cup\{\bondins{\alpha_1}{\alpha_2}\}) \mid 
                           E_C(\bullet t)=\{({\cal{X}}_1,{\cal{Y}}_1),({\cal{X}}_2,{\cal{Y}}_2)\} ;$\\ 
                           $\{\type(\alpha_1),\type(\alpha_2)\}\in F_R(t,p) ;
                           t\in T_R^{BC1} \cup T_R^{BC2}; (t,p)\in Domain(F_R)\}$ \\
Description of output arcs for BC1 and BC2 transitions. They describe the transfer of the molecules containing the instances of bases and bonds, 
obtained from the input places (BC2) or place (BC1) and the new bond. Types of the instances in the new bond should be
consistent with the label of the arc in the RPN. 
                 \\  $\;\cup$\\
                   $\{\bm{(p,t)}\mapsto \idle  \mid  (t,p)\in Domain(F_R); t\in T_R\}$\\ 
Description of arcs opposite to input arcs from the original RPN. An~idle token is transferred.\\
                   $\;\cup$\\
                   $\{\bm{(h_{jk},t_{i})}\mapsto cnt_l \mid 
                    t_i\in T_R ; \{i,l\}=\{j,k\} ; t_l\in dpc(t_i) \}$\\
Description of arc from connection history place to a transition. The~value obtained from that place is represented by variable $cnt_l$.\\
                   $\;\cup$\\
                   $\{\bm{(t_{i},h_{jk})}\mapsto E_C(h_{jk},t_i)+1\mid t_i\in T_R ; \{i,l\}=\{j,k\} ; t_l\in dpc(t_i)\}$\\
Description of the arc from a transition to its connection history place. It describes the transfer of the value obtained from that place (by the opposite arc)
increased by 1.\\
                   $\;\cup$\\
                   $\{\bm{(h_{i},t_{i})}\mapsto H_l\mid  
                   H_l\in 2^{(\Nat_b\times T_R \times T_R \times 2^{\written{A}})}\}$\\
Description of the arc from transition history place to the transition. The value obtained from that place is represented by variable $H_l$ and it contains the whole history of transition $t_i$ (a set of 4-tuples).\\
                   $\;\cup$\\
                   $\{\bm{(t_{i},h_{i})}\mapsto E_C(h_i,t_i)\cup
                   \bigcup_{t_l\in dpc(t_i)}
                   \{(E_C(h_{jk},t_i)+1,t_l,t_i,\{\bondins{\alpha_1}{\alpha_2} \})\}\}$\\
                   where \\
                           $  t_i\notin T_R^{TRN} ; \{i,l\}=\{j,k\} ;$
                           $\bondins{\type(\alpha_1)}{\type(\alpha_2)} \in F_R(t_i,t_i\bullet)\}$\\Description of the arc from BC1 or BC2 transition to its transition history place.
The value obtained from the transition history place is transferred back (its described by $E_C(h_i,t_i)$) and a new 4-tuple is added for every transition from $dpc(t_{i})$. Each tuple 
consists of 4 components: the first is a number of current 
execution of $t_{i}$ in the sequence of executions of $t_{i}$ and $t_{l}$ - this value is obtained from $h_{il}$ or $h_{li}$, the next two components are identifiers of transitions and the last one is the description of the bond created during the considered execution.
\\
                   $\;\cup$\\
                   $\{\bm{(t_{i},h_{i})}\mapsto E_C(h_i,t_i)
                   \cup\bigcup_{t_l\in dpc(t_i)}\{E_C(h_{jk},t_i)+1,t_l,t_i,\{\alpha\})\}$ \\
                   where\\
                           $t_i\in T_R^{TRN}; \{i,l\}=\{j,k\} ;
                           \type(\alpha) \in F_R(t_i,t_i\bullet)\}$
\\Description of the arc from TRN transition to its transition history place.
It is very similar to the previous one, except for the last component of the 4-tuples - in 
this case it is the description of the base instances which were transferred during the considered execution.\\                                
$\;\cup$\\   
$\{\bm{(h_{jk},tr_{i})}\mapsto cnt_l\mid  t_i\in T_R ; \{i,l\}=\{j,k\} ; t_l\in dpc(t_i)\}$\\
Description of the arc from transition history counter place of $t_{i}$ to its reversing transition 
$tr_{i}$. The value obtained from the place is represented by $cnt_{l}$ and it is a number of
executions of transitions $t_{i}$ and $t_{l}$.\\   
$\;\cup$\\ 
$\{\bm{(tr_{i},h_{jk})}\mapsto E_C(h_{jk},tr_i)-1\mid t_i\in T_R ; \{i,l\}=\{j,k\} ; t_l\in dpc(t_i) \}$\\
Description of the arc from reversing transition $tr_{i}$ to connection history place of $t_{i}$. It describes the transfer of the value obtained from the connection history place 
by the transition $tr_{i}$ decreased by one.\\
$\;\cup$\\
 $\{\bm{(h_{j},tr_{i})}\mapsto H_j\mid  (H_j\in 2^{(\Nat_b\times T_R \times T_R \times 2^{\written{A}})\\ }
 ; (t_j \in dph(t_i) \lor j = i ))\}$\\
Description of the arc from transition history place of $t_{i}$ to its reversing transition. The value obtained from that place is represented by variable $H_j$ and it contains the whole history of transition $t_i$ (a set of 4-tuples). 
 \\
$\;\cup$\\
$\{\bm{(p,tr_{i})}\mapsto 2`\idle  \mid   (p \in rin(t_i) ; 
\forall_{t_j \in T_R} p \notin t_{j}\bullet) \}$
\\
Description of the arc between a place to a reversing transition. The place has to be in a set
$rin(t_i)$ and it cannot be an input place to any transition. Then two idle tokens are transferred from the place.
\\
$\;\cup$\\
$\{\bm{(p,tr_{i})}\mapsto (1`({\cal{X}},{\cal{Y}}) ++ 1`\idle)  \mid  (p \in rin(t_i) ; 
\exists_{t_j \in T_R} p \in t_{j}\bullet) \}$\\
Description of the arc between a place to a reversing transition. The place has to be in a set
$rin(t_i)$ and it has to be an input place to some transition from the net. Then a molecule and
an idle token are transferred from the place (during execution the molecule also can be an idle token).
\\
$\cup$\\
$\{\bm{(tr_{i},h_{j})}\mapsto updateExtHist(k_j \in \mathit{Var}[G_C(tr_i)], E_C(h_j, tr_i)) $\\
where \\
$t_i \in T_R ; t_j \in dph(t_i) ; (b(k_j), t_{j}, t_{i}, Y) \in b(E_C(h_i, tr_i)) ;$\\
$ updateExtHist(k_j, E_C(h_j, tr_i)) =  \bigcup_{(k, t_{g\neq i}, t_{j}, X) \in b(E_C(h_j, tr_i))}
(k, t_{g}, t_{j}, X) $\\
$\cup\bigcup_{(k < k_j, t_{i}, t_{j}, X) \in b(E_C(h_j, tr_i))}
(k, t_{i}, t_{j}, X) \cup \bigcup_{(k > k_j, t_{i}, t_{j}, X) \in b(E_C(h_j, tr_i))}
(k-1, t_{i}, t_{j}, X)\}$\\
Description of the arc between the reversing transition of $t_{i}$ and history places of other transitions.
It contains calling of \textit{updateExtHist()} function. The first argument of the function is 
number $k_{j}$ which is the first component of 4-tuple from history of transition $t_{j}$ which
have been binded during evaluation of the $tr_{i}$ guard. The second argument of \textit{updateExtHist()} are elements of $t_{j}$ history and the function edits them:
elements not related to the pair $t_{i}$ and $t_{j}$ are not changed, elements related to $t_{i}$ and $t_{j}$ 
with the first component $k$ smaller than $k_{j}$ are also not changes, elements related to $t_{i}$ and $t_{j}$ 
with $k$ greater than $k_{j}$ are adjusted by decreasing $k$ by 1.
\\
$\cup$\\
$\{\bm{(tr_{i},h_{i})}\mapsto $\\
$updateIntHist( K=\{k_j \in \mathit{Var}[G_C(tr_i)] \mid t_j \in dpc(t_i) \}, E_C(h_i, tr_i))$\\
where \\
$t_i \in T_R ; \forall_{k_j \in K} (b(k_j), t_{j}, t_{i}, Y) \in b(E_C(h_i, tr_i)) ;$\\
$ updateIntHist( K, E_C(h_i, tr_i)) =  
\bigcup_{(k < k_j, t_{j}, t_{i}, X) \in b(E_C(h_i, tr_i))}
(k, t_{j}, t_{i}, X) $
$\cup\bigcup_{(k > k_j, t_{j}, t_{i}, X) \in b(E_C(h_i, tr_i))}
(k-1, t_{j}, t_{i}, X)\}$\\
$\cup$\\
Description of the arc between the reversing transition of $t_{i}$ and history places of transition $t_{i}$.
It contains calling of \textit{updateIntHist()} function. The first argument of the function are 
numbers $k_{j}$ which are the first components of 4-tuple from $t_{i}$ history related to pairs
$t_{i}$ and $t_{j} \in dpc(t_{i})$ which
have been binded during evaluation of the $tr_{i}$ guard. The second argument of \textit{updateIntHist()} are elements of $t_{i}$ history and the function edits them:
elements related to $t_{i}$ and $t_{j}$ 
with the first component $k$ smaller than $k_{j}$ are not changes, elements related to $t_{i}$ and $t_{j}$ 
with $k$ greater than $k_{j}$ are adjusted by decreasing $k$ by 1.
\\
$\{\bm{(tr_{i},p)}\mapsto (1`({\cal{X}}_1,{\cal{Y}}_1) ++ 1`(\es, \es)) $\\
where\\
$t_i \in T_R^{TRN} ; p \in rout(t_i) ; \{p\} = t\bullet ; $\\
having $({\cal{X}}_g, {\cal{Y}}_g) = E_C(p_g, tr_i)$ and $\alpha \in \mathit{Var}[G_C(tr_i)]$:\\
$({\cal{X}}_1,{\cal{Y}}_1) = (\es, \es) $ if\\
$t \neq max((\bigcup_{t_j \in (dph(t_i) \cup t_i)} b(E_C(h_j, tr_i))
|_{con(\alpha, \bigcup_{p_g \in rin(t_i)} ({\cal{X}}_g \cup {\cal{Y}}_g))}) ;$
${\cal{X}}_1\cup {\cal{Y}}_1 = con(\alpha, \bigcup_{p_g \in rin(t_i)} ({\cal{X}}_g \cup {\cal{Y}}_g))$ if\\ 
$t = max((\bigcup_{t_j \in (dph(t_i) \cup t_i)} b(E_C(h_j, tr_i))
|_{con(\alpha, \bigcup_{p_g \in rin(t_i)} ({\cal{X}}_g \cup {\cal{Y}}_g))})
\}$\\
Description of the arc between reversing transition of transporting transition $t_{i}$ and it 
output place $p$. Transition $t_{i}$ transported base instance $\alpha$ in the execution which is 
withdrewed in the current execution of $tr_{i}$ - value of $\alpha$ is evaluated by the $tr_{i}$ guard.
Place $p$ is an output place of some transition $t$. Molecules obtained by $tr_{i}$ from its input place
$p_{g}$ is denoted by $({\cal{X}}_g, {\cal{Y}}_g)$.
Transition $tr_{i}$ transports an idle token and $({\cal{X}}_1,{\cal{Y}}_1)$ token which can an idle token or a molecule. 
The $({\cal{X}}_1,{\cal{Y}}_1)$ token is an idle token if $t$ is not the maximal (last) transition of transitions from $dph(t_{i})$ based on
histories obtained from places $h_{j}$ among those transitions which used the molecule containing $\alpha$. The $({\cal{X}}_1,{\cal{Y}}_1)$ token is equal to the molecule containing $\alpha$ if $t$ is the maximal one.
\\
$\cup$\\
$\{\bm{(tr_{i},p)}\mapsto (1`({\cal{X}}_1,{\cal{Y}}_1) ++ 1`({\cal{X}}_2, {\cal{Y}}_2)) $\\
where\\
$t_i \in (T_R^{BC1} \cup T_R^{BC2}) ; p \in rout(t_i) ; \{p\} = t\bullet;$\\
having $({\cal{X}}_g, {\cal{Y}}_g) = E_C(p_g, tr_i)$ and $\bondins{\alpha_1}{\alpha_2} \in \mathit{Var}[G_C(tr_i)]$:\\
$({\cal{X}}_1, {\cal{Y}}_1) = (\es, \es)$ if\\
$t \neq max((\bigcup_{t_j \in (dph(t_i) \cup t_i)} b(E_C(h_j, tr_i))
|_{con(\alpha_1, \bigcup_{p_g \in rin(t_i)} ({\cal{X}}_g \cup {\cal{Y}}_g)\setminus \{\bondins{\alpha_1}{\alpha_2}\})});$
${\cal{X}}_1\cup {\cal{Y}}_1 = con(\alpha_1, \bigcup_{p_g \in rin(t_i)} ({\cal{X}}_g \cup {\cal{Y}}_g)\setminus \{\bondins{\alpha_1}{\alpha_2}\})$
if \\
$t = max((\bigcup_{t_j \in (dph(t_i) \cup t_i)} b(E_C(h_j, tr_i))
|_{con(\alpha_1, \bigcup_{p_g \in rin(t_i)} ({\cal{X}}_g \cup {\cal{Y}}_g)\setminus \{\bondins{\alpha_1}{\alpha_2}\})});$
${\cal{X}}_2\cup {\cal{Y}}_2 = (\es, \es)$ if \\
$( t \neq max((\bigcup_{t_j \in (dph(t_i) \cup t_i)} b(E_C(h_j, tr_i))
|_{con(\alpha_2, \bigcup_{p_g \in rin(t_i)} ({\cal{X}}_g \cup {\cal{Y}}_g)\setminus \{\bondins{\alpha_1}{\alpha_2}\})})$ 
$\lor ({\cal{X}}_1,{\cal{Y}}_1)= ({\cal{X}}_2,{\cal{Y}}_2));$\\
${\cal{X}}_2\cup {\cal{Y}}_2 = con(\alpha_2, \bigcup_{p_g \in rin(t_i)} ({\cal{X}}_g \cup {\cal{Y}}_g)\setminus \{\bondins{\alpha_1}{\alpha_2}\})$ if\\
$(t = max((\bigcup_{t_j \in (dph(t_i) \cup t_i)} b(E_C(h_j, tr_i))
|_{con(\alpha_2, \bigcup_{p_g \in rin(t_i)} ({\cal{X}}_g \cup {\cal{Y}}_g)\setminus \{\bondins{\alpha_1}{\alpha_2}\})});$\\
$({\cal{X}}_1,{\cal{Y}}_1) \neq ({\cal{X}}_2,{\cal{Y}}_2))$
$  \}$\\
Description of the arc between reversing transition of BC1 or BC2 transition $t_{i}$ and it 
output place $p$. Transition $t_{i}$ created a bond $\bondins{\alpha_1}{\alpha_2}$ in the execution which is 
withdrewed in the current execution of $tr_{i}$ - value of $\bondins{\alpha_1}{\alpha_2}$ is evaluated by the $tr_{i}$ guard. Molecules obtained by $tr_{i}$ from its input place
$p_{g}$ is denoted by $({\cal{X}}_g, {\cal{Y}}_g)$.
Place $p$ is an output place of some transition $t$.
Transition $tr_{i}$ transports two tokens: $({\cal{X}}_1,{\cal{Y}}_1)$ and $({\cal{X}}_2,{\cal{Y}}_2)$ - both can be an idle tokens. 
The $({\cal{X}}_1,{\cal{Y}}_1)$ token is an idle token if $t$ is not the maximal (last) transition of transitions from $dph(t_{i})$ based on
histories obtained from places $h_{j}$ among those transitions which used the molecule containing $\alpha_1$ after breaking bond $\bondins{\alpha_1}{\alpha_2}$. The $({\cal{X}}_1,{\cal{Y}}_1)$ token is equal to the molecule containing $\alpha_1$ after breaking $\bondins{\alpha_1}{\alpha_2}$ if $t$ is the maximal one. The same for $({\cal{X}}_2,{\cal{Y}}_2)$ but then we consider molecule containing $\alpha_2$. $({\cal{X}}_1,{\cal{Y}}_1)$ cannot be equal to $({\cal{X}}_2,{\cal{Y}}_2)$.

\item[{\bm{ $I_{C}$}=}]$\{\bm{p}\mapsto ConCom(M_0(p))++(\bound - \#ConCom(M_0(p)))`(\es, \es)\mid  p\in P_R\}\;\cup$
\\
                   $\{\bm{h_{i}}\mapsto \emptyset\mid t_i\in T_R\}\;\cup$\\
				   $\{\bm{h_{ij}}\mapsto 0\mid t_i,t_j \in T_R ; t_i\trord t_j ; t_j\in dpc(t_i)\}$\\
Initial expressions of places: places from $P_{R}$ contain molecules from the initial 
marking of $N_{R}$ and idle tokens (to fulfill $\bound$ strong safeness),
places $h_{i}$ contain empty sets and places $h_{ij}$ contain $0$. For technical reasons those
initial values in places are set by initialization transition $t_{0}$. That transition
is part of $dph(t)$ for every $t \in T_{R}$, is executed at the initial marking and cannot be reversed. It is added to the net so that the $max$ always exists.
\end{itemize}
%\todo{alg dla CPN - opisac dokladnie co sie przypisuje do strzalek, guardow itd, z uzyciem jezyka CPN,
%ze szczegolnym uwzglednienem punktow synchronizacji typu (a,n)(b,m).}
%\todo{finish this def, pokazac ze istnieje element max i jak go liczymy}
%$blockT: 2^{P_C} \times \written{A} \times T_R \rightarrow T_R \cup \{ \lambda \}$, where
% $blockT(LH, \alpha, t)$ for $LH$ being a local history of transition $t$, $\alpha \in \written{A}$, 
%$t \in T_R$ such that $\type(\alpha) \in F(t, t\bullet)$ and $\alpha$ such that
%there exists a tuple $(n, t_j, t, R) \in M(h)$ with $\alpha \in R$ ($h$
%is a history place for $t$),
%returns 

\section{State Equivalence}
\label{equivalence}
A state in an RPN is a pair $\state{M_R}{H_R}$, where $M_R$ is a marking and 
$H_R$ is a history function. According to definitions: 
\begin{itemize}
\item $M_R : P_R \rightarrow 2^{\written{A} \cup \written{B}}$,
\item ${H:T\rightarrow 2^{\Nat\times{2^{P\times \cal{A}}}}}$.
\end{itemize}
A state in a CPN generated from RPN is a marking $M_C$ of coloured Petri net,
%$M_C: P \rightarrow 2^{(2^{\written{A}} \times 2^{\written{B}})} \cup 2^{2^{(\Nat_b\times T_R \times T_R \times 2^{\written{A}})}} \cup 2^{\Nat_b}$ 
$M_C: P \rightarrow \Nat^{\Sigma_C}$
is a function that assigns multisets of tokens to places
consistently with $C_C$, as follows: 
\begin{itemize}
\item $M_C(p) \in \Nat^{(2^{\written{A}} \times 2^{\written{B}})} $, for $p \in P_R$,
\item $M_C(h_{i}) \in \Nat^{2^{(\Nat_b\times T_R \times T_R \times 2^{\written{A}})}}$,
for $t_i \in T_R$,
\item $M_C(h_{ij}) \in \Nat^{\Nat_b}$, for $t_i,t_j\in T_R ; i < j ; t_j\in dpc(t_i)$.
\end{itemize}

\begin{remark}
In both formalisms, namely RPNs and CPNs, the history of the whole performed execution has to be stored. In~RPNs it is conducted explicitly, with the use of history function: ${H:T\rightarrow 2^{\Nat\times{2^{P\times \cal{A}}}}}$ -- for transition $t\in T_R$, history $H(t)$ contains a set of pairs of the shape $(k_X,X)$, where $X$ is a set of pairs $\{(p_1,\alpha_1), (p_2,\alpha_2),\ldots,(p_l,\alpha_l)\}$. 
On the other hand, in CPNs the history is scattered among transitions history places. Every transition history place contains elements belonging to the following Cartesian product: $\Nat_b\times T_R\times T_R\times 2^{\cal{A}}$, i.e. elements of the form $(k_Y, t_i,t_j,Y)$.
Note that during the transformation we cannot just replace $X$ with $Y$ (nor the other way round), because they are of different types. 
Having a base instance belonging to the set $Y\subseteq\cal{A}$, we are able to find a suitable place by indicating the last transition which has been using the given base instance and utilizing its only output place (see Section~\ref{cpn-rpn} for details). 
On the other hand, having $X$, we are able to obtain the set of $Y$ by summing up the components located at the second coordinate of each pair: $Y=\bigcup_{(p,\alpha)\in X}\{\alpha\}$ (see Section~\ref{rpn-cpn}). 
\end{remark}

\subsection{Transformation of states from RPN to CPN}
\label{rpn-cpn}

Now, we focus on transformation of states from an RPN to a CPN. This process is
quite straightforward - the formulas presented below have to be used.

Let assume that $N_{R}$ is an RPN $(P_{R},T_{R}, F_{R}, A_{R}, B_{R})$
and $N_C(N_R)$ is a~CPN $(P_{C}, T_{C}, D_{C}, \Sigma_{C}, V_{C}, C_{C},
G_{C}, E_{C}, I_{C})$. Then to each state in $N_R$ we assign a single state in   $N_C(N_R)$ as follows:

\begin{itemize}
\item $M_C(p)({\cal{X}}) = 1$ for ${\cal{X}} \in ConCom(M_R(p)), p \in P_{R}$, \\
$M_C(p)((\es, \es)) = \bound - \#ConCom(M_R(p)), p \in P_{R}$. The number of tokens in every
 place $p \in P_{R}$ is equal to $\bound$. If a place at marking $M_{R}$ contains tokens,
 then equivalent tokens have to be present at marking $M_{C}$. To obtain a fixed number of tokens $\bound$, we fulfill the place with idle tokens.
\item $M_C(h_{ij}) = \#H_R(t_i) + \#H_R(t_j)$  for a connection 
history place $h_{ij}$. The content of such a place describes how many times transitions
$t_{i}$ and $t_{j}$ have been executed. Each execution of a transition
is related to one element in its $H_{R}$, hence it is sufficient to to sum up the elements in  $H_{R}$ for both transitions. 
\item $M_C(h_i) = $\\
$\bigcup_{(k,Y) \in H_R(t_i), t_j \in dpc(t_i)}
(\#{\{k'\mid{(k',Z) \in (H_R(t_i)} \cup H_R(t_j))\land (k' < k) \} + 1}, t_j, t_i, X)$, where $X=\bigcup_{(p,\alpha)\in Y}\{\alpha\}$, for a transition
history place $h_i$. One element of such a place is a 4-tuple and describes one execution of $t_{i}$ in the context of $t_{j}$. 
Last three components of that 4-tuple are easy to obtain - there are 
names of transitions and set $X$, which is obtainable from $H_{R}(t_{i})$  as a sum of all instances constituting the second coordinate of each pair $(p,\alpha)\in X$.  
To obtain the first component, we have to calculate how many times transitions $t_{i}$ 
and $t_{j}$ have been executed before the execution involving $X$ occurred. 
Those executions are indicated by elements of $H_{R}(t_{i})$ and $H_{R}(t_{j})$
with first components smaller than the first component in the element including $X$.
\end{itemize}

\subsection{Transformation of states from CPN to RPN for backtracking}
\label{cpn-rpn}

In the following section, we deliberate about the state transformation from a CPN to an RPN. It turns out that it is not as straightforward as the other way round. The main issue we face is that a single state of coloured Petri net constructed according the rules of transformation may correspond to many states of the initial reversible Petri net. The following example illustrates this phenomenon.

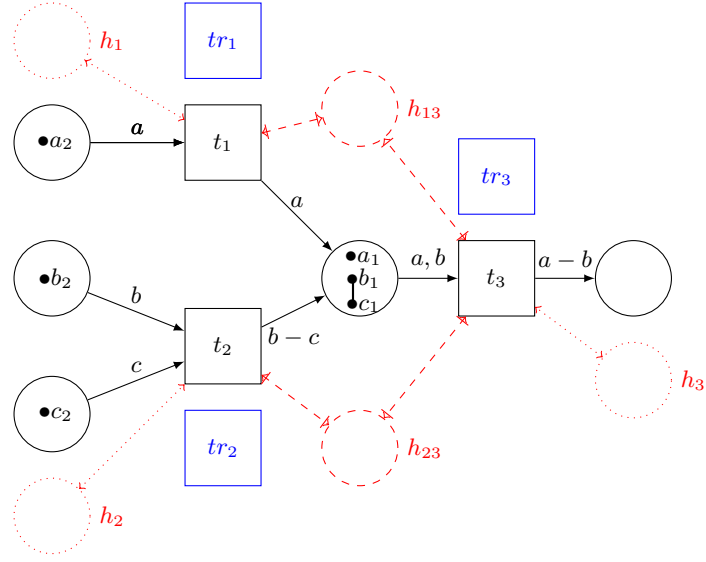
\begin{figure}
\begin{center}
\begin{tikzpicture}[scale=0.9]
%\node[circle,draw,minimum size=1cm] (p0) at (-1,4) {};
\node[circle,draw,minimum size=1cm] (p1) at (1.5,4) {};
\node at (1.55,4) {$\bullet a_2$};
\node[circle,draw,minimum size=1cm] (p2) at (1.5,2) {};
\node at (1.55,2) {$\bullet b_2$};
\node[circle,draw,minimum size=1cm] (p3) at (1.5,0) {};
\node at (1.55,0) {$\bullet c_2$};
\node[circle,draw,minimum size=1cm] (p4) at (6,2) {};
\node[circle,draw,minimum size=1cm] (p5) at (10,2) {};

\node[circle,dotted,red,draw,minimum size=1cm,label=right:\textcolor{red}{$h_1$}] (h1) at (1.5,5.5) {};
\node[circle,dotted,red,draw,minimum size=1cm,label=right:\textcolor{red}{$h_2$}] (h2) at (1.5,-1.5) {};
\node[circle,dotted,red,draw,minimum size=1cm,label=right:\textcolor{red}{$h_3$}] (h3) at (10,0.5) {};

\node[circle,dashed,red,draw,minimum size=1cm,label=right:\textcolor{red}{$h_{13}$}] (h13) at (6,4.5) {};
\node[circle,dashed,red,draw,minimum size=1cm,,label=right:\textcolor{red}{$h_{23}$}] (h23) at (6,-0.5) {};

\node at (6.05,2.3) {$\bullet a_1$};
\node (tok1) at (6.05,2) {$\bullet {b_1}$};
\node (tok2) at (6.05,1.6) {$\bullet {c_1}$};
\draw[thick] (5.9,2) -- (5.9,1.6) ;

%\node[draw,minimum size=1cm] (t0) at (-1,2) {$t_0$};
\node[draw,minimum size=1cm] (t1) at (4,4) {$t_1$};
\node[draw,blue,minimum size=1cm] (tr1) at (4,5.5) {$tr_1$};
\node[draw,minimum size=1cm] (t2) at (4,1) {$t_2$};
\node[draw,blue,minimum size=1cm] (tr2) at (4,-0.5) {$tr_2$};
\node[draw,minimum size=1cm] (t3) at (8,2) {$t_3$};
\node[draw,blue,minimum size=1cm] (tr3) at (8,3.5) {$tr_3$};

%\draw[-latex] (p0) to node [left] {$a,b,c$} (t0) ;
%\draw[-latex] (t0) to node [above] {$a$} (p1) ;
%\draw[-latex] (t0) to node [above] {$b$} (p2) ;
%\draw[-latex] (t0) to node [above] {$c$} (p3) ;

\draw[-latex] (p1) to node [above] {$a$} (t1) ;
\draw[-latex] (p1) to node [above] {$a$} (t1) ;
\draw[-latex] (p1) to node [above] {$a$} (t1) ;

\draw[-latex] (p2) to node [above] {$b$} (t2) ;
\draw[-latex] (p3) to node [above] {$c$} (t2) ;
\draw[-latex] (t1) to node [above] {$a$} (p4) ;
\draw[-latex] (t2) to node [below,inner sep=6pt] {$b-c$} (p4) ;
\draw[-latex] (p4) to node [above] {$a,b$} (t3) ;
\draw[-latex] (t3) to node [above] {$a-b$} (p5) ;

\draw[dashed,red,<|-|>] (t2) -- (h23) ;
\draw[dashed,red,<|-|>] (t3) -- (h23) ;
\draw[dashed,red,<|-|>] (t1) -- (h13) ;
\draw[dashed,red,<|-|>] (t3) -- (h13) ;

\draw[dotted,red,<->] (h1) -- (t1) ;
\draw[dotted,red,<->] (h2) -- (t2) ;
\draw[dotted,red,<->] (h3) -- (t3) ;

\end{tikzpicture}
\end{center}
\caption{Coloured Petri Net constructed on the basis of the reversible Petri net depicted in Figure~\ref{f:rpn-ex} and  according to out-of-causal-order semantics.}
\label{f:cpn-ex}
\end{figure}

\begin{example}
\label{exp3}
Let us construct a coloured Petri net $N_C$ on the basis of the~\RPN $N_R$ depicted in Figure~\ref{f:rpn-ex}.
At the very beginning, we add reverse transitions $tr_1,tr_2,tr_3$ to the net - they are depicted blue.
For the reasons of transparency and clarity, 
here we only mark the reversing transitions, without specifying all their connections.
We want to start with adding history places, first we add one transition history place for each transition  $t_1,t_2,t_3$ -- respectively: $h_1,h_2,h_3$. The places are marked dotted red. 
Next we need to compute the sets of places: 
$h_{13}, h_{23}$ - out-of-causal dependency counters 
and $h_{1}, h_{2}, h_{3}$ - out-of-causal dependency histories, for $t_1,t_2, t_3$ obtaining the following: 
\begin{itemize}
\item $dpc_{OOC}(t_1)=dph_{OOC}(t_1)=\{t_3\}$
\item $dpc_{OOC}(t_2)=dph_{OOC}(t_2)=\{t_3\}$
\item $dpc_{OOC}(t_3)=dph_{OOC}(t_3)=\{t_1,t_2\}$.
\end{itemize}
Therefore we need to add the following connection history places: $h_{13}$, $h_{23}$ - those places are depicted dashed red.
Now we should add the arcs between transitions and places  $h_1,h_2,h_3$ and $h_{13}, h_{23}$, according to the transformation described in Section~\ref{sec-transformation}. 
They are marked by dashed and dotted arrows.
Assume that, after the execution of the sequence $t_1t_2$, we obtain marking~$M_C'$, depicted
in Figure~\ref{f:cpn-ex}. The content of history places is as follows:
\begin{itemize}
\item $M_C'(h_1)=\{(1,3,1,\{a_1\})\}$
\item $M_C'(h_2)=\{(1,3,2,\{b_1,c_1\})\}$
\item $M_C'(h_3)=\emptyset$
\item $M_C'(h_{13})=1$
\item $M_C'(h_{23})=1$.
\end{itemize}
On the other hand, exactly the same marking one can obtain after the execution of $t_2t_1$. 
This is because in a history place of a transition we only store information concerning other transitions being in relation $\trord$ with it. In our example the two transitions $t_1$ and $t_2$ might be considered ``independent'', as they are not in the relation, namely $\lnot(t_1\trord t_2\lor t_2\trord t_1)$. 
The order of occurrences of the transitions $t_1$ and $t_2$ is not stored anywhere in the net.
For comparison, let us look again at the example shown in the Figure~\ref{f:rpn-ex}, 
considering only the black, solid part, which is the starting RPN. The markings obtained after $t_1t_2$ and after $t_2t_1$ are equal but, according to the semantics presented in Definition~\ref{forw}, the history function returns different values. 
\end{example}

\begin{remark}
Let us note that the above example would look significantly different when considering backtracking semantics, because in that case for a given transition $t$ the set of dependency histories $dph(t)$ would contain all transitions different from $t$.
\end{remark}

The above observations allows us formulate an extremely important 
\mbox{{\bf conclusion}}:\\
One state of the coloured Petri net created from a given \RPN as a result of transformation described above may correspond to many states of the original reversing Petri net.
Even when, for casual and out-of-casual-order semantics, there exists no one-to-one
correspondence between states in a CPN and corresponding RPN, for backtracking it is possible to 
obtain states in RPNs on the basis of states in CPNs.
It can be done quite
straightforwardly, by using the ensuing formulas. 
To every state in CPN $N_C(N_R)$ we assign a single state in RPN $N_R$ as follows:
\begin{itemize}
\item $M_R(p) = \bigcup_{({\cal{X}},{\cal{Y}}) \in M_C(p)}{\cal{X}} \cup {\cal{Y}} $, for $p \in P_R$; in the RPN places contain the same base and bond instances as in the CPN.
\\
\item $H_R(t_i)= \bigcup_{(k, j, i, X) \in M_C(h_i)} [K(X): \bigcup_{\alpha \in X}(t_\alpha \bullet, \alpha)]$ where for a particular set $X$: \\
$K(X) = 1 + \Sigma_{(k_j, j, i, X) \in M_C(h_i)}(k_j - 1) - $\\
$\#\{(k_g, j, i, Y) \in M_C(h_i)  \mid  (k_g < k_f) \land ((k_f, j, i, X)\in M_C(h_i))\} \cdot \frac{\#dpc(t_i) - 1}{\#dpc(t_i)}$,\\
$t_\alpha = max((\bigcup_{t_j \in (dph(t_i) \cup \{t_i\})} M_C(h_j))
|_{con(\alpha, \bigcup_{p \in rin(t_i)} M_C(p))})$.\footnote{Note that in above equations $i$ is
fixed as the number of transition $t_{i}$.}  

Calculating the value of the history function in RPNs is a bit tricky. It
contains two elements: a number in the sequence of executions and a~set of pairs: 
a place from which a base instance has been taken and the instance. Those base instances 
are included in markings of transition history places and can be obtained from the last components of 4-tuples. 
Using the partial order of transitions the last transition (before the considered execution of $t_{i}$), which has been using the given base
instance, can be indicated - hence its output place must be the place from which $t_{i}$ 
obtained the instance. The most difficult to compute is $K(X)$, which denotes the index of execution of the transition $t_{i}$ related to instances included in $X$ in the sequence of all executions of transitions. In CPNs this information is splitted among 4-tuples in marking of the transition history place. The first component in each 4-tuple: $k_{j}$ means that the considered execution
of $t_{i}$ (related to $X$) was $k_{j}$-th in the sequence of executions of transitions $t_{i}$ and $t_{j}$. Hence, there had to occur $k_{j} - 1$ executions before the considered one. However, among those $k_{j} - 1$ executions for each 4-tuple, also previous executions of $t_{i}$ are included, and they are included in each 4-tuple. That is why
we have to calculate how many times $t_{i}$ was executed before the considered execution and subtract the redundant information. 
Each previous execution of transition $t_{i}$ is described by $\#dcp(t_{i})$ elements of $M_{C}(h_{i})$. Hence, to obtain the number of previous executions we have to multiply by
$\frac{\#dpc(t_i) - 1}{\#dpc(t_i)}$.
Number $1$ is added to the obtained value to include the considered execution of $t_{i}$.
\end{itemize}
\subsection{Transformation of states from CPN to RPN - all modes}

The transformations described in Sections~\ref{rpn-cpn}~and~\ref{cpn-rpn} determine the \emph{correspondence} between states of an RPN and states of a corresponding CPN.
If $M_C$ is a state in the CPN obtained from $M_R$ in the RPN as a result of transformation described in Section~\ref{rpn-cpn}, or $M_R$ in the RPN is composed according to description presented in Section~\ref{cpn-rpn} from $M_C$ in the CPN, then we say that $M_R$ \emph{corresponds} to $M_C$ and $M_C$ \emph{corresponds} to $M_R$. Have in mind that, in backtracking semantics the \textit{correspondence} is one-to-one, while in 
causal and out-of-causal might be many-to-one (see Example~\ref{exp3}). To determine corresponding states in all semantics the following theorem should be used.

The following fact follows directly from the transformation described in Section~\ref{cpn-rpn}.
\begin{lemma}
For each state $M_C$ in the CPN there exists at least one reachable state $\state{M_R}{H_R}$ in RPN to which $M_C$ is assigned.
\label{states}
\end{lemma}
%\begin{proof}
%The initial marking generated by $I_C$ in CPN is assigned to the initial marking in RPN.
%Let assume that marking $M_C$ in CPN is assigned to $\state{M_R}{H_R}$ in RPN. $M_C$ can be transformed to a new state $M_{C}'$ 
%only by firing $t_i$ or $tr_j$. Analogically, in RPN in state $\state{M_R}{H_R}$ transition $t_i$ can be executed
%(if it has been executed in $M_C$) or $t_j$ can be reversed (if $tr_j$ has been executed in $M_C$) which would transform 
%to a new state $\state{M'_R}{H'_R}$ and which is assigned to $M'_C$. 
%\\
%\\Version 2:\\
%Let assume that CPN is in state $M_C$. Then, using formulas in Section~\ref{cpn-rpn}, a state $\state{M_R}{H_R}$ in RPN
%would be generated. Then $M_C$ is assigned to $\state{M_R}{H_R}$.
%\end{proof}

Now, we are ready to prove the main theorem of the paper.
\begin{theorem}
Let $N_R$ be an RPN and $N_C(N_R)$ the equivalent CPN. Then $\state{M_{C_0}}{H_0 = \es}$ 
(an initial marking of the RPN with
empty history) is equivalent to CPN marking obtained as a result of $I_C$/$t_0$. Moreover, if $\state{M_R}{H_R}$
is a state reachable in $N_R$ and $M_C$ is a corresponding marking of $N_C$ then $M_C$ is reachable in $N_C$.
In addition, if $t_i \in T_R$ is enabled in $N_R$ at $\state{M_R}{H_R}$ and 
its execution leads to $\state{M_R'}{H_R'}$, then
$t_i$ is enabled in $N_C$ at $M_C$ and its execution leads to $M_C'$ which corresponds to $\state{M_R'}{H_R'}$.
If $t_i \in T_R$ can be reversed at $\state{M_R}{H_R}$ and its reverse leads to $\state{M_R'}{H_R'}$, then
$tr_i$ is enabled in $N_C$ at $M_C$ and its execution leads to $M_C'$ which corresponds to $\state{M_R'}{H_R'}$.
\end{theorem}
\begin{proof}
Let $N_R$ be an RPN and $N_C(N_R)$ the equivalent CPN.
In the initial marking of $N_C(N_R)$ we assign to each place $p \in P_R$
a set of tokens $ConCom(M_{C_0}(p))$ supplemented by the proper number of idle tokens $(\es, \es)$. 
Moreover, $M_C(h_i) = \es$ since history $H_0$ is empty and $M_C(h_{ij}) = 0$, 
since all $H_R(t_i) = \es$.

Let $\state{M_R}{H_R}$ be a reachable state, such that $t_i$ is enabled at $M_R$. Let $M_C$ be
a marking corresponding to $\state{M_R}{H_R}$.\\
\\
\textbf{Part A: forward executions}\\ 
\\
\textbf{Case 1:} $t_i \in T^{TRN}_R$. \\
In this case, in the RPN we have one input arc and one output arc for $t_i$, with the same inscriptions 
equal to $a \in A$.
In the CPN, transition $t_i$ has an input arc labelled with single $({\cal{X}},{\cal{Y}})$.
By Definition~\ref{forward}(1), if $t_i$ is enabled, then in $S(\bullet t_i)$ we have only one instance $\alpha$ of type $a$.
Moreover, $\alpha \in M_C(\bullet t_i)$, hence one can choose a binding $b$ in which $\alpha \in b({\cal{X}})$.
We need to proceed with all additional negative inscriptions on the input arc. 
Note that, $b(({\cal{X}},{\cal{Y}})) = con(\alpha, M_C(\bullet t_i))$, hence by Definition~\ref{forward}(2,3)
$(\overline{A} \cup \overline{B}) \cap \type({\cal{X}} \cup {\cal{Y}}) = \es$.
This way the guard function of $t_i$ with binding $b$ returns true and $t_i$ is enabled in the CPN.

After the execution of $t_i$ in the RPN the contents of two places $\bullet t_i, t_i \bullet$ change,
$M_R'(\bullet t_i) = M_R(\bullet t_i)\setminus\connected(\alpha,M_R(\bullet t_i))$ and
$M_R'(t_i \bullet) = M_R(t_i \bullet) \cup \connected(\alpha,M_R(\bullet t_i))$.
On the other hand, in the CPN $M_C'(\bullet t_i) = M_C(\bullet t_i) - b(({\cal{X}},{\cal{Y}})) + (\es, \es)$, while
$M_C'(t_i \bullet) = M_C(t_i \bullet) - (\es, \es) + b(({\cal{X}},{\cal{Y}}))$. 
Moreover, $M_C'(h_i) = M_C(h_i) \cup \bigcup_{t_l\in dpc(t_i), i < l}\{b(E_C(h_{il},t_i))+1,t_l,t_i,\{\alpha\})
\cup \bigcup_{t_l\in dpc(t_i), i > l}\{b(E_C(h_{li},t_i))+1,t_l,t_i,\{\alpha\})$,
while for every $t_l \in dpc(t_i)$ we have 
$M_C'(h_{il}) = M_C(h_{il}) + 1$ for $i < l$  and 
$M_C'(h_{li}) = M_C(h_{li}) + 1$ for $i > l$.
The contents of the other places in both nets do not change. We are left with the determination of the value of history.
In the RPN we have
$H_R'(t_i) = H_R(t_i)\cup\{( k=max\{k'\mid(k',S') \in H(t'),t' \in T\}+1,\{(\bullet t_i, \alpha)\})\}$.
Note that, $M_C'$ is assigned to $M_R'$ since $M_C$ is assigned to $M_R$ and:
\begin{itemize}
\item $M_C'(p)= M_C(p)$ for $p \in P_R \setminus \{\bullet t_i, t_i \bullet \}$;
\item $M_C'(\bullet t_i)({\cal{X}}) = 1$ for ${\cal{X}} \in ConCom(M_R(\bullet t_i))\setminus\connected(\alpha,M_R(\bullet t_i))=$ \\
$ConCom(M_R(\bullet t_i)\setminus \connected(\alpha,M_R(\bullet t_i)))=ConCom(M_R'(\bullet t_i))$, \\
$M_C'(\bullet t_i)((\es, \es)) = M_C(\bullet t_i)((\es, \es)) + 1 = \#ConCom(M_R(\bullet t_i)) + 1 = \bound - \#ConCom(M_R'(\bullet t_i))$;
\item $M_C'(t_i \bullet)({\cal{X}}) = 1$ for ${\cal{X}} \in ConCom(M_R(t_i \bullet))\cup\connected(\alpha,M_R(\bullet t_i))=$\\
$ConCom(M_R(t_i \bullet)\cup \connected(\alpha,M_R(t_i \bullet)))=ConCom(M_R'(t_i \bullet))$, \\
$M_C'(t_i \bullet)((\es, \es)) = M_C(t_i \bullet)((\es, \es)) - 1 = \#ConCom(M_R(t_i \bullet)) - 1 = \bound - \#ConCom(M_R'(t_i \bullet))$;
\item $M_C'(h_{jl}) = M_C(h_{jl})$ for $j\neq i, l\neq i $;
\item $M_C'(h_{il}) = M_C(h_{il})+1 = \#H_R(t_i) + 1 + \#H_R(t_l) = \#H_R'(t_i) + \#H_R'(t_l)$, for
$i<l ; t_l\in dpc(t_i)$
\item $M_C'(h_{li}) = M_C(h_{li})+1 = \#H_R(t_i) + 1 + \#H_R(t_l) = \#H_R'(t_i) + \#H_R'(t_l)$, for
$i>l ; t_l\in dpc(t_i)$
\item $M_C'(h_j) = M_C(h_j)$ for $j\neq i$;
\item $M_C'(h_i) = $\\
$M_C(h_i) \cup \bigcup_{t_l\in dpc(t_i), i < l}\{b(E_C(h_{il},t_i))+1,t_l,t_i,\{\alpha\}) \cup $\\
$\bigcup_{t_l\in dpc(t_i), i > l}\{b(E_C(h_{li},t_i))+1,t_l,t_i,\{\alpha\}) = $\\
$\bigcup_{(k, \{ (p_\alpha', \alpha')\}) \in H_R(t_i), t_l \in dpc(t_i)}
\{(\#\{k'\mid\{ (k', \{ (p_\alpha'', \alpha'')\}) \in H_R(t_i) \cup H_R(t_l)\}\land k' < k \} + 1, t_l, t_i, \{ \alpha'\})\}\cup$\\
$\bigcup_{t_l\in dpc(t_i), i < l}\{b(E_C(h_{il},t_i))+1,t_l,t_i,\{\alpha\})\cup$\\
$\bigcup_{t_l\in dpc(t_i), i > l}\{b(E_C(h_{li},t_i))+1,t_l,t_i,\{\alpha\})=$\\
$\bigcup_{(k, \{ (p_\alpha', \alpha')\}) \in H_R'(t_i), t_l \in dpc(t_i)}
\{(\#\{k'\mid\{ (k', \{ (p_\alpha'', \alpha'')\}) \in H_R'(t_i) \cup H_R'(t_l)\}\land
k' < k \} + 1, t_l, t_i, \{ \alpha'\})\}$, where $p_{\alpha}$ denotes a place from which $\alpha$
is obtained. 
\end{itemize}
\textbf{Case 2:} $t_i \in T^{BC2}_R$. \\
In this case, in the RPN we have $\bullet t_i = \{p_1, p_2\}$, hence
it has two input arcs with inscriptions, precisely: $a_1$ arc from $p_1$, $a_2$ arc from $p_2$, and one output arc with inscription $\{\bondpair{a_1} {a_2}\}$, for $a_1, a_2 \in A$.
In the CPN, transition $t_i$ has two input arcs: one labelled with $({\cal{X}}_1,{\cal{Y}}_1)$, 
and the other with $({\cal{X}}_2,{\cal{Y}}_2)$.
By Definition~\ref{forward}(1), in $S(\bullet t_i)$ two instances are present: $\alpha_1$
of type $a_1$ and $\alpha_2$ of type $a_2$.
Moreover, $\alpha_1 \in M_C(p_1)$ and $\alpha_2 \in M_C(p_2)$, hence one can choose a binding $b$ in which 
$\alpha_1 \in b({\cal{X}}_1)$ and $\alpha_2 \in b({ \cal{X}}_2 )$. 
We need to proceed with all additional negative inscriptions on the input arc. 
Note that, $b(({\cal{X}}_1,{\cal{Y}}_1)) = con(\alpha_1, M_C(p_1))$  and
$b(({\cal{X}}_2,{\cal{Y}}_2)) = con(\alpha_2, M_C(p_2))$
hence by Definition~\ref{forward}(2,3)
$(\overline{A} \cup \overline{B}) \cap \type({\cal{X}}_1 \cup {\cal{Y}}_1) 
\cap \type({\cal{X}}_2 \cup {\cal{Y}}_2)= \es$.
This way the guard function of $t_i$ with binding $b$ returns true and $t_i$ is enabled in the CPN.

After the execution of $t_i$ in the RPN, the contents of three places $p_1, p_2 \in \bullet t_i$ and $t_i \bullet$ change,
$M_R'(p_1) = M_R(p_1)\setminus \connected(\alpha_1,M_R(p_1)) $,
$M_R'(p_2) = M_R(p_2)\setminus \connected(\alpha_2,M_R(p_2)) $
 and
$M_R'(t_i \bullet) = M_R(t_i \bullet) \cup \connected(\alpha_1,M_R(p_1)) 
\cup \connected(\alpha_2,M_R(p_2)) \cup \{ \bondpair{\alpha_1} {\alpha_2} \}$.
On the other hand, in the CPN 
$M_C'(p_1) = M_C(p_1) - b(({\cal{X}}_1,{\cal{Y}}_1)) + (\es, \es)$,
$M_C'(p_2) = M_C(p_2) - b(({\cal{X}}_2,{\cal{Y}}_2)) + (\es, \es)$, 
 while
$M_C'(t_i \bullet) = M_C(t_i \bullet) - (\es, \es) + b(({\cal{X}}_1\cup {\cal{X}}_2,{\cal{Y}}_1\cup {\cal{Y}}_2\cup\{\bondins{\alpha_1}{\alpha_2}\}))$. 
Moreover, $M_C'(h_i) = M_C(h_i) \cup \bigcup_{t_l\in dpc(t_i), i < l}\{b(E_C(h_{il},t_i))+1,t_l,t_i,\{\bondpair{\alpha_1} {\alpha_2}\})
\cup \bigcup_{t_l\in dpc(t_i), i > l}\{b(E_C(h_{li},t_i))+1,t_l,t_i,\{\bondpair{\alpha_1} {\alpha_2}\})$,
while for every $t_l \in dpc(t_i)$ we have 
$M_C'(h_{il})$ for $i < l$ is equal to $M_C(h_{il}) + 1$ and 
$M_C'(h_{li})$ for $i > l$ is equal to $M_C(h_{li}) + 1$.
The contents of the other places in both nets do not change. We are left with the determination of the value of history.
In the RPN we have
$H_R'(t_i) = H_R(t_i)\cup\{( k=max\{k'\mid(k',S') \in H(t'),t' \in T\}+1,\{(p_1, \alpha_1), (p_2, \alpha_2)\})\}$.
Note that, $M_C'$ is assigned to $M_R'$ since $M_C$ is assigned to $M_R$ and:
\begin{itemize}
\item $M_C'(p)= M_C(p)$ for $p \in P_R \setminus \{p_1, p_2, t_i \bullet \}$;
\item $M_C'(p_1)({\cal{X}}) = 1$ for ${\cal{X}} \in ConCom(M_R(p_1))\setminus\connected(\alpha_1,M_R(p_1))=$ \\
$ConCom(M_R(p_1)\setminus \connected(\alpha_1,M_R(p_1)))=ConCom(M_R'(p_1))$, \\
$M_C'(p_1)((\es, \es)) = M_C(p_1)((\es, \es)) + 1 = \#ConCom(M_R(p_1)) + 1 = \bound - \#ConCom(M_R'(p_1))$;
\item $M_C'(p_2)({\cal{X}}) = 1$ for ${\cal{X}} \in ConCom(M_R(p_2))\setminus\connected(\alpha_2,M_R(p_2))=$ \\
$ConCom(M_R(p_2)\setminus \connected(\alpha_2,M_R(p_2)))=ConCom(M_R'(p_2))$, \\
$M_C'(p_2)((\es, \es)) = M_C(p_2)((\es, \es)) + 1 = \#ConCom(M_R(p_2)) + 1 = \bound - \#ConCom(M_R'(p_2))$;
\item $M_C'(t_i \bullet)({\cal{X}}) = 1$ for ${\cal{X}} \in ConCom(M_R(t_i \bullet))\cup\connected(\alpha_1,M_R(p_1)) \cup \connected(\alpha_2,M_R(p_2)) \cup \{\bondpair{a_1} {a_2} \}=$\\
$ConCom(M_R(t_i \bullet)\cup (\connected(\alpha_1,M_R(p_1)) \cup \connected(\alpha_2,M_R(p_2)) \cup \{\bondpair{a_1} {a_2} \} ))=ConCom(M_R'(t_i \bullet))$, \\
$M_C'(t_i \bullet)((\es, \es)) = M_C(t_i \bullet)((\es, \es)) - 1 = \#ConCom(M_R(t_i \bullet)) - 1 = \bound - \#ConCom(M_R'(t_i \bullet))$;
\item $M_C'(h_{jl}) = M_C(h_{jl})$ for $j\neq i, l\neq i $;
\item $M_C'(h_{il}) = M_C(h_{il})+1 = \#H_R(t_i) + 1 + \#H_R(t_l) = \#H_R'(t_i) + \#H_R'(t_l)$, for
$i<l ; t_l\in dpc(t_i)$
\item $M_C'(h_{li}) = M_C(h_{li})+1 = \#H_R(t_i) + 1 + \#H_R(t_l) = \#H_R'(t_i) + \#H_R'(t_l)$, for
$i>l ; t_l\in dpc(t_i)$
\item $M_C'(h_j) = M_C(h_j)$ for $j\neq i$;
\item $M_C'(h_i) = $\\
$M_C(h_i) \cup \bigcup_{t_l\in dpc(t_i), i < l}\{b(E_C(h_{il},t_i))+1,t_l,t_i,\{\bondpair{\alpha_1} {\alpha_2}\}) \cup $\\
$\bigcup_{t_l\in dpc(t_i), i > l}\{b(E_C(h_{li},t_i))+1,t_l,t_i,\{\bondpair{\alpha_1} {\alpha_2}\}) = $\\
$\bigcup_{(k, \{ (p_{\alpha_{1}}', {\alpha_{1}}'), (p_{\alpha_{2}}', {\alpha_{2}}')\}) \in H_R(t_i), t_l \in dpc(t_i)}
\{(\#\{k'\mid\{ (k', \{(p_{\alpha_{1}}'', {\alpha_{1}}''), (p_{\alpha_{2}}'', {\alpha_{2}}'')\}) \in H_R(t_i) \cup H_R(t_l)\}\land k' < k \} + 1, t_l, t_i, \{ \bondpair{\alpha_1'} {\alpha_2'}\})\}\cup$\\
$\bigcup_{t_l\in dpc(t_i), i < l}\{b(E_C(h_{il},t_i))+1,t_l,t_i,\{\bondpair{\alpha_1} {\alpha_2}\})\cup$\\
$\bigcup_{t_l\in dpc(t_i), i > l}\{b(E_C(h_{li},t_i))+1,t_l,t_i,\{\bondpair{\alpha_1} {\alpha_2}\})=$\\
$\bigcup_{(k, \{ (p_{\alpha_{1}}', {\alpha_{1}}'), (p_{\alpha_{2}}', {\alpha_{2}}')\}) \in H_R'(t_i), t_l \in dpc(t_i)}
\{(\#\{k'\mid\{ (k', \{ (p_{\alpha_{1}}', {\alpha_{1}}'), (p_{\alpha_{2}}', {\alpha_{2}}')\}) \in H_R'(t_i) \cup H_R'(t_l)\}\land
k' < k \} + 1, t_l, t_i, \{\bondpair{\alpha_1'} {\alpha_2'}\})\}$, where $p_{\alpha}$ denotes a place from which $\alpha$
is obtained. \\
\end{itemize} 
\textbf{Case 3:} $t_i \in T^{BC1}_R$. \\
In this case, in the RPN $t_i$  has one input arc with inscription $\{a_1, a_2\}$ and
one output arc with inscription $\{\bondpair{a_1} {a_2}\}$ for $a_1, a_2 \in A$
In the CPN transition $t_i$ has an input arc labelled with $1`({\cal{X}}_1,{\cal{Y}}_1) ++ 1`({\cal{X}}_2,{\cal{Y}}_2)$.
By Definition~\ref{forward}(1), in $S(\bullet t_i)$ two instances are present: $\alpha_1$
of type $a_1$ and $\alpha_2$ of type $a_2$.
Moreover, $\alpha_1, \alpha_2 \in M_C(\bullet t_i)$, hence 
we have two possibilities of binding:\\
\textbf{Subcase A:} \\
$\alpha_1$ and $\alpha_2$ are included in the same molecule and 
one can choose a binding $b$ in which 
$\alpha_1, \alpha_2 \in b({\cal{X}}_1)$ and $({\cal{X}}_2,{\cal{Y}}_2) =  (\es, \es)$.
We need to proceed with all additional negative inscriptions on the input arc. 
Note that, $b(({\cal{X}}_1,{\cal{Y}}_1)) = con(\alpha_1, M_C(\bullet t_i)) = con(\alpha_2, M_C(\bullet t_i))$
hence by Definition~\ref{forward}(2,3)
$(\overline{A} \cup \overline{B}) \cap \type({\cal{X}}_1 \cup {\cal{Y}}_1) = \es$.
This way the guard function of $t_i$ with binding $b$ returns true and $t_i$ is enabled in the CPN.

After the execution of $t_i$ in the RPN the contents of two places $\bullet t_i, t_i \bullet$ change,
$M_R'(\bullet t_i) = M_R(\bullet t_i)\setminus \connected(\alpha_1,M_R(\bullet t_i) )$
 and
$M_R'(t_i \bullet) = M_R(t_i \bullet) \cup \bigcup_{ \alpha' \in S(\bullet t_i)}\connected(\alpha',M_R(\bullet t_i)) \cup \{ \bondpair{\alpha_1} {\alpha_2} \}$.
On the other hand, in the CPN 
$M_C'(\bullet t_i) = M_C(\bullet t_i) - b(({\cal{X}}_1,{\cal{Y}}_1)) - (\es, \es) + 2`(\es, \es)$, while
$M_C'(t_i \bullet) = M_C(t_i \bullet) - (\es, \es) + b(({\cal{X}}_1,{\cal{Y}}_1\cup\{\bondins{\alpha_1}{\alpha_2}\}))$. 
Moreover, $M_C'(h_i) = M_C(h_i) \cup \bigcup_{t_l\in dpc(t_i), i < l}\{b(E_C(h_{il},t_i))+1,t_l,t_i,\{\bondpair{\alpha_1} {\alpha_2}\})
\cup \bigcup_{t_l\in dpc(t_i), i > l}\{b(E_C(h_{li},t_i))+1,t_l,t_i,\{\bondpair{\alpha_1} {\alpha_2}\})$,
while for every $t_l \in dpc(t_i)$ we have 
$M_C'(h_{il})$ for $i < l$ is equal to $M_C(h_{il}) + 1$ and 
$M_C'(h_{li})$ for $i > l$ is equal to $M_C(h_{li}) + 1$.
The contents of the other places in both nets do not change. We are left with the determination of the value of history:
$H_R'(t_i) = H_R(t_i)\cup\{( k=max\{k'\mid(k',S') \in H(t'),t' \in T\}+1,\{(\bullet t_i, \alpha_1), (\bullet t_i, \alpha_2)\})\}$.
Note that, $M_C'$ is assigned to $M_R'$ since $M_C$ is assigned to $M_R$ and:
\begin{itemize}
\item $M_C'(p)= M_C(p)$ for $p \in P_R \setminus \{\bullet t_i, t_i \bullet \}$;
\item $M_C'(\bullet t_i)({\cal{X}}) = 1$ for ${\cal{X}} \in ConCom(M_R(\bullet t_i))\setminus\connected(\alpha_1,M_R(\bullet t_i))=$ \\
$ConCom(M_R(\bullet t_i)\setminus \connected(\alpha_1,M_R(\bullet t_i)))=ConCom(M_R'(\bullet t_i))$, \\
$M_C'(\bullet t_i)((\es, \es)) = M_C(\bullet t_i)((\es, \es)) - 1 + 2 = \#ConCom(M_R(\bullet t_i)) + 1 = \bound - \#ConCom(M_R'(\bullet t_i))$;
\item $M_C'(t_i \bullet)({\cal{X}}) = 1$ for ${\cal{X}} \in ConCom(M_R(t_i \bullet))\cup\connected(\alpha_1,M_R(\bullet t_i)) \cup \{\bondpair{\alpha_1} {\alpha_2} \}=$\\
$ConCom(M_R(t_i \bullet)\cup \connected(\alpha_1,M_R(t_i \bullet)))=ConCom(M_R'(t_i \bullet))$, \\
$M_C'(t_i \bullet)((\es, \es)) = M_C(t_i \bullet)((\es, \es)) - 1 = \#ConCom(M_R(t_i \bullet)) - 1 = \bound - \#ConCom(M_R'(t_i \bullet))$;
\item $M_C'(h_{jl}) = M_C(h_{jl})$ for $j\neq i, l\neq i $;
\item $M_C'(h_{il}) = M_C(h_{il})+1 = \#H_R(t_i) + 1 + \#H_R(t_l) = \#H_R'(t_i) + \#H_R'(t_l)$, for
$i<l ; t_l\in dpc(t_i)$
\item $M_C'(h_{li}) = M_C(h_{li})+1 = \#H_R(t_i) + 1 + \#H_R(t_l) = \#H_R'(t_i) + \#H_R'(t_l)$, for
$i>l ; t_l\in dpc(t_i)$
\item $M_C'(h_j) = M_C(h_j)$ for $j\neq i$;
\item $M_C'(h_i) = $\\
$M_C(h_i) \cup \bigcup_{t_l\in dpc(t_i), i < l}\{b(E_C(h_{il},t_i))+1,t_l,t_i,\{\bondpair{\alpha_1} {\alpha_2}\}) \cup $\\
$\bigcup_{t_l\in dpc(t_i), i > l}\{b(E_C(h_{li},t_i))+1,t_l,t_i,\{\bondpair{\alpha_1} {\alpha_2}\}) = $\\
$\bigcup_{(k, \{ (p_{\alpha_{1}}', {\alpha_{1}}'), (p_{\alpha_{1}}', {\alpha_{2}}')\}) \in H_R(t_i), t_l \in dpc(t_i)}
\{(\#\{k'\mid\{ (k', \{(p_{\alpha_{1}}'', {\alpha_{1}}''), (p_{\alpha_{1}}'', {\alpha_{2}}'')\}) \in H_R(t_i) \cup H_R(t_l)\}\land k' < k \} + 1, t_l, t_i, \{ \bondpair{\alpha_1'} {\alpha_2'}\})\}\cup$\\
$\bigcup_{t_l\in dpc(t_i), i < l}\{b(E_C(h_{il},t_i))+1,t_l,t_i,\{\bondpair{\alpha_1} {\alpha_2}\})\cup$\\
$\bigcup_{t_l\in dpc(t_i), i > l}\{b(E_C(h_{li},t_i))+1,t_l,t_i,\{\bondpair{\alpha_1} {\alpha_2}\})=$\\
$\bigcup_{(k, \{ (p_{\alpha_{1}}', {\alpha_{1}}'), (p_{\alpha_{1}}', {\alpha_{2}}')\}) \in H_R'(t_i), t_l \in dpc(t_i)}
\{(\#\{k'\mid\{ (k', \{ (p_{\alpha_{1}}', {\alpha_{1}}'), (p_{\alpha_{1}}', {\alpha_{2}}')\}) \in H_R'(t_i) \cup H_R'(t_l)\}\land
k' < k \} + 1, t_l, t_i, \{\bondpair{\alpha_1'} {\alpha_2'}\})\}$, where $p_{\alpha}$ denotes a place from which $\alpha$
is obtained. 
\end{itemize} 
\textbf{Subcase B:} \\
$\alpha_1$ and $\alpha_2$ are included in different molecules, hence
one can choose a binding $b$ in which 
$\alpha_1 \in b({\cal{X}}_1)$ and $\alpha_2 \in b({ \cal{X}}_2 )$. 
We need to proceed with all additional negative inscriptions on the input arc. 
Note that, $b(({\cal{X}}_1,{\cal{Y}}_1)) = con(\alpha_1, M_C(\bullet t_i)),
b(({\cal{X}}_2,{\cal{Y}}_2)) = con(\alpha_2, M_C(\bullet t_i))$
hence by Definition~\ref{forward}(2,3)
$(\overline{A} \cup \overline{B}) \cap \type({\cal{X}}_1 \cup {\cal{Y}}_1) 
\cap \type({\cal{X}}_2 \cup {\cal{Y}}_2) = \es$.
This way the guard function of $t_i$ with binding $b$ returns true and $t_i$ is enabled in the CPN.

After the execution of $t_i$ in the RPN the contents of two places $\bullet t_i, t_i \bullet$ change,
$M_R'(\bullet t_i) = M_R(\bullet t_i)\setminus (\connected(\alpha_1,M_R(\bullet t_i)) \cup \connected(\alpha_2,M_R(\bullet t_i)) )$
 and
$M_R'(t_i \bullet) = M_R(t_i \bullet) \cup \bigcup_{ \alpha' \in S(\bullet t_i)}\connected(\alpha',M_R(\bullet t_i)) \cup \{ \bondpair{\alpha_1} {\alpha_2} \}$.
On the other hand, in the CPN 
$M_C'(\bullet t_i) = M_C(\bullet t_i) - b(({\cal{X}}_1,{\cal{Y}}_1)) 
- b(({\cal{X}}_2,{\cal{Y}}_2)) 
+ 2`(\es, \es)$, while
$M_C'(t_i \bullet) = M_C(t_i \bullet) - (\es, \es) +
 b(({\cal{X}}_1 \cup {\cal{X}}_2,{\cal{Y}}_1\cup {\cal{Y}}_2 \cup\{\bondins{\alpha_1}{\alpha_2}\}))$. 
Moreover $M_C'(h_i) = M_C(h_i) \cup \bigcup_{t_l\in dpc(t_i), i < l}\{b(E_C(h_{il},t_i))+1,t_l,t_i,\{\bondpair{\alpha_1} {\alpha_2}\})
\cup \bigcup_{t_l\in dpc(t_i), i > l}\{b(E_C(h_{li},t_i))+1,t_l,t_i,\{\bondpair{\alpha_1} {\alpha_2}\})$,
while for every $t_l \in dpc(t_i)$ we have 
$M_C'(h_{il})$ for $i < l$ is equal to $M_C(h_{il}) + 1$ and 
$M_C'(h_{li})$ for $i > l$ is equal to $M_C(h_{li}) + 1$.
The contents of the other places in both nets do not change. We are left with the determination of the value of history:
$H_R'(t_i) = H_R(t_i)\cup\{( k=max\{k'\mid(k',S') \in H(t'),t' \in T\}+1,\{(\bullet t_i, \alpha_1), (\bullet t_i, \alpha_2)\})\}$.
Note that, $M_C'$ is assigned to $M_R'$ since $M_C$ is assigned to $M_R$ and:
\begin{itemize}
\item $M_C'(p)= M_C(p)$ for $p \in P_R \setminus \{\bullet t_i, t_i \bullet \}$;
\item $M_C'(\bullet t_i)({\cal{X}}) = 1$ for ${\cal{X}} \in ConCom(M_R(\bullet t_i))\setminus(\connected(\alpha_1,M_R(\bullet t_i))\cup \connected(\alpha_2,M_R(\bullet t_i)) )=$ \\
$ConCom(M_R(\bullet t_i)\setminus (\connected(\alpha_1,M_R(\bullet t_i)) \cup \connected(\alpha_2,M_R(\bullet t_i)) )=ConCom(M_R'(\bullet t_i))$, \\
$M_C'(\bullet t_i)((\es, \es)) = M_C(\bullet t_i)((\es, \es)) + 2 = \#ConCom(M_R(\bullet t_i)) + 2 = \bound - \#ConCom(M_R'(\bullet t_i))$;
\item $M_C'(t_i \bullet)({\cal{X}}) = 1$ for ${\cal{X}} \in ConCom(M_R(t_i \bullet))\cup\connected(\alpha_1,M_R(\bullet t_i)) 
\cup\connected(\alpha_2,M_R(\bullet t_i))
\cup \{\bondpair{\alpha_1} {\alpha_2} \}=$\\
$ConCom(M_R(t_i \bullet)\cup \connected(\alpha_1,M_R(t_i \bullet)  ))=ConCom(M_R'(t_i \bullet))$, \\
$M_C'(t_i \bullet)((\es, \es)) = M_C(t_i \bullet)((\es, \es)) - 1 = \#ConCom(M_R(t_i \bullet)) - 1 = \bound - \#ConCom(M_R'(t_i \bullet))$
\item Changes of markings of other places are the same as in the previous situation.\\
\end{itemize} 
\textbf{Part B: Reverse executions -- equivalence of enabledness}\\

Note that, in case of reverse executions, we have to deal with three different semantics, 
and the notion of transition enabledness differs depending on the chosen one. Therefore, we need to investigate three possible situations. \\
\\
\textbf{Case 1:} \\
Backtracking. Notice, that in the RPN transition $t$ is enabled to be reversed
when it is the last forward executed transition in the execution of the net, that is 
according to Definition~\ref{br-enb}, if it has the greatest number 
in $H(t')$ for $t' \in T$. 
In the CPN, the last executed transition $t$ can be established
as the one with the values on the first position in elements of its history place
equal to values obtained from respective connection history places related to $t$.
Note that, by the construction of connection history places, those values are the greatest possible. Hence, by the transformation transition $t$ is $bt-$enabled. \\
\\
\textbf{Case 2:} \\
Causal reversing. According to Definition~\ref{co-enabled} in the RPN transition $t$ is 
$co$-enabled if there is no other transition $t'$ that has been executed after $t$ 
(with value $k'$ from the history higher that $k$ assigned to $t$) and has used any token instance 
$\alpha_i$ utilised by $t$. In other words any other transition $t'$ dependent on $t$ 
has not been fired, and the dependency is based on token instances used by transitions. 
In the CPN, $t$ can be reversed, if token instances utilised by $t$ 
are present in places from $rin(t)$, which is for causal semantics the set of output places 
of $t$. Since $t$ can be reversed, any other transition $t'$ such that $t \bullet \cap \bullet t' \neq \es$ 
using the same token instances has not been executed. Once more, we may say
that any other transition $t'$ dependent on $t$ 
has not been fired, and the dependency is based mostly on the structure of the net.
However, since cycles are not allowed in RPNs, both dependency approaches are equivalent. 
We refer readers interested in cycles in RPNs and how they impact dependency between transition in 
causal reversing to~\cite{BG}. 
Notice, that the first part of the guards of reversing transitions corresponds to the choice of
the execution of $t$ to be reversed and collect all information from the 
history place related to that chosen execution. This part is fulfilled after any
execution of $t$. The same holds for out-of-causal-order semantics.\\
\\
\textbf{Case 3:} \\
Out-of-causal-order reversing. In the RPN, transition $t$ is $o-$enabled if it was executed,
i.e. its history $H(t)$ is not empty (Definition~\ref{o-enb}).
In the CPN, similarly to causal semantics case, $t$ can be reversed if token instances 
transferred by $t$ are present in places from $rin(t)$. 
For out-of-causal-order semantics,
$rin(t)$ consists of all places to which token instances 
used by $t$ could be transported. Hence, this condition is naturally fulfilled.
Then the only condition necessary to satisfy is 
the first part of the guard, which is achieved simply by any execution of $t$.\\
\\
\textbf{Part C: Reverse executions -- equivalence of markings}\\
\\
According to Definition~\ref{co-def} for causal semantics, the  
change in marking $M_{R}'$ during reversing is the same as for backtracking
semantics, which is presented in Definition~\ref{br-def}.
Moreover, the change in marking $M_{R}'$ during reversing for backtracking 
can be described by the formula for out-of-causal-order reversing - Definition~\ref{oco-def}.
In that case
let $q$ be the only output place of $t$ and
 $\alpha \in M_{R}'(t \bullet) = M_{R}'(q)$. 
We are faced with two possible situations:
  either
$\bullet t = last(C_{\alpha, q}, H_R) \bullet$ or 
$C_{\alpha,q}\subseteq M_0(\bullet t)$.
Hence, in the following paragraph, we can focus only on the formula 
presented in Definition~\ref{oco-def}.

Notice, that according to Remark~\ref{prec}, 
$t$, being the last transition according to the partial order defined by
$\prec$, is equal to the last transition indicated by Definition~\ref{last}.
If $\first{C,H}$ equals $\bot$  the maximal transition equals the initial transition $t_0$.
\\
\\
\textbf{Case 1:} 
$t_i \in T^{TRN}_R$\\
Assume, that during its execution, transition $t_i$ which is reversed has transported token instance
$\alpha$ of type $a$.
Then, $\beffect{t_i,S} = \es$ and $\lbrace \alpha \rbrace = S$. During reversing,
only the contents of two places change, namely place $q$, such that 
$\alpha \in M_{R}(q)$ 
and place $p$ such that either $p = {\first{C_{\alpha,q},  H_R'}} \bullet$ or 
$C_{\alpha, q} \in M_{{R0}}(p)$ for 
${\first{C_{\alpha,q},  H_R'}} = \bot$. We use the 
abbreviation $C_{\alpha,q} = \connected(\alpha,M_{R}(q))$ from Definition~\ref{oco-def}.
\begin{itemize}
\item $M_C'(p')= M_C(p')$ for $p' \in P_R \setminus \{p, q\}$;
\item $M_C'(q)({\cal{X}}) = 1$ for ${\cal{X}} \in ConCom(M_R(q))\setminus\connected(\alpha,M_R(q))=$ \\
$ConCom(M_R(q)\setminus \connected(\alpha,M_R(q)))=ConCom(M_R'(q))$, \\
$M_C'(q)((\es, \es)) = M_C(q)((\es, \es)) + 1 = \#ConCom(M_R(q)) + 1 = \bound - \#ConCom(M_R'(q))$;
\item $M_C'(p)({\cal{X}}) = 1$ for ${\cal{X}} \in ConCom(M_R(p))\cup\connected(\alpha,M_R(q))=$\\
$ConCom(M_R(p)\cup \connected(\alpha,M_R(q)))=ConCom(M_R'(p))$, \\
$M_C(p)((\es, \es)) = M_C'(p)((\es, \es)) - 1 = \#ConCom(M_R'(p)) - 1 = \bound - \#ConCom(M_R(p))$;
\item $M_C'(h_{jl}) = M_C(h_{jl})$ for $j\neq i, l\neq i $;
\item $M_C'(h_{il}) = M_C(h_{il})-1 = \#H_R(t_i) - 1 + \#H_R(t_l) = \#H_R'(t_i) + \#H_R'(t_l)$, for
$i<l ; t_l\in dpc(t_i)$
\item $M_C'(h_{li}) = M_C(h_{li})-1 = \#H_R(t_i) - 1 + \#H_R(t_l) = \#H_R'(t_i) + \#H_R'(t_l)$, for
$i>l ; t_l\in dpc(t_i)$
\item $M_C'(h_i) = \bigcup_{(k < k_j, t_{j}, t_{i}, \{\alpha'\}) \in M_C(h_i),
(k_j, t_{j}, t_{i}, \{\alpha\}) \in M_C(h_i)}
(k, t_{j}, t_{i}, \{\alpha'\}) \cup$\\
$\bigcup_{(k > k_j, t_{j}, t_{i}, \{\alpha'\}) \in M_C(h_i), (k_j, t_{j}, t_{i}, \{\alpha\}) \in M_C(h_i)}
(k-1, t_{j}, t_{i}, \{\alpha'\})\} = $\\
$\bigcup_{(k, \{ (p_\alpha', \alpha')\}) \in H_R'(t_i), t_l \in dpc(t_i)}
\{(\#\{k'\mid\{ (k', \{ (p_\alpha'', \alpha'')\}) \in H_R'(t_i) \cup H_R'(t_l)\}\land
k' < k \} + 1, t_l, t_i, \{ \alpha'\})\}$, where $p_{\alpha}$ denotes a place from which $\alpha$ is obtained.
The last equality holds because $H_R'$ is modified according to Definition~\ref{co-def}.
\item $M_C'(h_j) = M_C(h_j)$ for $t_j \notin dph(t_i)$
\item For $t_j \in dph(t_i)$ and $t_j \in T^{TRN}$ we have:\\
$M_C'(h_j) = \bigcup_{(k, t_{g\neq i}, t_{j}, \{\alpha'\}) \in M_C(h_j)}
(k, t_{g}, t_{j}, \{\alpha'\}) \cup$\\
$\bigcup_{(k < k_j, t_{i}, t_{j}, \{\alpha'\}) \in M_C(h_j), (k_j, t_{j}, t_{i}, \{\alpha\}) \in M_C(h_i)}
(k, t_{i}, t_{j}, \{\alpha'\}) \cup$\\
$ \bigcup_{(k > k_j, t_{i}, t_{j}, \{\alpha'\}) \in M_C(h_j), (k_j, t_{j}, t_{i}, \{\alpha\}) \in M_C(h_i)}
(k-1, t_{i}, t_{j}, \{\alpha'\})\}=$\\
$\bigcup_{(k, \{ (p_\alpha', \alpha')\}) \in H_R'(t_j), t_l \in dpc(t_j)}
\{(\#\{k'\mid\{ (k', \{ (p_\alpha'', \alpha'')\}) \in H_R'(t_j) \cup H_R'(t_l)\}\land
k' < k \} + 1, t_l, t_j, \{ \alpha'\})\}$, where $p_{\alpha}$ denotes a place from which $\alpha$ is obtained. 
The last equality holds because $H_R'$ is modified according to Definition~\ref{co-def}.
\item For $t_j \in dph(t_i)$ and $t_j \in T^{BC1} \cup T^{BC2}$ we have:\\
$M_C'(h_j) = \bigcup_{(k, t_{g\neq i}, t_{j}, \{\bondpair{\alpha_1'} {\alpha_2'}\}) \in M_C(h_j)}
(k, t_{g}, t_{j}, \{\bondpair{\alpha_1'} {\alpha_2'}\}) \cup$\\
$\bigcup_{(k < k_j, t_{i}, t_{j}, \{\bondpair{\alpha_1'} {\alpha_2'}\}) \in M_C(h_j), (k_j, t_{j}, t_{i}, \{\alpha\}) \in M_C(h_i)}
(k, t_{i}, t_{j}, \{\bondpair{\alpha_1'} {\alpha_2'}\}) \cup$\\
$ \bigcup_{(k > k_j, t_{i}, t_{j}, \{\bondpair{\alpha_1'} {\alpha_2'}\}) \in M_C(h_j), (k_j, t_{j}, t_{i}, \{\alpha\}) \in M_C(h_i)}
(k-1, t_{i}, t_{j}, \{\bondpair{\alpha_1'} {\alpha_2'}\})=$\\
$\bigcup_{(k, \{ (p_{\alpha_{1}}', {\alpha_{1}}'), (p_{\alpha_{1}}', {\alpha_{2}}')\}) \in H_R'(t_i), t_l \in dpc(t_i)}
\{(\#\{k'\mid\{ (k', \{ (p_{\alpha_{1}}', {\alpha_{1}}'), (p_{\alpha_{1}}', {\alpha_{2}}')\}) \in H_R'(t_i) \cup H_R'(t_l)\}\land
k' < k \} + 1, t_l, t_i, \{\bondpair{\alpha_1'} {\alpha_2'}\})\}$, where $p_{\alpha}$ denotes a place from which $\alpha$
is obtained.
The last equality holds because $H_R'$ is modified according to Definition~\ref{co-def}.

\end{itemize}

\textbf{Case 2:} $t_i \in T^{BC2}_R$\\
Assume, that during its execution transition $t_i$ which is reversed has 
created a bond 
$\bondpair{\alpha_1} {\alpha_2}$, where $\alpha_1$ is
of type $a_1$ and $\alpha_2$ of type $a_2$.
Then, $\beffect{t_i,S} =\{\bondpair{\alpha_1} {\alpha_2}\}$ and $\lbrace \alpha_1, \alpha_2 \rbrace = S$, $\connected(\alpha_1,M_R(q)) = \connected(\alpha_2,M_R(q))$ 
because in $M_R(q)$ $\alpha_1$ and $\alpha_2$ are bonded. 
During reversing,
only the contents of three places change, namely: 
place $q$ for $\bondpair{\alpha_1} {\alpha_2} \in M_{R}'(q)$, and place $p_1$ 
such that either $p_1 = {\first{C_{\alpha_1,q},  H_R}} \bullet$ or $C_{\alpha_1, q} \in M_{{R0}}(p_1)$ if ${\first{C_{\alpha_1,q},  H_R}} = \bot$ and place $p_2$ such that either
$p_2 = {\first{C_{\alpha_2,q},  H_R}} \bullet$ or $C_{\alpha_2, q} \in M_{{R0}}(p_2)$ if
${\first{C_{\alpha_2,q},  H_R}} = \bot$. We use the 
abbreviation $C_{\alpha,q} = \connected(\alpha,M_{R}'(q) \setminus \beffect{t_i,S})$ defined in Definition~\ref{oco-def}.
Here, markings of two places $p_1$ and $p_2$ change, in contrast to the case
$t_i \in T^{TRN}_R$ (only one place).
Since reversing of $t_i \in T^{BC2}_R$, always results in splitting 
of molecule containing the bond 
$\bondpair{\alpha_1} {\alpha_2}$ (in $M_R'$ located in place $q$) 
into two parts: the first containing $\alpha_1$, which goes back to $p_1$ in $M_R$ and 
the second containing $\alpha_2$, goes back to $p_2$ in $M_R$.
\begin{itemize}
\item $M_C'(p')= M_C(p')$ for $p' \in P_R \setminus \{p_1, p_2, q\}$;
\item $M_C'(q)({\cal{X}}) = 1$ for ${\cal{X}} \in ConCom(M_R(q))\setminus\connected(\alpha_1,M_R(q))=$ \\
$ConCom(M_R(q)\setminus \connected(\alpha,M_R(q)))=ConCom(M_R'(q))$, \\
$M_C'(q)((\es, \es)) = M_C(q)((\es, \es)) + 1 = \#ConCom(M_R(q)) + 1 = \bound - \#ConCom(M_R'(q))$;
\item $M_C'(p_1)({\cal{X}}) = 1$ for ${\cal{X}} \in ConCom(M_R(p_1))\cup\connected(\alpha_1,M_R(q) \setminus \{\bondpair{\alpha_1} {\alpha_2}\})=$\\
$ConCom(M_R(p_1)\cup \connected(\alpha_1,M_R(q) \setminus \{\bondpair{\alpha_1} {\alpha_2}\}))=ConCom(M_R'(p_1))$, \\
$M_C'(p_1)((\es, \es)) = M_C(p_1)((\es, \es)) - 1 = \#ConCom(M_R(p_1)) - 1 = \bound - \#ConCom(M_R'(p_1))$;
\item $M_C'(p_2)({\cal{X}}) = 1$ for ${\cal{X}} \in ConCom(M_R(p_2))\cup\connected(\alpha_2,M_R(q) \setminus \{\bondpair{\alpha_1} {\alpha_2}\})=$\\
$ConCom(M_R(p_2)\cup \connected(\alpha_2,M_R(q)\setminus \{\bondpair{\alpha_1} {\alpha_2}\}))=ConCom(M_R'(p_2))$, \\
$M_C'(p_2)((\es, \es)) = M_C(p_2)((\es, \es)) - 1 = \#ConCom(M_R(p_2)) - 1 = \bound - \#ConCom(M_R'(p_2))$;
\item $M_C'(h_{jl}) = M_C(h_{jl})$ for $j\neq i, l\neq i $;
\item $M_C'(h_{il}) = M_C(h_{il})-1 = \#H_R(t_i) - 1 + \#H_R(t_l) = \#H_R'(t_i) + \#H_R'(t_l)$, for
$i<l ; t_l\in dpc(t_i)$
\item $M_C'(h_{li}) = M_C(h_{li})-1 = \#H_R(t_i) - 1 + \#H_R(t_l) = \#H_R'(t_i) + \#H_R'(t_l)$, for
$i>l ; t_l\in dpc(t_i)$
\item $M_C'(h_i) = \bigcup_{(k < k_j, t_{j}, t_{i}, \{\bondpair{\alpha_1'} {\alpha_2'}\}) \in M_C(h_i),
(k_j, t_{j}, t_{i}, \{\bondpair{\alpha_1} {\alpha_2}\}) \in M_C(h_i)}
(k, t_{j}, t_{i}, \{\bondpair{\alpha_1'} {\alpha_2'}\}) \cup$\\
$\bigcup_{(k > k_j, t_{j}, t_{i}, \{\bondpair{\alpha_1'} {\alpha_2'}\}) \in M_C(h_i), (k_j, t_{j}, t_{i}, \{\bondpair{\alpha_1} {\alpha_2}\}) \in M_C(h_i)}
(k-1, t_{j}, t_{i}, \{\bondpair{\alpha_1'} {\alpha_2'}\}
)\} = $\\
$\bigcup_{(k, \{ (p_{\alpha_{1}}', {\alpha_{1}}'), (p_{\alpha_{1}}', {\alpha_{2}}')\}) \in H_R'(t_i), t_l \in dpc(t_i)}
\{(\#\{k'\mid\{ (k', \{ (p_{\alpha_{1}}', {\alpha_{1}}'), (p_{\alpha_{1}}', {\alpha_{2}}')\}) \in H_R'(t_i) \cup H_R'(t_l)\}\land
k' < k \} + 1, t_l, t_i, \{\bondpair{\alpha_1'} {\alpha_2'}\})\}$ where $p_{\alpha}$ denotes a place from which $\alpha$ is obtained.
The last equality holds because $H_R'$ is modified according to Definition~\ref{co-def}.
\item $M_C'(h_j) = M_C(h_j)$ for $t_j \notin dph(t_i)$
\item For $t_j \in dph(t_i)$ and $t_j \in T^{TRN}$ we have:\\
$M_C'(h_j) = \bigcup_{(k, t_{g\neq i}, t_{j}, \{\alpha'\}) \in M_C(h_j)}
(k, t_{g}, t_{j}, \{\alpha'\}) \cup$\\
$\bigcup_{(k < k_j, t_{i}, t_{j}, \{\alpha'\}) \in M_C(h_j), (k_j, t_{j}, t_{i}, \{\bondpair{\alpha_1} {\alpha_2}\}) \in M_C(h_i)}
(k, t_{i}, t_{j}, \{\alpha'\}) \cup$\\
$ \bigcup_{(k > k_j, t_{i}, t_{j}, \{\alpha'\}) \in M_C(h_j), (k_j, t_{j}, t_{i}, \{\bondpair{\alpha_1} {\alpha_2}\}) \in M_C(h_i)}
(k-1, t_{i}, t_{j}, \{\alpha'\})\}=$\\
$\bigcup_{(k, \{ (p_\alpha', \alpha')\}) \in H_R'(t_j), t_l \in dpc(t_j)}
\{(\#\{k'\mid\{ (k', \{ (p_\alpha'', \alpha'')\}) \in H_R'(t_j) \cup H_R'(t_l)\}\land
k' < k \} + 1, t_l, t_j, \{ \alpha'\})\}$, where $p_{\alpha}$ denotes a place from which $\alpha$ is obtained.
The last equality holds because $H_R'$ is modified according to Definition~\ref{co-def}.
\item For $t_j \in dph(t_i)$ and $t_j \in T^{BC1} \cup T^{BC2}$ we have:\\
$M_C'(h_j) = \bigcup_{(k, t_{g\neq i}, t_{j}, \{\bondpair{\alpha_1'} {\alpha_2'}\}) \in M_C(h_j)}
(k, t_{g}, t_{j}, \{\bondpair{\alpha_1'} {\alpha_2'}\}) \cup$\\
$\bigcup_{(k < k_j, t_{i}, t_{j}, \{\bondpair{\alpha_1'} {\alpha_2'}\}) \in M_C(h_j), (k_j, t_{j}, t_{i},\{\bondpair{\alpha_1} {\alpha_2}\}) \in M_C(h_i)}
(k, t_{i}, t_{j}, \{\bondpair{\alpha_1'} {\alpha_2'}\}) \cup$\\
$ \bigcup_{(k > k_j, t_{i}, t_{j}, \{\bondpair{\alpha_1'} {\alpha_2'}\}) \in M_C(h_j), (k_j, t_{j}, t_{i}, \{\bondpair{\alpha_1} {\alpha_2}\}) \in M_C(h_i)}
(k-1, t_{i}, t_{j}, \{\bondpair{\alpha_1'} {\alpha_2'}\})=$\\
$\bigcup_{(k, \{ (p_{\alpha_{1}}', {\alpha_{1}}'), (p_{\alpha_{1}}', {\alpha_{2}}')\}) \in H_R'(t_i), t_l \in dpc(t_i)}
\{(\#\{k'\mid\{ (k', \{ (p_{\alpha_{1}}', {\alpha_{1}}'), (p_{\alpha_{1}}', {\alpha_{2}}')\}) \in H_R'(t_i) \cup H_R'(t_l)\}\land
k' < k \} + 1, t_l, t_i, \{\bondpair{\alpha_1'} {\alpha_2'}\})\}$, where $p_{\alpha}$ denotes a place from which $\alpha$
is obtained.
The last equality holds because $H_R'$ is modified according to Definition~\ref{co-def}.
\end{itemize}
\textbf{Case 3:} $t_i \in T^{BC1}_R$ \\
Assume, that during its execution transition $t_i$ which is reversed, 
has created a bond 
$\bondpair{\alpha_1} {\alpha_2}$, where $\alpha_1$ is
of type $a_1$ and $\alpha_2$ of type $a_2$.
Here, two situations are possible. In the first, after reversing of $t_i$ which results 
in breaking the bond $\bondpair{\alpha_1} {\alpha_2}$, we have 
$\connected(\alpha_1,M_R(q)\setminus \{\bondpair{\alpha_1} {\alpha_2} \}) = \connected(\alpha_2,M_R(q)\setminus \{\bondpair{\alpha_1} {\alpha_2} \})$ - token instances 
$\alpha_1$ and $\alpha_2$ after breaking $\bondpair{\alpha_1} {\alpha_2}$ are still connected.
Now, only the contents of two places change, namely $q$ 
for $\bondpair{\alpha_1} {\alpha_2} \in M_{R}(q)$ 
and $p$ such that either $p = {\first{C_{\alpha_1,q},  H_R'}} \bullet$ or $C_{\alpha_1, q} \in M_{{R0}}(p)$ if
${\first{C_{\alpha,q},  H_R'}} = \bot$, where 
$C_{\alpha_1,q} = \connected(\alpha_1,M_{R}(q) \setminus \{\bondpair{\alpha_1} {\alpha_2} \})$.
This situation is analogous to $t_i \in T^{TRN}_R $.
In the second possibility, after reversing of $t_i$ which results 
in breaking bond $\bondpair{\alpha_1} {\alpha_2}$, we have 
$\connected(\alpha_1,M_R(q)\setminus \{\bondpair{\alpha_1} {\alpha_2} \}) \neq \connected(\alpha_2,M_R(q)\setminus \{\bondpair{\alpha_1} {\alpha_2} \})$ - 
molecule containing  $\alpha_1$ and $\alpha_2$ is split into two molecules,
the first containing $\alpha_1$ and the second $\alpha_2$.
This situation is analogous to $t_i \in T^{BC2}_R$.

\end{proof}

It remains for us to prove the simple fact.
\begin{lemma}
The following operations are correct:\\
$ConCom(M_R(\bullet t_i))\setminus\connected(\alpha,M_R(\bullet t_i))=ConCom(M_R(\bullet t_i)\setminus \connected(\alpha,M_R(\bullet t_i)))$\\
$ConCom(M_R(\bullet t_i))\cup\connected(\alpha,M_R(\bullet t_i))=ConCom(M_R(\bullet t_i)\cup \connected(\alpha,M_R(\bullet t_i)))$
\end{lemma}
\begin{proof}
Function $ConCom()$ returns a set of connected components of its argument. The 
molecule $\connected(\alpha,M_R(\bullet t_i))$ is a connected
component by definition. Subtraction of $\connected(\alpha,M_R(\bullet t_i))$ is an operation on
one connected component from $M_R(\bullet t_i)$ -- it removes one connected component from $M_R(\bullet t_i)$ or a part of it. Hence, 
the subtracting before or after applying $ConCom()$ function will not change the result of $ConCom()$.
Analogously for the addition of $\connected(\alpha,M_R(\bullet t_i))$.
\end{proof}

%----------------------------------------------------------
% !TEX root = translationRPN2CPN-short.tex
% Software tool description

\section{Software}

The transformation of RPNs to CPNs
described in this paper has been implemented in a 
java application \textbf{RPNEditor}, which
may be downloaded at the webpage
\mbox{\texttt{https://www.mat.umk.pl/\textasciitilde folco/rpneditor}}.

The application provides a graphical interface for displaying and edition 
of low-level RPNs based on an open source framework
\textit{Java Universal Network/Graph Framework}~\cite{jung}.
A net prepared in the application can be stored in an 
XML file containing all the details related to the net.

\begin{figure}[ht]
\vspace*{-3mm}
    \begin{center}
    \includegraphics[height=4cm]{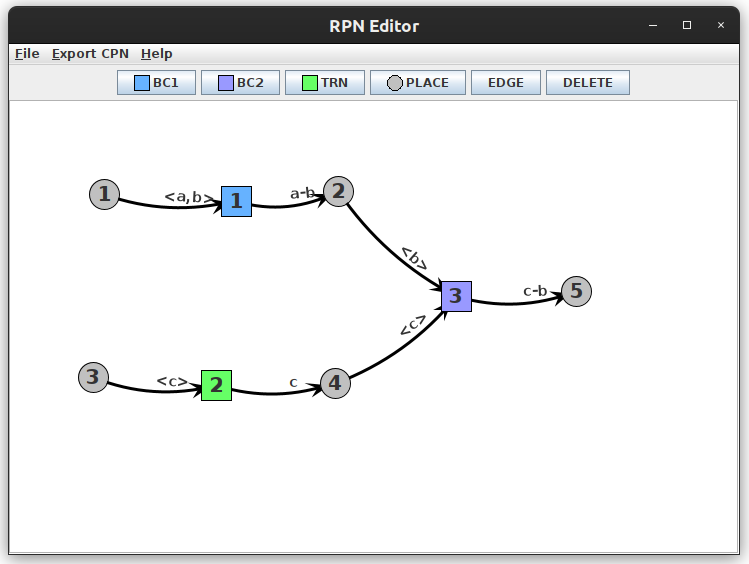}
    \hspace{0.25cm}
    \includegraphics[height=4cm]{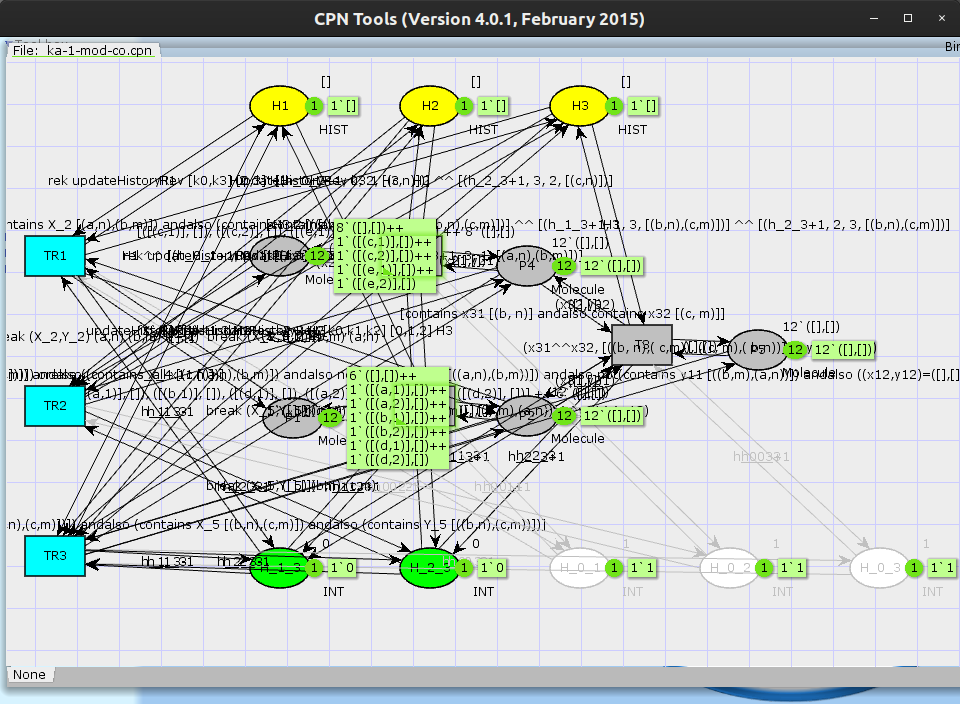}
    \end{center}
\vspace*{-6mm}
    \caption{The RPN created using RPNEditor (left)
    and its translation to CPN opened in CPN-Tools software (right).}
    \label{fig:screenshot}
\vspace*{-4mm}
\end{figure}

The primary functionality of RPNEditor
is the translation from RPNs to CPNs.
An RPN prepared in the application (alternatively loaded from
a XML file) may be
transformed into a CPN and stored in the format required
by CPN-Tools software \cite{cpn-tools}.

CPN-Tools is a software with a well-established reputation in the community
of Coloured Petri Nets.
It contains many useful tools allowing investigation of properties and
behaviour of CPNs.
Therefore, we decided to produce the output in the CPN-Tools readable format.
On the other hand, low-level RPNs are not widely known.
Moreover, construction of a correct RPN requires a few essential conditions
to be satisfied.
Hence, providing an RPN editor equipped with the net correctness validation
was necessary.

The translation from RPNs to CPNs is possible for all three reversing semantics, 
the desired one has to be chosen from the menu.
The resulting CPN behaviour corresponds to the one of the original
RPN according to the semantics chosen 
and may be directly simulated using CPN-Tools.
During the transformation,
several new structural elements are introduced: 
transition and connection history places, reversing transitions, etc.
Moreover, the simulation of the original RPN behaviour requires the usage
of a number of variables representing bases, bonds, molecules, etc.
Therefore, it is necessary to define specialised colours.
They include, among other, tuples stored at transition history places and
containing the log of transition execution, representation of bases and bonds,
lists of bases and bonds comprising a molecule, etc.

In the case of a transition with no dependence
(i.e. it does not process effects of other transitions and
its effect is not processed by any other transition)
a transition history place would remain empty.
As a consequence, the reversing of such a transition would not be possible.
Avoiding such problems was one of the reasons for introducing the initial transition
(denoted as $t_0$).
It is not a part of the original net and may be executed only once just before
the net computation starts.
Its sole purpose is to produce the initial marking.
After that neither its execution nor reversal are possible.
Since $t_0$ produces the initial content of all the places,
it is dependent with all other transitions,
namely $t_0 \in dpc(t)$ for each transition $t\in T_R$.
Therefore, as a result of the translation, the connection history places
for each transition paired with $t_0$ are produced.
To avoid creation of nodes and edges, which are not used during the actual computation,
and losing the readability
we decided to make $t_0$ 
virtual and not to put it directly in the resulting net.
Instead, the initial marking of the CPN is set to the one
just after execution of $t_0$.
However, the traces of its initial execution may be observed at the history
places of all other transitions.

The arrangement and location of all elements of the input RPN
may be freely edited by the user
resulting in the unpredictability of the processed reversing Petri net shape.
Therefore, to facilitate the readability of the resulting CPN
all the newly introduced objects are located around the RPN area.
The reversing transitions are placed to the right,
transition history places above,
and connection history places below the original net.
Moreover, we decided
to distinguish all types of objects with different colours:
the original net structure - grey, transition history places - yellow, 
connection history places - green, reversing transitions - blue,
and connection history places related to the initial transition $t_0$ -- light grey.

%In the file produced by the translation (as seen in CPN-Tools) 
%the original net structure (transitions and places) are gray,
%transition history places are yellow, connection history places are green,
%and reversing transitions are blue.
%\begin{color}{red}
%Moreover, the artificial initial transition $t_0$ and all related components
%(connection history places, arcs connecting $t_0$ with places from the original net)
%are light gray.
%\end{color}

The newly-created RPN objects (transitions and places) are automatically assigned
with unique identifiers (starting from~1).
Since one is allowed to create, move, and delete them in any order, the resulting RPN
may not preserve the topological order of transition identifiers. % required by the transformation.
Therefore, before the transformation, all transitions are renamed according to their
topological order.
Moreover, the translation is not possible before RPN is completed (i.e. all transitions
have the required number of input and output places, and all the edges have correct labels).

%----------------------------------------------------------
\section {Conclusions}
In this paper we have presented and formally proved the correctness of a translation from 
reversing Petri nets with multiple tokens to coloured Petri nets.
Building upon previous work, we have enhanced the translation by
lifting restrictions on token uniqueness and refining the transformation 
process. The resulting transformation accommodates all three semantics of backtracking, causal-order, and out-of-causal-order reversibility, 
providing a unified approach. Additionally, an  automated transformation 
algorithm for RPNs to CPNs has been introduced along with the accompanied 
modelling tool providing the potential
for analysing RPN models using CPN tools. 

Our primary objectives for future work include optimizing the 
transformation process from reversing Petri nets to coloured Petri nets,
and extending it to capture cycles as well as towards controlled
reversibility~\cite{RC19,DBLP:series/lncs/0001ACPPU20}. Furthermore,
we aim to assess the scalability of the approach and the developed tool
for handling complex systems, whereby we foresee the application of the 
framework to model and analyse case studies arising within computer science and beyond.


\begin{thebibliography}{99}

\bibitem{BGMPPP}
K.~Barylska, A.~Gogolinska, {\L}.~Mikulski, A.~Philippou, M.~Pi\k{a}tkowski, and K.~Psara.
\newblock Reversing Computations Modelled by Coloured Petri Nets.
\newblock In {\em Proceedings of ATAED 2018}, 91--111, 2018.

\bibitem{RC2022}
K.~Barylska, A.~Gogolinska, {\L}.~Mikulski, A.~Philippou, M.~Pi\k{a}tkowski, and K.~Psara.
\newblock Formal Translation from Reversing Petri Nets to Coloured Petri Nets
\newblock In {\em Proceedings of RC 2022}, LNCS 13354, pages 172--186. Springer 2022.

\bibitem{BG}
K.~Barylska and A.~Gogolinska.
\newblock Acyclic and Cyclic Reversing Computations in Petri Nets.
\newblock {\em Fundamenta Informaticae}, 184(4), 273–296, 2021.

\bibitem{PetriNets}
K.~Barylska, M.~Koutny, {\L}.~Mikulski, and M.~Pi\k{a}tkowski.
\newblock Reversible computation vs. reversibility in {P}etri nets.
\newblock {\em Science of Computer Programming}, 151:48--60, 2018.

\bibitem{BoundedPNs}
K.~Barylska, {\L}.~Mikulski, M.~Pi\k{a}tkowski, M.~Koutny, and E.~Erofeev.
\newblock Reversing transitions in bounded {P}etri nets.
\newblock {\em Fundamenta Informaticae}, 157:341--357, 2018. 

\bibitem{CardelliL11}
L.~Cardelli and C.~Laneve.
\newblock Reversible structures.
\newblock In {\em Proceedings of CMSB 2011}, pages 131--140. {ACM}, 2011.

\bibitem{cpn-tools}
\newblock CPN Tools project website, http://cpntools.org/.

\bibitem{RCCS}
V.~Danos and J.~Krivine.
\newblock Reversible communicating systems.
\newblock In {\em Proceedings of CONCUR 2004}, LNCS 3170, pages 292--307.
  Springer, 2004.

\bibitem{TransactionsRCCS}
V.~Danos and J.~Krivine.
\newblock Transactions in {RCCS}.
\newblock In {\em Proceedings of CONCUR 2005}, LNCS 3653, pages 398--412.
  Springer, 2005.


\bibitem{investigating}
D. de Frutos{-}Escrig, M. Koutny, and {\L}. Mikulski,
 \newblock Investigating Reversibility of Steps in Petri Nets
 \newblock {CoRR, abs/2110.10535},
 2021.

\bibitem{ASPtoRPNs}
Y. Dimopoulos, E. Kouppari, A. Philippou, and K. Psara.
 \newblock Encoding Reversing {P}etri Nets in Answer Set Programming
  \newblock In {\em Proceedings of {RC} 2020}, LNCS 12227, pages 264--271. Springer, 2020.
  
\bibitem{CPN}
K.~Jensen and L.~M. Kristensen.
\newblock Coloured Petri Nets - Modelling and Validation of Concurrent
  Systems.
\newblock Springer, 2009.

\bibitem{jung}
{JUNG}: {{J}ava {U}niversal {N}etwork/{G}raph {F}ramework,\\}
\newblock \texttt{http://jung.sourceforge.net/}

%\bibitem{LocalRev}
%S.~Kuhn and I.~Ulidowski.
%\newblock A calculus for local reversibility.
%\newblock In {\em Proceedings of RC 2016}, LNCS 9720, %pages 20--35. Springer,  2016.

\bibitem{DBLP:series/lncs/0001ACPPU20}
S. Kuhn, B. Aman, G. Ciobanu, A. Philippou, K. Psara, and I. Ulidowski.
\newblock Reversibility in Chemical Reactions.
\newblock {\em Reversible Computation: Extending Horizons of Computing - Selected Results of the {COST} Action {IC1405}}, LNCS 1270, pages 151--176. Springer, 2020.


%\bibitem{Landauer}
%R.~Landauer.
%\newblock Irreversibility and heat generation in the computing process.
|%\newblock {\em {IBM} Journal of Research and Development}, 5(3):183--191, 1961.

%\bibitem{LaneseLMSS13}
%I.~Lanese, M.~Lienhardt, C.~A. Mezzina, A.~Schmitt, and J.~Stefani.
%\newblock Concurrent flexible reversibility.
%\newblock In {\em Proceedings of ESOP 2013}, LNCS 7792, pages 370--390.
%  Springer, 2013.

\bibitem{LaneseMS16}
I.~Lanese, C.~A. Mezzina, and J.~Stefani.
\newblock Reversibility in the higher-order {\(\pi\)}-calculus.
\newblock {\em Theoretical Computer Science}, 625:25--84, 2016.

\bibitem{RPlaceTrans}
 H. C. Melgratti, C. A. Mezzina, and I. Ulidowski.
 \newblock Reversing Place Transition Nets.
\newblock {\em Logical Methods in Computer Science},
	16(4), 2020.

\bibitem{RON}
   H. C. Melgratti, C. A. Mezzina, I. Phillips, G. M. Pinna, and I. Ulidowski
   \newblock Reversible Occurrence Nets and Causal Reversible Prime Event Structures
    \newblock In {\em Proceedings of {RC} 2020},
	LNCS 12227, pages 35--53. Spinger 2020.
 
\bibitem{Unbounded}
{\L}. Mikulski and I. Lanese.
\newblock Reversing Unbounded {P}etri Nets.
\newblock In {\em Proceedings of {PETRI} {NETS} 2019}, {LNCS} 11522, pages {213--233}.
Springer, 2019.
 
%\bibitem{Murata}
%T.~Murata. 
%\newblock Petri nets: Properties, analysis and applications. 
%\newblock In {\em Proceedings of the IEEE 77.4}, 541--580, 1989.

\bibitem{PP}
A.~Philippou and K.~Psara.
\newblock Reversible computation in Petri nets,
\newblock In {\em Proceedings of RC 2018}, LNCS 11106, pages 84--101. Springer 2018.

\bibitem{netsWithBonds}
	A. Philippou and K. Psara,
	\newblock Reversible computation in nets with bonds
    \newblock {\em Journal of Logical and Algebraic Methods in Programming},
	124:100718, 2022.

\bibitem{ExpressPP}
A. Philippou and K. Psara.
\newblock Token Multiplicity in Reversing Petri Nets Under the Individual Token Interpretation.
\newblock In {\em Proceedings of {EXPRESS/SOS} 2022}, EPTCS 368, pages 131--150, 2022

\bibitem{coll}
A. Philippou and  K. Psara.
\newblock A collective interpretation semantics for reversing Petri nets.
\newblock {\em Theoretical Computer Science}, 924: 148-170, 2022.

\bibitem{RC19}
	A. Philippou, K. Psara, and H. Siljak.
	\newblock Controlling Reversibility in Reversing {P}etri Nets with Application to Wireless Communications
    \newblock In {\em Proceedings of {RC} 2019}, LNCS 11497, pages
	238--245.
	Springer, 2019.
 
%\bibitem{Algebraic}
%I.~Phillips and I.~Ulidowski.
%\newblock Reversing algebraic process calculi.
%\newblock In {\em Proceedings of FOSSACS 2006}, {LNCS} 3921, pages 246--260.
%  Springer, 2016.

\bibitem{phillips2007reversing}
I. Phillips and I. Ulidowski.
\newblock Reversing algebraic process calculi.
\newblock {\em Journal of Logic and Algebraic Programming}, 73(1-2):70--96.
Elsevier, 2007.

%\bibitem{ERK}
%I.~Phillips, I.~Ulidowski, and S.~Yuen.
%\newblock A reversible process calculus and the modelling of the {ERK}
%  signalling pathway.
%\newblock In {\em Proceedings of RC 2012}, {LNCS} 7581, pages
%  218--232. Springer, 2012.

%\bibitem{Bonding}
%I.~Phillips, I.~Ulidowski, and S.~Yuen.
%\newblock Modelling of bonding with processes and events.
%\newblock In {\em Proceedings of RC 2013}, {LNCS} 7947, pages 141--154. Springer,
%  2013.
  
\bibitem{KP-2020}
K. Psara.
\newblock Reversible Computation in Petri Nets
\newblock PhD Thesis, Department of Computer Science, University of Cyprus, 2020.
\bibitem{CPNtools}
A.~V. Ratzer, L.~Wells, H.~M. Lassen, M.~Laursen, J.~F. Qvortrup, M.~S.
  Stissing, M.~Westergaard, S.~Christensen, and K.~Jensen.
\newblock {CPN} tools for editing, simulating, and analysing coloured {P}etri
  nets.
\newblock In {\em Proceedings of
  {ICATPN} 2003}, {LNCS} 2679, pages 450--462. Springer, 2003.

\bibitem{PN}
W.~Reisig.
\newblock {\em Understanding Petri Nets - Modeling Techniques, Analysis
  Methods, Case Studies}.
\newblock Springer, 2013.

\bibitem{ConRev}
I.~Ulidowski, I.~Phillips, and S.~Yuen.
\newblock Concurrency and reversibility.
\newblock In {\em Proceedings of RC 2014}, LNCS 8507, pages 1--14. Springer,
  2014.
  

\end{thebibliography}
\end{document}